\documentclass[fleqn,usenatbib]{mnras}
\usepackage{newtxtext,newtxmath}

\usepackage[T1]{fontenc}
\usepackage{ae,aecompl}

\usepackage{graphicx}
\usepackage{amsmath}
\usepackage{siunitx}[=v2]
\usepackage{orcidlink}

\newcommand{\HI}{\text{H\textsc{i}}}
\usepackage{url}
\usepackage{caption}
\usepackage[normalem]{ulem}
\usepackage[usenames, dvipsnames]{xcolor}

\title{DOWSING, Dust Observations With SCUBA-2 In Nearby Galaxies: I. Dust in M\,81} % Max 45 characters

\author[Uzuner et al.]{
M. Uzuner$^1$\thanks{E-mail: UzunerM@cardiff.ac.uk}\orcidlink{0000-0002-4387-7684},
M. W. L. Smith$^1$\orcidlink{0000-0002-3532-6970},
S. Eales$^1$\orcidlink{0009-0003-0635-0175},
A. Chung$^{2}$\orcidlink{0000-0003-1440-8552},
E. J. Chung$^{3}$\orcidlink{0000-0003-0014-1527},
D. L. Clements$^{4}$\orcidlink{0000-0002-9548-5033},
\newauthor
I. De Looze$^{5}$,
D. J. Eden$^{6,7}$,
L. Fanciullo$^{8}$\orcidlink{0000-0001-9930-9240},
L. C. Ho$^{9,10}$\orcidlink{0000-0001-6947-5846},
F. Kemper$^{11,12,13}$\orcidlink{0000-0003-2743-8240},
K. Pattle$^{14}$\orcidlink{0000-0002-8557-3582},
\newauthor
M. Rela\~no$^{15,16}$\orcidlink{0000-0003-1682-1148},
A. Saintonge$^{17,18}$\orcidlink{0000-0003-4357-3450},
X. Tang$^{19}$\orcidlink{0000-0002-4154-4309},
S. A. Van Der Giessen$^{20}$\orcidlink{0009-0001-4144-9635},
J. Wang$^{21}$\orcidlink{0000-0003-4874-0369},
\newauthor
T. G. Williams$^{22}$,
C. D. Wilson$^{23}$
\\
$^1$ Cardiff Hub for Astrophysics Research \& Technology, School of Physics \& Astronomy, Cardiff University, Queens Buildings, Cardiff CF24 3AA, UK\\
$^{2}$ Department of Astronomy, Yonsei University, 50 Yonsei-ro, Seodaemun-gu, Seoul, 03722, Republic of Korea\\
$^3$ Korea Astronomy and Space Science Institute, 776 Daedeokdae-ro, Yuseong-gu, Daejeon 34055, Republic of Korea\\
$^{4}$ Blackett Lab, Imperial College London, Prince Consort Road, London SW7 2AZ, UK\\
$^{5}$ Department of Physics and Astronomy, Universiteit Gent, Proeftuinstraat 86 N3, B-9000 Ghent, Belgium\\
$^{6}$ Global Banking School, Universal Square, Devonshire Street North, Manchester, M12 6JH, UK\\
$^{7}$ Armagh Observatory and Planetarium, College Hill. Armagh, BT61 9DB, UK\\
$^{8}$ Tamkang University, 151 Yingzhuan Rd., Tamsui Dist., New Taipei City 251301, Taiwan, R.O.C.\\
$^{9}$ Kavli Institute for Astronomy and Astrophysics, Peking University, Beijing 100871, China\\
$^{10}$ Department of Astronomy, School of Physics, Peking University, Beijing 100871, China\\
$^{11}$ Institut de Ciències de l'Espai (ICE, CSIC), Can Magrans, s/n, E-08193 Cerdanyola del Vallès, Barcelona, Spain\\
$^{12}$ ICREA, Pg. Lluís Companys 23, E-08010 Barcelona, Spain\\
$^{13}$ Institut d'Estudis Espacials de Catalunya (IEEC), E-08860 Castelldefels, Barcelona, Spain\\
$^{14}$ Department of Physics and Astronomy, University College London, Gower Street, London WC1E 6BT, UK\\ % Kate Pattle
$^{15}$ Dept. Física Teórica y del Cosmos, E-18071 Granada, Spain\\ % M. Relano
$^{16}$ Instituto Universitario Carlos I de Física Te\'{o}rica y Computacional, Universidad de Granada, 18071, Granada, Spain\\ % M. Relano
$^{17}$ Department of Physics \& Astronomy, University College London, London, WC1E 6BT, UK\\ % A.Saintonge
$^{18}$ Max Planck Institute for Radio Astronomy, Auf dem H\"{u}gel 69, 53121 Bonn, Germany\\ % A.Saintonge
$^{19}$ Xinjiang Astronomical Observatory, Chinese Academy of Sciences, Urumqi 830011, PR China\\
$^{20}$ Université Paris-Saclay, Université Paris Cité, CEA, CNRS, AIM, 91191, Gif-sur-Yvette, France\\
$^{21}$ Department of Astronomy, Xiamen University, Xiamen, Fujian 361005, China\\ % J. Wang
$^{22}$ UK ALMA Regional Centre Node, Jodrell Bank Centre for Astrophysics, Department of Physics and Astronomy, The University of Manchester,\\ Oxford Road, Manchester M13 9PL, UK\\
$^{23}$ Department of Physics \& Astronomy, McMaster University, 1280 Main Street W, Hamilton, ON, L8S 4M1, Canada
}

\date{Accepted XXX. Received YYY; in original form ZZZ}
\pubyear{\the\year{}}
\begin{document}
\label{firstpage}
\pagerange{\pageref{firstpage}-\pageref{lastpage}}
\maketitle
%!!!!!!!!!!!!!!!!!!!!!!!!!!!!!!!!!!!!!!!!!!!!!!!

%% ABSTRACT
\begin{abstract}
We present a resolved far-infrared/sub-millimetre study of dust in the nearby spiral galaxy M\,81 using \textit{Herschel} PACS/SPIRE and JCMT SCUBA-2 observations from 70 to \SI{850}{\um}. We model the dust emission using both single-temperature modified blackbody (SMBB) fits and point process mapping (PPMAP), comparing pixel-by-pixel SED-fitting with image cubes of dust surface density as a function of position and temperature. For fixed $\beta$\,=\,1.9 and a dust opacity of $\kappa_{\SI{350}{\um}}$\,=\,1.92\,$\mathrm{cm^2\,g^{-1}}$, SMBB gives an integrated dust mass of $(1.77\pm0.09)\times10^{7}M_\odot$, while PPMAP gives $(2.65\pm0.13)\times10^{7}M_\odot$. The higher PPMAP mass is expected because real lines of sight contain dust over a range of temperatures, whereas SMBB assumes a single effective temperature and can underestimate the dust mass. Both methods show enhanced dust surface densities along the spiral arms and a radially declining dust temperature, with the warmest dust in the central region. Variable-$\beta$ fits indicate only moderate spatial variation in $\beta$ for well-constrained pixels, but the strong $T$--$\beta$ degeneracy motivates a fixed emissivity index for the main analysis. Combining the dust maps with \HI\ data, we find a positive radial gas-to-dust ratio gradient, with $d\log_{10}({\rm GDR})/dR=0.065\,{\rm dex\,kpc^{-1}}$. Comparison with M81's metallicity gradient implies a declining dust-to-metals ratio, potentially consistent with more efficient grain growth in the inner disc. The star formation rate is modest, $\sim$0.27\,M$_\odot\,{\rm yr}^{-1}$, and is strongly concentrated along the spiral arms. These results suggest that the dust, gas, and star-forming structures of M\,81 are strongly shaped by its spiral arms, while the central region hosts the warmest dust in the disc.
\end{abstract}

%% KEYWORDS: 

\section{Introduction}

Dust in the interstellar medium (ISM) is crucial for various astrophysical processes, acting as a coolant in star-forming regions, heating the gas in photodissociation regions \citep{Draine1978}, facilitating molecular hydrogen formation \citep{Gould&Salpeter1963}, and indirectly supporting star and planet formation. Dust particles are produced from stellar outflows and supernovae, and they absorb and re-radiate about half of the universe's emitted energy within the far-infrared (FIR) and sub-millimetre \citep[sub-mm;][]{Lagache2005}, making this wavelength range essential for studying the cold universe.

Continuum studies of FIR and sub-mm emission in nearby galaxies face strong observational limitations. Over most of the FIR regime the atmosphere is effectively opaque, making ground-based imaging impractical and space missions essential for sensitive surveys. In the sub-mm, by contrast, a small number of atmospheric windows permit ground-based observations; however, they remain technically demanding because atmospheric transmission is poor at these wavelengths, the sky background is bright and spatially extended, and low-surface-brightness discs require long integrations and careful removal of large-scale atmospheric fluctuations. Space-based facilities have therefore played a central role in building large FIR-to-sub-mm datasets. In particular, the \textit{Herschel Space Observatory} \citep{Pilbratt2010} provided sensitive imaging from 70 to \SI{500}{\um}, substantially improving wavelength coverage over earlier space missions. It also delivered higher angular resolution, for example \SI{6}{\arcsec} at \SI{70}{\um} with PACS compared to \SI{18}{\arcsec} with \textit{Spitzer} MIPS. These capabilities also contributed to legacy data sets such as the DustPedia \citep{Clark2026} database \citep{Clark2018}, which compiles multiwavelength observations of 875 nearby galaxies from a range of surveys. Nonetheless, although \textit{Herschel} transformed FIR studies of nearby galaxies, its wavelength coverage ends at \SI{500}{\um}, and the cold-dust-sensitive higher-wavelength SPIRE bands have poorer angular resolution than the shorter-wavelength PACS bands. These limitations motivate complementary ground-based sub-mm and mm observations \citep{HASHTAG1_2020Li, Katsioli2023, Pantoni2026}.

\textit{Herschel} alone provides limited leverage on the Rayleigh-Jeans tail of the dust spectral energy distribution (SED), which in the optically thin limit is commonly approximated by a modified blackbody, $S_{\nu}$\,$\propto$\, $B_{\nu}(T_d)$\,$\times$\,$\nu^{\beta}$ where $\beta$ is the dust emissivity index that controls the long-wavelength slope. Longer-wavelength sub-mm data at \SI{850}{\um} from instruments such as the James Clerk Maxwell Telescope's (JCMT) SCUBA-2 camera \citep{Holland2013} extend the wavelength coverage beyond \textit{Herschel}'s \SI{500}{\um} band, helping to constrain the long-wavelength slope more precisely and improving estimates of dust mass and temperature. This enables mapping of cold-dust properties on sub-kpc scales, even in galaxies beyond the Local Group, using pixel-by-pixel SED fitting. This approach reveals how dust physics and the interstellar radiation field (ISRF) vary across different galactic environments \citep[e.g.][]{KINGFISH,Dustpedia}. 

Radial variations in dust properties within nearby galaxies are shown in multiple resolved studies \citep{Smith2012, Tabatabaei2014, Kirkpatrick2014, Galametz2016, Relano2018, Chiang2018, Vilchez2019, vanderGiessen2024, Park2024}, but the physical origin of these variations remains uncertain. For example, radial analyses of the KINGFISH sample demonstrate that the dust temperature or equivalently the strength of the ISRF heating the majority of the dust generally decline with galactocentric radius, while the apparent emissivity index $\beta$ shows substantial diversity among galaxies, including decreasing, increasing, and non-monotonic radial behaviour \citep{Hunt2015}. Importantly, \citet{Hunt2015} argue that radial trends in $\beta$ do not uniquely imply intrinsic changes in grain emissivity: physically motivated dust models with fixed intrinsic emissivities can reproduce apparently shallow long-wavelength slopes, because temperature mixing and variations in the ISRF intensity distribution broaden the SED and can bias single-temperature modified blackbody (SMBB) fits toward low apparent $\beta$, particularly in cool outer discs where the ISRF intensity is low. Consequently, spatial variations in $\beta$ are likely to reflect a combination of changing radiation-field conditions and genuine radial changes in grain chemical/physical properties (e.g. composition, structure, mantling) driven by gradients in ISM conditions, but the relative importance of these effects remains uncertain.

We address this through the \textit{Dust Observations with SCUBA-2 in Nearby Galaxies} (DOWSING) survey, a new programme targeting 12 nearby, large-angular-size galaxies with existing \textit{Herschel} continuum observations. The SCUBA-2 data complement these observations by providing higher-resolution imaging at \SI{450}{\um} and extending the wavelength coverage to \SI{850}{\um}, with effective beam FWHMs of approximately \SI{9.8}{\arcsec} and \SI{14.6}{\arcsec} \citep{Dempsey2013}, respectively. The main objectives of the survey are to study the causes of variations in the $\beta$ parameter, to measure the variation in the gas-to-dust ratio and the CO-to-H$_2$ conversion factor in each individual galaxy, and to investigate the presence of CO-dark molecular gas (H$_2$ that cannot be traced by CO, as expected where H$_2$ survives but CO is photodissociated; \citealt{Wolfire2010}) in high-mass galaxies beyond the Local Group, where its contribution remains less well-constrained than in the Milky Way \citep{Grenier2005, Planck2011, Langer2014}. SCUBA-2 has been obtaining deep images of these galaxies at 450 and \SI{850}{\um} since 2020, and $\sim$60\% of the survey has been completed. When the survey is finished, SCUBA-2 images in combination with available \textit{Herschel} images will be used to create high-resolution dust maps of these 12 galaxies. 

This paper presents the first DOWSING results: resolved dust maps of M\,81 (NGC\,3031), a nearby grand-design spiral of Hubble type SA(s)ab \citep{Freedman1994} at a distance of D\,=\,3.63\,Mpc \citep{Freedman2001}. By combining 70-\SI{500}{\um} \textit{Herschel} data and 450 and \SI{850}{\um} SCUBA-2 data, we derive pixel-by-pixel maps of dust surface density $\Sigma_{\mathrm{d}}$, dust temperature $T_{\mathrm{d}}$, and emissivity index $\beta$, and investigate how dust properties vary in M\,81. The dust properties in M\,81 have been previously studied using \textit{Herschel} images by \citet{Bendo2010}, \citet{Davies2010}, and \citet{Bendo2012}. However, a limitation of \textit{Herschel}-only studies is the weak determination of the Rayleigh-Jeans tail of the SED, which introduces degeneracies between dust temperature and emissivity index and can hide very cold dust. Adding long-wavelength data (like \SI{850}{\um} in our work) simultaneously helps to find the SED slope better and improves dust mass estimates, while still offering the same angular resolution as \textit{Herschel}-\SI{250}{\um} band. In addition, we investigate the application of Point Process MAPping (PPMAP; \citealt{Marsh2015}), which is not limited by the need to convolve all input data to a common resolution. As a result, the derived dust-property maps can achieve substantially higher angular resolution, approaching that of the PACS \SI{70}{\um} data in our study. Section\,\ref{sec:Data} summarises the observations and data reduction. Section\,\ref{sec:SED_fitting} details the SED modelling. Section\,\ref{sec:PPMAP} introduces the PPMAP method and presents the corresponding PPMAP-derived $\Sigma_{\mathrm{d}}$ and $T_{\mathrm{d}}$ maps. Section\,\ref{sec:Discussion} presents and discusses the results, and the conclusions are provided in Section\,\ref{sec:Conclusion}.

\section{Observation and Data Reduction} 
\label{sec:Data}

\subsection{SCUBA-2 450 and \SI{850}{\um} data}

The DOWSING sample consists of nearby galaxies from the DustPedia catalogue \citep{Dustpedia}, which were selected to have \textit{Herschel} observations, velocities v\,<\,3000\,km\,s$^{-1}$, large optical angular sizes ($D_{25}$\,>\,\SI{1}{\arcmin}), and $\ge$\,5$\sigma$ \textit{Wide-field Infrared Survey Explorer} (\textit{WISE}) detections. The targets were chosen to have the largest apparent dust discs that are observable with the JCMT, excluding Local Group systems M\,31 and M\,33, for which deep SCUBA-2 observations already exist \citep{2019Williams, Smith2021_HASHTAG}, and the Small and Large Magellanic Clouds, which are not observable from the JCMT site. SCUBA-2 observing time was awarded for 12 galaxies, six of which were fully observed before the current pause in SCUBA-2 operations. Here we present the analysis of one of the first DOWSING targets, M\,81.

Sub-mm continuum maps of M\,81 were obtained with the SCUBA-2 instrument \citep{Holland2013} on the 15\,m JCMT at \SI{450}{\um} (FWHM\,$\simeq$\,\SI{9.8}{\arcsec}) and \SI{850}{\um} (FWHM\,$\simeq$\,\SI{14.6}{\arcsec}). We executed 2\,$\times$\,\SI{30}{\arcmin} PONG scans in JCMT band-2 weather (where the opacity is 0.05\,<\,$\tau_{225}$\,<\,0.08), reaching a 1$\sigma$ rms noise level of 3\,mJy\,beam$^{-1}$ (at \SI{850}{\um}) and 44\,mJy\,beam$^{-1}$ (at \SI{450}{\um}), comparable to the depths of the M\,31 \citep{Smith2021_HASHTAG} and M\,33 \citep{2019Williams} legacy surveys. All observation areas were selected to ensure sufficient background region for effective data reduction. To reduce the raw SCUBA-2 data and create maps, we followed the steps in \cite{Smith2021_HASHTAG}. Final maps were created with pixel sizes of \SI{2}{\arcsec} and \SI{4}{\arcsec} for the 450 and \SI{850}{\um} maps, respectively.

A major challenge for ground-based sub-mm observations with SCUBA-2 is the removal of spatially and temporally variable atmospheric emission. This is generally solved by Fourier filtering; however, this method causes a loss in real signals on large angular scales (the low-k Fourier components). Space telescopes do not suffer from the same atmospheric emission variations, but they lack the angular resolution that we have in ground-based telescopes. To recover the large-scale emission filtered out of the ground-based maps while keeping the SCUBA-2 resolution, we used the "feathering" technique \citep{Bajaja1979, Clark2021}. In this approach, the maps are combined in Fourier space so that the satellite data provide the extended emission that is filtered out of the SCUBA-2 observations, while the SCUBA-2 data provide the finer structure at higher angular resolution. Following \cite{Smith2021_HASHTAG}, we feathered the SCUBA-2 \SI{450}{\um} map with the corresponding \textit{Herschel} \SI{500}{\um} map, and the SCUBA-2 \SI{850}{\um} map with the \textit{Planck} 353\,GHz ($\sim$\SI{850}{\um}) map.

\subsection{Ancillary data}

\begin{figure}
\centering
\includegraphics[width=\columnwidth]{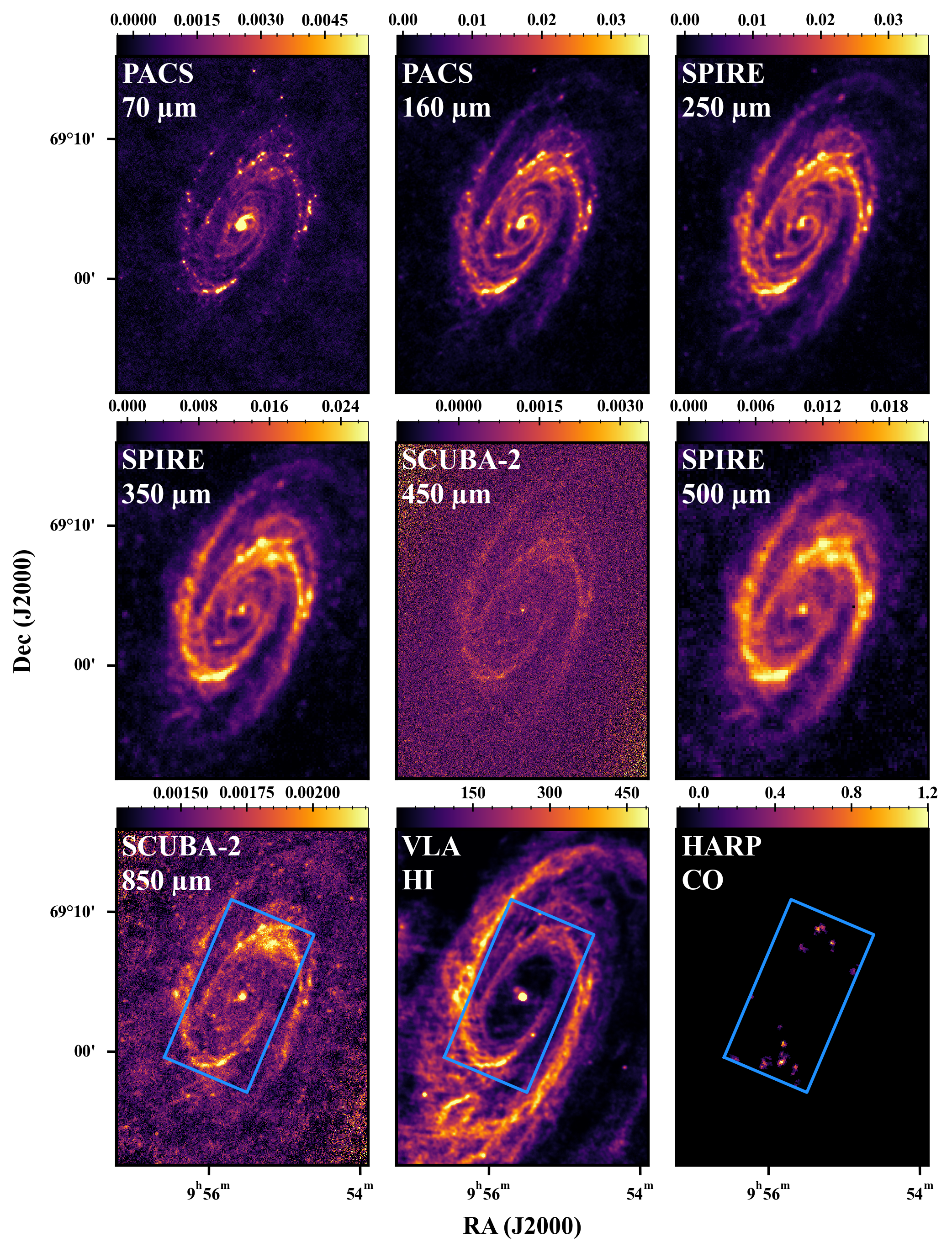}
\caption{M\,81 maps at their native angular resolutions. The 70 and \SI{160}{\um} maps are from \textit{Herschel}/PACS, while the 250, 350, and \SI{500}{\um} maps are from \textit{Herschel}/SPIRE. The 450 and \SI{850}{\um} maps were obtained with JCMT/SCUBA-2 as part of the DOWSING Survey. The \HI\ map is taken from \citet{Walter2008}, and the CO(\textit{J}\,=\,3\,--\,2) map shows the integrated-intensity map produced using a $3\sigma$ clipping level \citep{Sanchez-Gallego2011}. The blue rectangular region of approximately \SI{13.2}{\arcmin}\,$\times$\,\SI{7.7}{\arcmin} represents where CO(\textit{J}\,=\,3\,--\,2) was mapped by \citet{Sanchez-Gallego2011}. The same area is also plotted on the \SI{850}{\um} and \HI\ maps for comparison. The Herschel and SCUBA-2 maps are in units of $\mathrm{Jy\,pixel^{-1}}$, the \HI\ map is in $\mathrm{Jy\,beam^{-1}\,m\,s^{-1}}$, and the CO map is in $\mathrm{K\,km\,s^{-1}}$. All maps are shown over a common 0.4$^\circ$\,$\times$\,0.3$^\circ$ field centred on RA\,=\,9h\,55m\,33.17s, Dec\,=\,+69$^\circ$\,03\arcmin\,55.1\arcsec.}
\label{fig:all_data}
\end{figure}

In addition to the new SCUBA-2 observations, we make use of archival multiwavelength imaging of M\,81. From the DustPedia database \citep{Dustpedia} we obtain \textit{Herschel}/PACS \citep{Poglitsch2010} 70 and \SI{160}{\um} maps, \textit{Herschel}/SPIRE \citep{Griffin2010} 250, 350, and \SI{500}{\um} maps, \textit{GALEX} \citep{Galex} FUV and NUV images, and \textit{Spitzer} IRAC \citep{Spitzer-IRAC} \SI{3.6}{\um} and MIPS \citep{Spitzer-MIPS} \SI{24}{\um} maps. We also use an H\textsc{i} map \citep{Walter2008} and a CO(\textit{J}\,=\,3\,--\,2) map \citep{Sanchez-Gallego2011} of M\,81 from the literature to constrain the neutral gas distribution. The maps used for the dust and gas analysis are shown in Figure\,\ref{fig:all_data} after being aligned with the galactic centre.

The PACS, SPIRE, \textit{GALEX}, and \textit{Spitzer} maps are taken directly from DustPedia \citep{Dustpedia}, which provide science-ready images with consistent backgrounds, astrometry, and photometric calibration. Full details of the original observations and reduction pipelines can be found in \citet{Clark2018}. 

For the atomic gas, we use the H\textsc{i} map of M\,81 (FWHM of $\sim$\SI{12.9}{\arcsec}\,$\times$\,\SI{12.4}{\arcsec}) from \citet{Walter2008}, observed with the Very Large Array \citep{Thompson1980}. For the molecular gas, we use the JCMT/HARP CO(\textit{J}\,=\,3\,--\,2) map from the Nearby Galaxies Legacy Survey \citep[NGLS;][]{Sanchez-Gallego2011, Wilson2012}, which has an angular resolution of \SI{14.5}{\arcsec}. We use the CO map constructed from the datacube using a sinc-sinc kernel and a $3\sigma$ clipping threshold, as delivered by the NGLS processing \citep{Sanchez-Gallego2011}. 

Since the SCUBA-2 \SI{850}{\um} bandpass includes the $^{12}$CO\,(\textit{J}\,=\,3\,--\,2) line at 345.796\,GHz, we correct the \SI{850}{\um} data for possible line contamination following \citet{Drabek2012} and \citet{Parsons2018}. The CO(\textit{J}\,=\,3\,--\,2) integrated intensity map is reprojected onto the SCUBA-2 \SI{850}{\um} grid, and an observation-specific line-to-continuum conversion factor was calculated from the median atmospheric opacity using the relation of \citet{Parsons2018}. The resulting contamination map is then subtracted from the \SI{850}{\um} continuum map to obtain the uncontaminated \SI{850}{\um} map. Across the disc of M\,81 the CO(\textit{J}\,=\,3\,--\,2) contribution is generally small: the median contamination is $\sim$0.6\% of the observed \SI{850}{\um} flux, with a maximum of $\sim$2\% in the brightest CO peaks. The correction is therefore only significant in a small number of compact regions; only $\leq$\,0.1\% of \SI{850}{\um} pixels have an estimated CO contamination >1\%.

\section{Spectral Energy Distribution Modelling}
\label{sec:SED_fitting}

Dust grains are not perfect blackbodies and they do not emit radiation as efficiently as a blackbody at all wavelengths. We therefore describe the dust thermal emission with an SMBB model \citep{Hildebrand1983}, in which the observed flux density $F_\nu$ is related to the dust mass in each pixel via
\begin{equation}
    F_{\nu} = \frac{M_{\rm d}}{D^2}\,\kappa_\nu\,B_\nu(T_{\rm d}),
    \label{eq:SMBB}
\end{equation}
where $M_{\rm d}$ is the dust mass in the pixel, $D$ is the distance to the galaxy, $\kappa_{\nu}$ is the dust mass absorption (opacity) coefficient with emissivity index $\beta$, and $B_\nu(T_{\rm d})$ is the Planck function at dust temperature $T_{\rm d}$. The dust mass absorption coefficient is often approximated as a power law in frequency,
\begin{equation}
    \kappa_\nu = \kappa_{\nu_0}\left(\frac{\nu}{\nu_0}\right)^{\beta},
    \label{eq:kappa}
\end{equation}
where $\kappa_{\nu_0}$ is the reference opacity at frequency $\nu_0$. We assume a distance to M\,81 of D\,=\,3.63\,Mpc \citep{Freedman2001} and fit this model to the SED of each individual pixel.

In SMBB fits, one of the dominant systematic uncertainties is $\kappa_\nu$, which links the observed FIR/sub-mm luminosity to the inferred dust mass. $\kappa_\nu$ is set by the grains' composition, density, size distribution and surface-to-volume ratio, but these physical properties can vary between environments and are difficult to measure directly. As a result, published estimates of $\kappa_\nu$ show substantial spread, depending on wavelength, grain properties, and derivation method \citep{Alton2004, Clark2019}. Constraints on $\kappa_\nu$ can be obtained from mineralogical models \citep{Jones2013}, from extinction-to-emission comparisons in galactic clouds or nearby galaxies \citep{Kim1994}, or from laboratory studies of pre-solar grain analogues \citep{Mutschke2013}, however, each approach requires assumptions about grain properties and radiative transfer. Therefore, SMBB dust masses should be interpreted with the adopted opacity in mind, particularly when comparing inferred dust masses between studies. In addition, SMBB fits assume a single dust temperature along each line of sight; in reality, dust is exposed to a range of ISRFs, so temperature mixing and the resulting $T$\,--\,$\beta$ degeneracy introduce additional systematic uncertainty in the inferred dust temperatures, emissivity indices and dust masses.

\begin{figure}
\centering
\includegraphics[width=\linewidth]{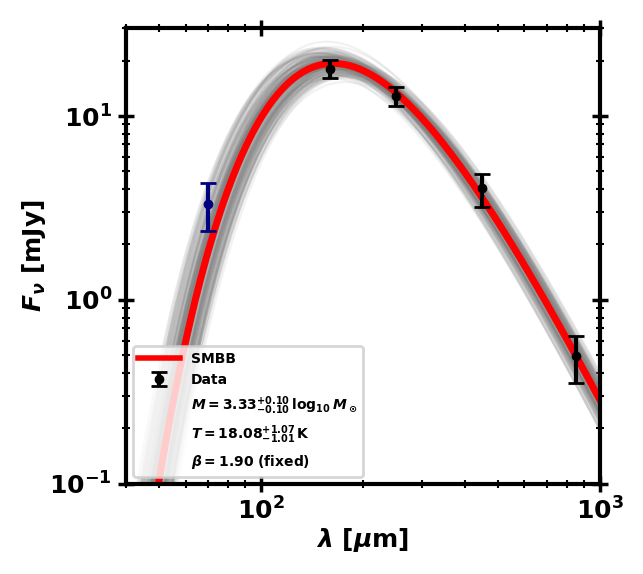}
\caption{Example SED for a pixel in the arm of M\,81. The blue data point at \SI{70}{\um} was used as an upper limit in the fitting process. The red line is the best-fit (median) model. Error bars include calibration uncertainties following \citealt{Ciesla2013}.}
\label{fig:SED_Example}
\end{figure}

Because $\kappa_\nu$ in Equation\,\ref{eq:SMBB} is a multiplicative factor, the poorly known value of $\kappa_\nu$ primarily affects the overall scale of the derived dust masses. Adopting a different $\kappa_\nu$ would rescale all masses by approximately a constant factor, while leaving relative spatial variations in $\Sigma_{\rm d}$ and correlations with other quantities (e.g. gas surface density or $\Sigma_{\rm SFR}$) essentially unchanged. For ease of comparison with previous studies, including \citet{Smith2012}, \citet{Smith2016}, and \citet{Pattle2023}, we adopt $\kappa_{\SI{350}{\um}}$\,=\,1.92\,$\mathrm{cm^2\,g^{-1}}$, following \citet{Draine2003}.

We performed the variable-$\beta$ SMBB fitting at two common resolutions. The first set of fits was carried out at the resolution of the \SI{250}{\um} SPIRE map. In this case, the 70, 160, 250, 450, and \SI{850}{\um} maps were convolved to the SPIRE \SI{250}{\um} PSF \citep[see][]{Aniano2011} and reprojected onto the SPIRE \SI{250}{\um} pixel grid. The \SI{350}{\um} and \SI{500}{\um} SPIRE maps were not included in this fit because their native beams are broader than the SPIRE \SI{250}{\um} beam. Each pixel was fitted independently; however, adjacent pixels are not independent measurements because the SPIRE \SI{250}{\um} pixel size (\SI{6}{\arcsec}) oversamples the \SI{18.2}{\arcsec} beam. For each band, a single background level was estimated from regions surrounding the galaxy and subtracted from the full map. The flux uncertainty in each pixel is calculated by combining the standard deviation of pixel values in the background regions with the calibration uncertainties. These calibration uncertainties include both an independent random term and a term that is correlated across bands. In most pixels, the uncertainty budget is dominated by the calibration terms. We adopted calibration uncertainties of 7.0\% for PACS and 5.5\% for SPIRE, following \citet{Clark2018}, and 10.0\% and 5.0\% for SCUBA-2 at \SI{450}{\um} and \SI{850}{\um}, respectively \citep{Dempsey2013}.

We fitted an SMBB model for each pixel with SNR\,$\ge$\,5 in at least one band. The fitting procedure is a Bayesian framework using PyMC, including explicit colour corrections and band-to-band correlated uncertainties (Smith et al., in prep.). Although the FIR/sub-mm SEDs are generally well described by an SMBB at wavelengths $\lambda$\,$\geq$\,\SI{160}{\um}, the \SI{70}{\um} band can include emission from warmer dust and stochastically heated small grains \citep{DraineLi2001_StochasticHeating}, which are not represented by the cold-dust SMBB component. In some regions, this can produce an excess relative to the extrapolation of the longer-wavelength fit, as also discussed by \citet{Bendo2010} and seen in other nearby galaxy studies (e.g. \citealt{Smith2010}). We therefore treat the \SI{70}{\um} flux as an upper limit in the SMBB fitting, rather than forcing the cold-dust model to reproduce this band. A two-temperature MBB model is not adopted here because the available maps do not provide sufficient short-wavelength coverage to constrain an additional warm component. An example SED is shown in Figure\,\ref{fig:SED_Example}.

In the first run, the emissivity index $\beta$ was allowed to vary, with three free parameters fitted in each pixel: dust mass, dust temperature, and $\beta$. We adopt weakly informative priors: a Student's t-distribution prior with $\nu$\,=\,8 and scale $\sigma$\,=\,1 for the dust mass parameter (centred automatically on the SED normalisation), a Gamma prior on the temperature with mode 12.5\,K and standard deviation 6.0\,K, and a Gamma prior on the emissivity index with mode $\beta$\,=\,2.0 and standard deviation 0.5. Posterior sampling is performed with Markov Chain Monte Carlo, and for each pixel we take the posterior median of each parameter as the point estimate.

\begin{figure}
\centering
\includegraphics[width=\linewidth]{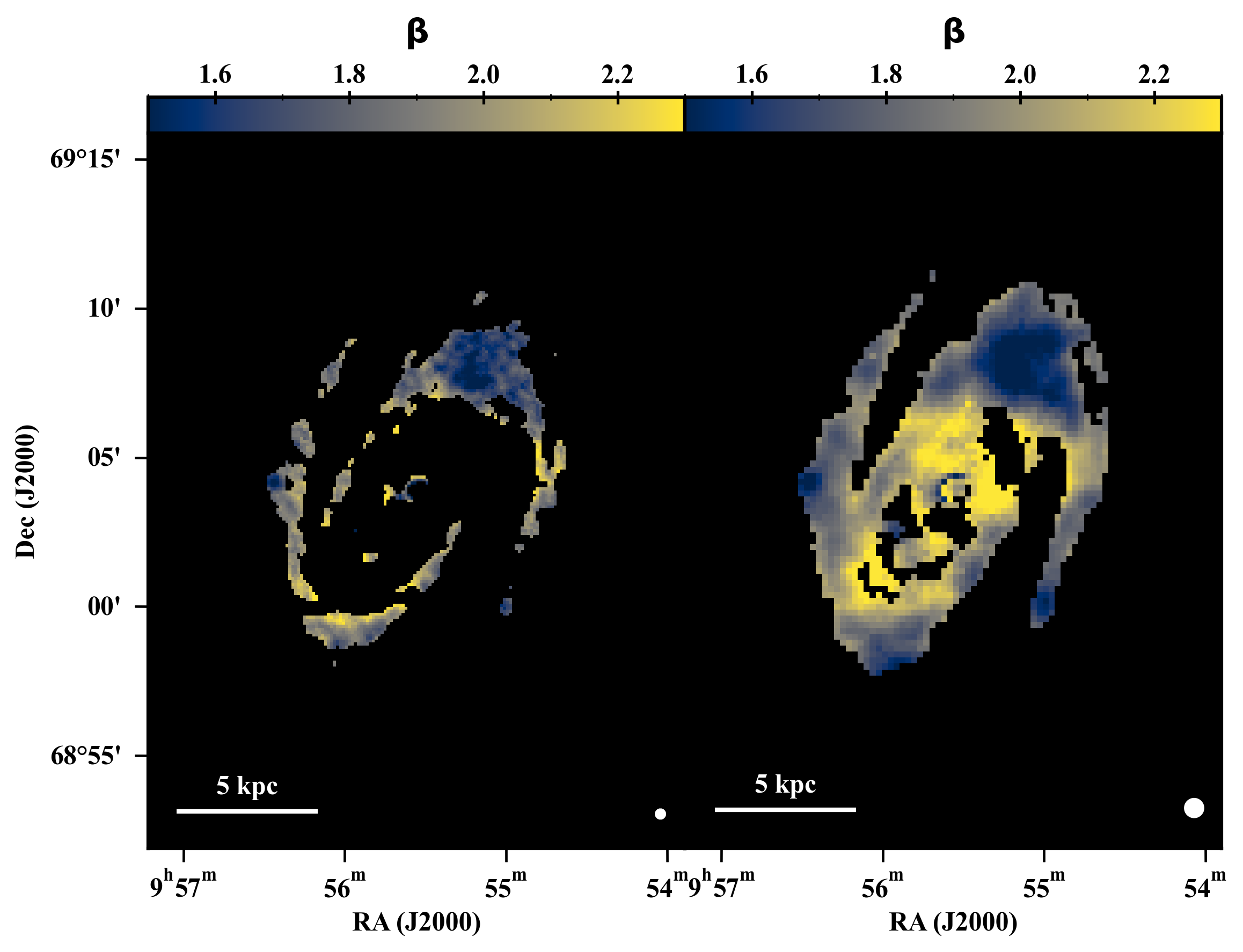}
\caption{Posterior-median emissivity index, $\beta$, maps of M\,81 derived from two variable-$\beta$ SMBB fits. The left panel uses the 70, 160, 250, 450, and \SI{850}{\um} maps convolved to the SPIRE \SI{250}{\um} resolution. The right panel uses the 70, 160, 250, 350, 450, 500, and \SI{850}{\um} maps convolved to the SPIRE \SI{500}{\um} resolution. The beam FWHMs of the two fits, \SI{18.2}{\arcsec} and \SI{36.3}{\arcsec}, are indicated by the white circles in the lower-right corner of each panel. Pixels are shown only where the posterior uncertainty on the fitted emissivity index satisfies $\sigma_{\beta}$\,$\leq$\,0.3.}
\label{fig:wcs_vb}
\end{figure}

The \SI{250}{\um}-resolution fit preserves more spatial detail, but the posterior uncertainty on $\beta$ is high over much of the low-surface-brightness disc. After applying the uncertainty cut $\sigma_\beta$\,$\leq$\,0.3, the well-constrained $\beta$ measurements are mainly restricted to the spiral arms. To test whether including the SPIRE \SI{350}{\um} and \SI{500}{\um} could significantly improve our $\beta$ mapping, we performed a second set of fits at the \SI{500}{\um} resolution, including all seven bands from 70 to \SI{850}{\um}. This fit provides stronger long-wavelength constraints on the SED shape. Larger pixel sizes and beam size also improve the SNR, although at the cost of spatial resolution, yielding a more spatially continuous $\beta$ map across the disc. The resulting variable-$\beta$ maps from the \SI{250}{\um} and \SI{500}{\um}-resolution fits are shown in Figure\,\ref{fig:wcs_vb}. Both fits recover the same broad spatial pattern, with lower values of $\beta$ ($\sim$\,1.5\,--\,1.8) preferentially found in the northern spiral-arm regions. In the \SI{500}{\um}-resolution fit, the well-constrained coverage extends farther into the central disc, where higher values of $\beta$ ($\sim$\,2.0\,--\,2.3) are generally recovered.

\begin{figure}
\centering
\includegraphics[width=\linewidth]{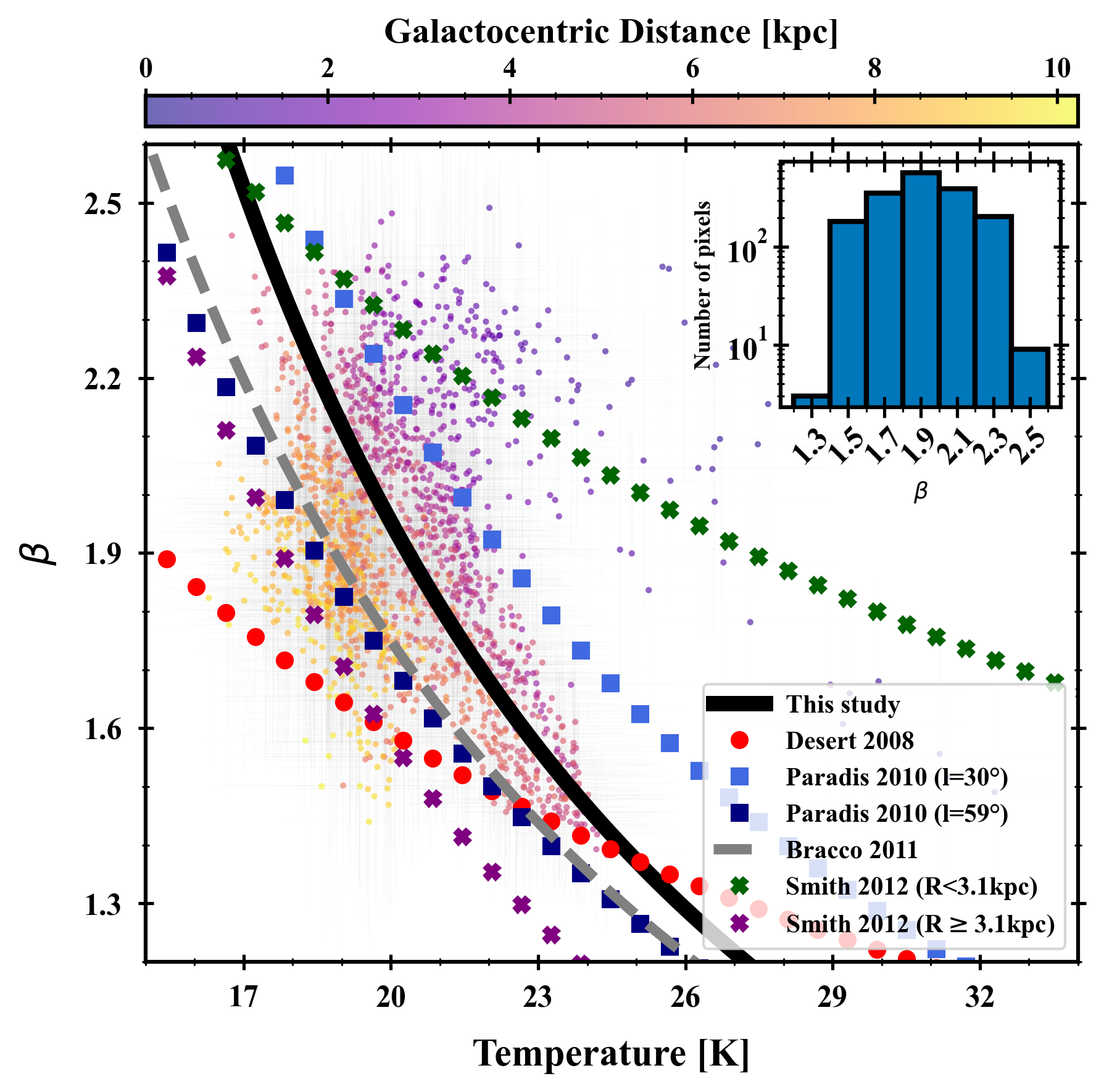}
\caption{Pixel-by-pixel $\beta$--$T_{\rm d}$ relation from the variable-$\beta$ SMBB fit at SPIRE \SI{500}{\um} resolution. Coloured points show posterior-median values for pixels satisfying $\sigma_{\rm T}$\,$\leq$\,3\,K and $\sigma_{\rm \beta}$\,$\leq$\,0.3. The solid black curve shows the best-fitting relation $\beta$\,=\,$\beta_0$\,$(T_{\rm d}/20$\,K$)^\alpha$, giving $\beta$\,=\,1.95\,$(T_{\rm d}/20\,{\rm K})^{-1.58}$. Dotted curves show literature $\beta$--$T_{\rm d}$ relations from \citet{Desert2008}, \citet{Paradis2010}, \citet{Bracco2011}, and \citet{Smith2012}. The inset histogram shows the distribution of fitted $\beta$ values. The colour scale shows the deprojected galactocentric distance of each pixel in the disc of M\,81.}
\label{fig:T_vs_B}
\end{figure}

Figure\,\ref{fig:T_vs_B} shows the relation between dust temperature and emissivity index for the well-constrained pixels, satisfying $\sigma_{\rm T}$\,$\leq$\,3\,K and $\sigma_{\rm \beta}$\,$\leq$\,0.3, in the \SI{500}{\um}-resolution variable-$\beta$ fit. The recovered emissivity indices typically lie in the range $\beta$\,$\sim$\,1.5\,--\,2.3, with the distribution peaking close to $\beta$\,$\sim$\,1.8\,--\,2.0, indicating only moderate spatial variation in M\,81 compared with the stronger variations reported for some other spiral galaxies (e.g. \citealt{Smith2012, Tabatabaei2014, Hunt2015}). The data show an apparent inverse relation between $T_{\rm d}$ and $\beta$, with warmer pixels generally associated with lower $\beta$ values. Unlike the HELGA analysis of M\,31 \citep{Smith2012}, which identified distinct inner and outer $T_{\rm d}$\,--\,$\beta$ branches separated at a transition radius of R\,$\simeq$\,3.1\,kpc, the M\,81 pixels show no clear bifurcation into radial branches. We therefore describe them using a single empirical relation. A small number of low-temperature, high-$\beta$ pixels occupy the upper part of the distribution; these are mainly associated with the central region, but do not form a distinct branch comparable to that seen in M\,31. An inverse $T_{\rm d}$\,--\,$\beta$ relation can arise from real changes in grain properties, such as composition, structure, or coagulation state, but it can also be produced by the degeneracy between $T_{\rm d}$ and $\beta$ in SMBB fitting. This degeneracy has been demonstrated in both resolved studies of M\,31 and global analyses of nearby galaxies \citep[e.g.][]{Smith2012, Lamperti2019}. This motivates our use of a fixed emissivity index for the final dust surface density and temperature maps.

We performed the final set of fits with $\beta$ held fixed at 1.9. This value is motivated by the median emissivity indices recovered from the well-constrained pixels in the variable-$\beta$ fits, after masking pixels with $\sigma_\beta$\,>\,0.3. In this fixed-$\beta$ run, we retained the same mass prior and updated the temperature prior to the global median value of $18.5$\,K, while keeping the same standard deviation of 6.0\,K.

In the nuclear region of M\,81 we detect excess at \SI{850}{\um} relative to the expectations of an SMBB model. This excess leads to formally poor and unstable fits if the \SI{850}{\um} point is forced to be explained by dust emission alone. To identify and treat the affected pixels in a consistent way, we adopted the following two-step procedure.

\begin{figure}
\centering
\includegraphics[width=\linewidth]{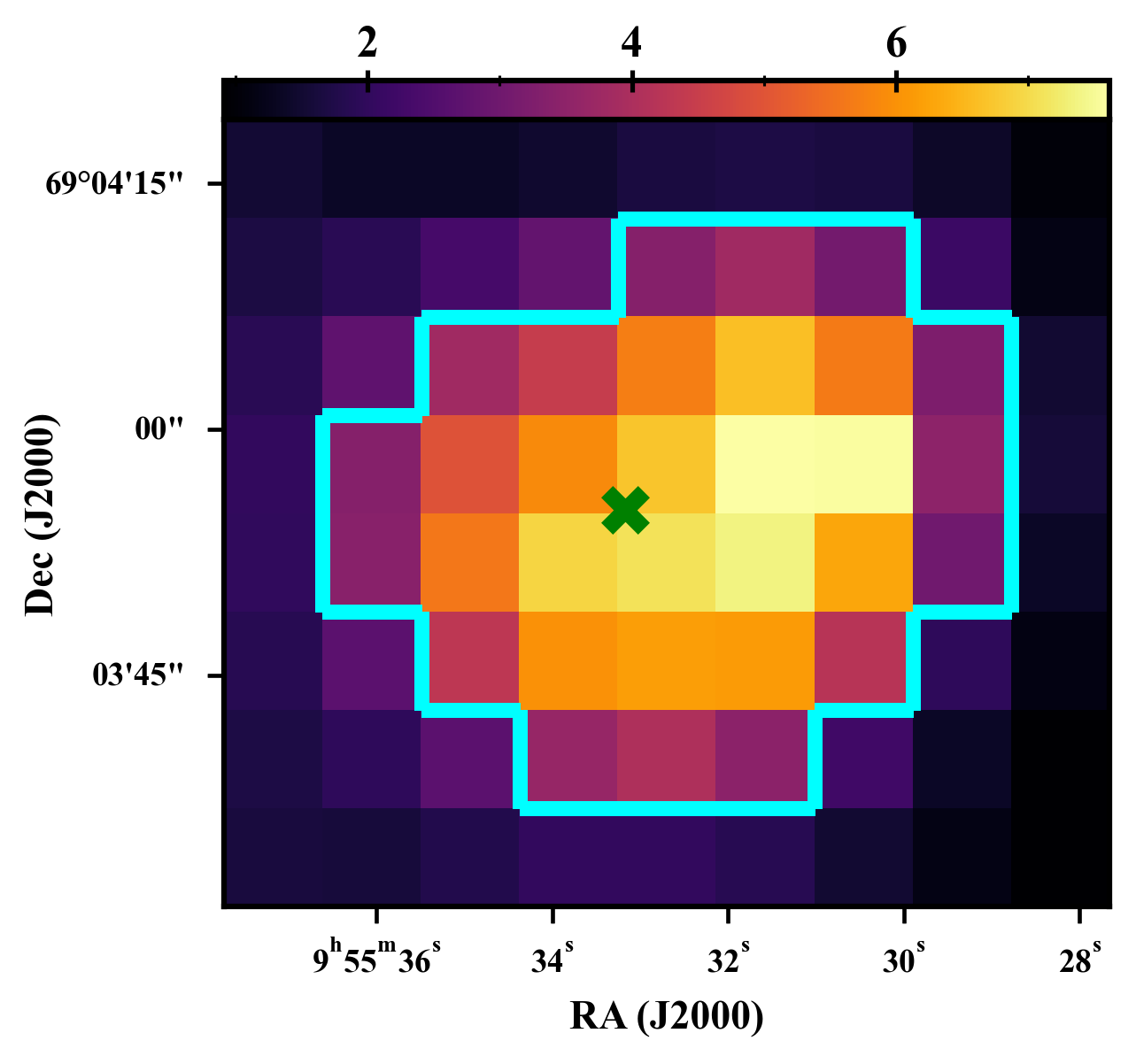}
\caption{Map of the ratio between the observed and predicted \SI{850}{\um} surface brightness, $R_{850}$\,=\,$F^{\rm obs}_{850}$\,/\,$F^{\rm pred}_{850}$, in the centre of M\,81, at the SPIRE \SI{250}{\um} resolution. The green marker indicates the galaxy centre. The blue contour shows pixels with $R_{850}$\,$\geq$\,3, which were subsequently fitted with an SMBB plus synchrotron model.}
\label{fig:850_excess}
\end{figure}

First, using the maps convolved to the SPIRE \SI{250}{\um} PSF and reprojected onto the corresponding pixel grid, we defined a central region of approximately 20\,$\times$\,20 pixels ($\sim$2.1\,$\times$\,2.1\,kpc) around the galaxy centre and performed an SMBB fit in this area using the same Bayesian framework, but with the \SI{850}{\um} data point excluded from the likelihood. In this run we used the same priors as the fixed-$\beta$ fit, but with the temperature prior updated to the median central temperature. From these fits, we predicted the \SI{850}{\um} flux expected from dust alone by extrapolating the best-fitting SMBB model to \SI{850}{\um}, and constructed a map of the ratio $R_{850}$\,=\,$F^{\rm obs}_{850}$\,/\,$F^{\rm pred}_{850}$. As a conservative empirical criterion, pixels with $R_{850}$\,$\ge$\,3 were flagged as candidate \SI{850}{\um} excess pixels (see Figure\,\ref{fig:850_excess}). Since M\,81 hosts a known compact radio nucleus \citep{Doi2005}, the spatial distribution of the excess supports its interpretation as likely nuclear non-thermal contamination rather than a disc-wide sub-mm excess.

\begin{figure}
\centering
\includegraphics{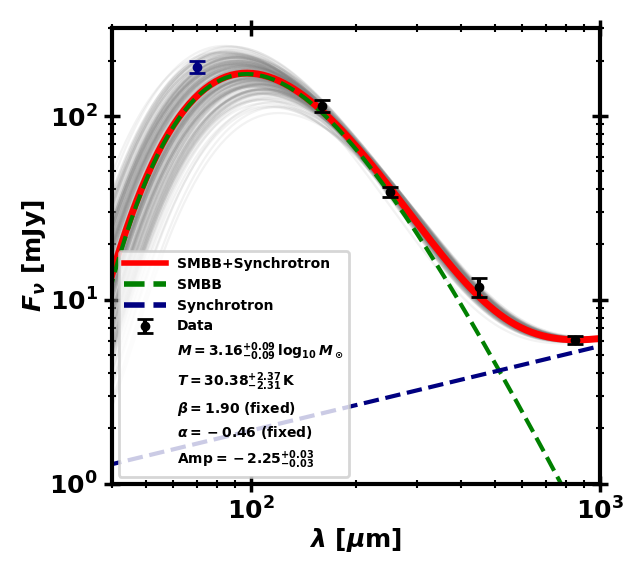}
\caption{Example SED for a pixel in the centre of M\,81, shown for synchrotron\,+\,SMBB. The blue data point at \SI{70}{\um} was used as an upper limit in the fitting process. The red line indicates the best-fit model, while the green and blue lines show the individual components of the combined model. Error bars include calibration uncertainties following \citealt{Ciesla2013}.}
\label{fig:SED_Example_with_Synch}
\end{figure}

\begin{figure}
\centering
\includegraphics[width=\linewidth]{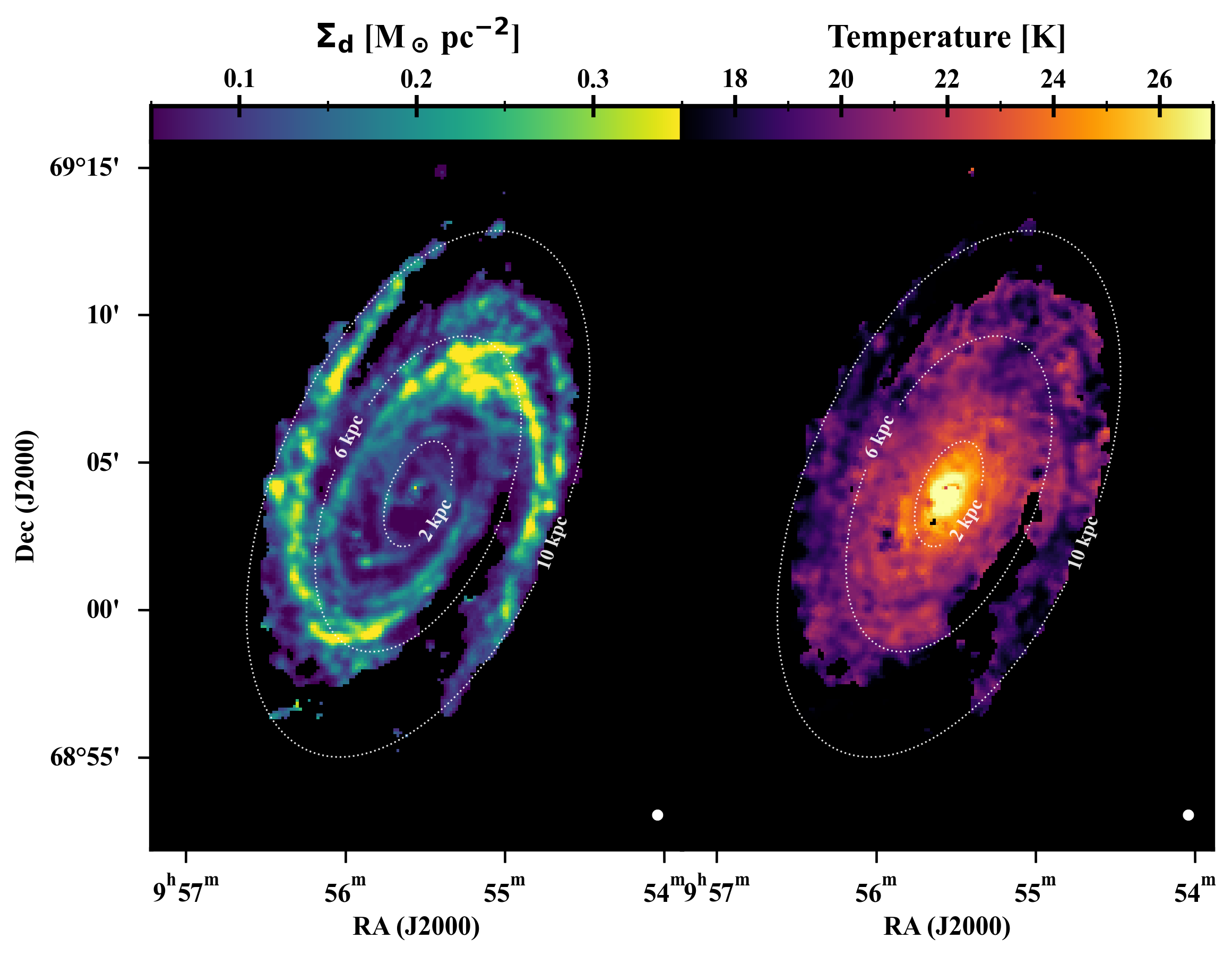}
\caption{Dust surface density (left) and temperature (right) maps in M\,81 derived from SMBB fitting with a fixed dust emissivity index of $\beta$\,=\,1.9. In the left and right panels, pixels are shown only where the uncertainties in dust surface density and temperature are below 0.1\,M$_\odot$\,pc$^{-2}$ and 3\,K, respectively. The fits use photometric data at 70, 160, 250, 450, and \SI{850}{\um}. The dotted ellipses correspond to deprojected galactocentric distances of 2, 6, and 10 kpc in the plane of M\,81's disc.}
\label{fig:MCMC_fb}
\end{figure}

In the second step, we re-fitted the flagged pixels with an SMBB\,+\,synchrotron model, fixing the spectral index to $\alpha$\,=\,0.46 \citep{Doi2005}. The dust parameters in this fit were assigned the same priors as in the central SMBB fit, and the 70, 160, 250, 450, and \SI{850}{\um} measurements were included in the likelihood. For each flagged pixel, we replaced the dust parameters from the original SMBB run with the corresponding posterior medians from the SMBB\,+\,synchrotron fit, including $\beta$ where it was allowed to vary. An example fit is shown in Figure\,\ref{fig:SED_Example_with_Synch}. After applying this correction to the flagged nuclear pixels, the resulting fixed-$\beta$ SMBB dust surface density and temperature maps are shown in Figure\,\ref{fig:MCMC_fb}.

\section{PPMAP} \label{sec:PPMAP}

\begin{figure*}
\centering
\includegraphics[width=\textwidth]{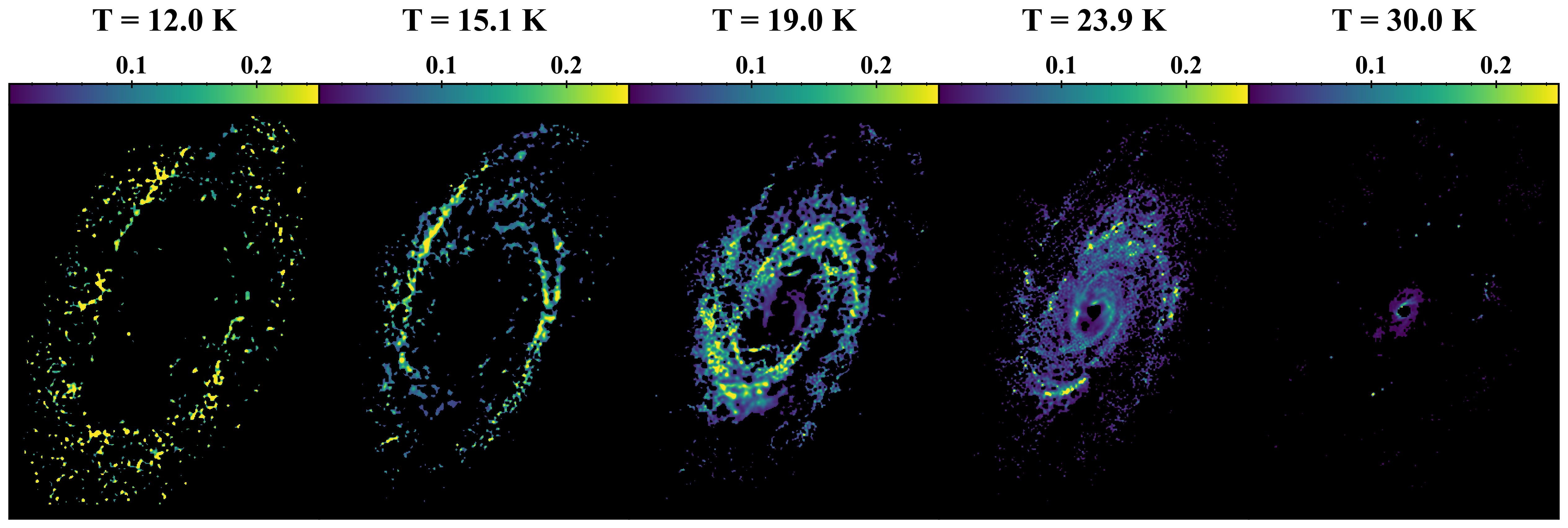}
\caption{Distribution of PPMAP dust mass surface density across the five logarithmically spaced temperature bins between 12 and 30\,K. Each panel shows the dust surface density for one temperature bin, in units of M$_\odot$\,pc$^{-2}$. Pixels with SNR\,$\leq$\,3 are masked separately in each slice.}
\label{fig:PPMAP_perT}
\end{figure*}

There are some limitations to the standard pixel-by-pixel fitting procedure. First, we assume a single dust temperature along each line of sight. Second, the method requires all maps to be convolved to the resolution of the lowest-resolution band, resulting in a loss of spatial information.

To address these limitations, PPMAP \citep{Marsh2015}, a Bayesian multidimensional fitting technique, was introduced. Unlike traditional methods, PPMAP does not assume a single, potentially unphysical dust temperature or a predefined temperature distribution. Instead, it models a discrete range of temperatures and dust emissivity indexes ($\beta$). Furthermore, in contrast to conventional SED modelling, PPMAP incorporates a PSF specific to each instrument. As a result, it does not require the input images to be convolved to a common resolution. This allows PPMAP to generate higher-resolution dust temperature and column density maps (approximately at the resolution of the highest resolution observation) than those produced using standard SED fitting methods.

PPMAP reconstructs the dust distribution using a stepwise Bayesian procedure. It begins by artificially increasing the noise assigned to the input maps until the observations provide essentially no information, such that the initial solution is determined by the adopted prior. The artificial noise is then gradually reduced to its measured level, while the inferred differential column density, which describes how the column density is distributed among the adopted temperature bins at each position, is updated at each step. The final output is an image cube containing the differential column density, converted here to dust-mass surface density, in each adopted temperature bin and at each position on the sky, rather than a single best-fitting temperature for every pixel \citep[for details, see][]{Marsh2015}.

\begin{figure}
\centering
\includegraphics[width=\linewidth]{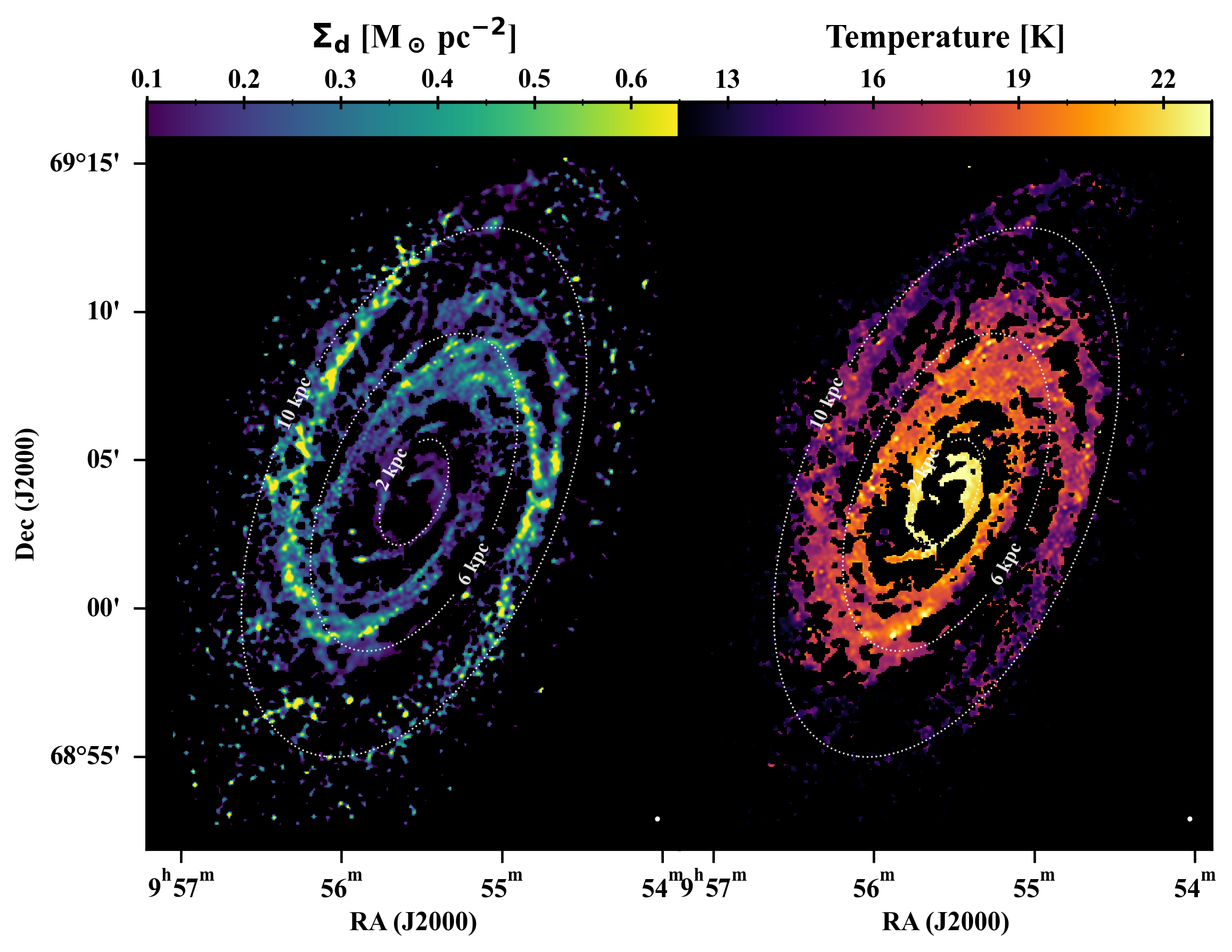}
\caption{Dust surface density (left) and dust-surface-density-weighted temperature (right) maps of M\,81 derived using PPMAP, with a fixed dust emissivity index of $\beta$\,=\,1.9. The maps have an effective angular resolution of approximately \SI{5.6}{\arcsec}, comparable to that of the PACS \SI{70}{\um} data, and use a temperature grid spanning 12.0\,--\,30.0\,K. Only PPMAP pixels with SNR\,$\geq$\,3 in the total dust mass surface density map, obtained by integrating over all temperature bins, are shown. The PPMAP analysis use the same photometric data as the standard approach, with the addition of 350 and \SI{500}{\um} bands. The dotted ellipses correspond to deprojected galactocentric distances of 2, 6, and 10 kpc in the plane of M\,81's disc.}
\label{fig:PPMAP_fb}
\end{figure}

For the PPMAP analysis, we include the 350 and \SI{500}{\um} maps in addition to the 70, 160, 250, 450, and \SI{850}{\um} maps which we used also for our standard approach. Figure\,\ref{fig:PPMAP_fb} presents the PPMAP results obtained for the whole galaxy. Initially, we selected 12 logarithmically spaced temperature bins from 8 to 50\,K, and assumed a fixed emissivity index of $\beta$\,=\,1.9, corresponding to the median value derived from the variable-$\beta$ SMBB. However, we found that the contribution of the highest four and the lowest three temperature bins to the total dust surface density was negligible (<\,1\% of total dust mass of M\,81). Moreover, including these additional temperature bins increased the associated uncertainties. Therefore, we excluded these temperature bins, and used only five temperature values: 12.0, 15.1, 19.0, 23.9, and 30.0\,K. 

The PPMAP dust surface density maps for the individual temperature bins are shown in Figure\,\ref{fig:PPMAP_perT}. In each bin, only pixels with a dust surface density detected at SNR\,$\geq$\,3 are retained. The morphology changes systematically with temperature. The 12.0 and 15.1\,K components are distributed over the extended disc, particularly along the outer spiral structure. The $\sim$19.0\,K component provides the most continuous view of the disc and traces the main spiral arms. In contrast, the warmer components become increasingly concentrated toward regions of stronger radiation field: the 23.9\,K component is strongest in the inner disc and bright spiral features, while the 30.0\,K component is largely restricted to the central region. This behaviour is consistent with previous studies of nearby spiral galaxies, which show that cold dust emission is generally more extended and warmer dust is preferentially found where the ISRF is stronger, including central regions, spiral arms, and star-forming regions \citep[e.g.][]{Whitworth2019}.

Summing the dust surface density in each temperature bin shows that most of the detected PPMAP dust mass is associated with the cold and intermediate-temperature bins. The 12.0, 15.1, and 19.0\,K bins contain 35.4\%, 23.2\%, and 27.3\% of the detected mass, respectively. Together, these three bins account for $\sim$85.9\% of the total PPMAP dust mass. The 23.9\,K bin contributes 13.5\% and the 30.0\,K component contributes only 0.7\% of the total PPMAP dust mass, respectively. 

\section{Results and Discussion}
\label{sec:Discussion}

\subsection{SED Results}

\begin{figure}
\centering
\includegraphics[width=\linewidth]{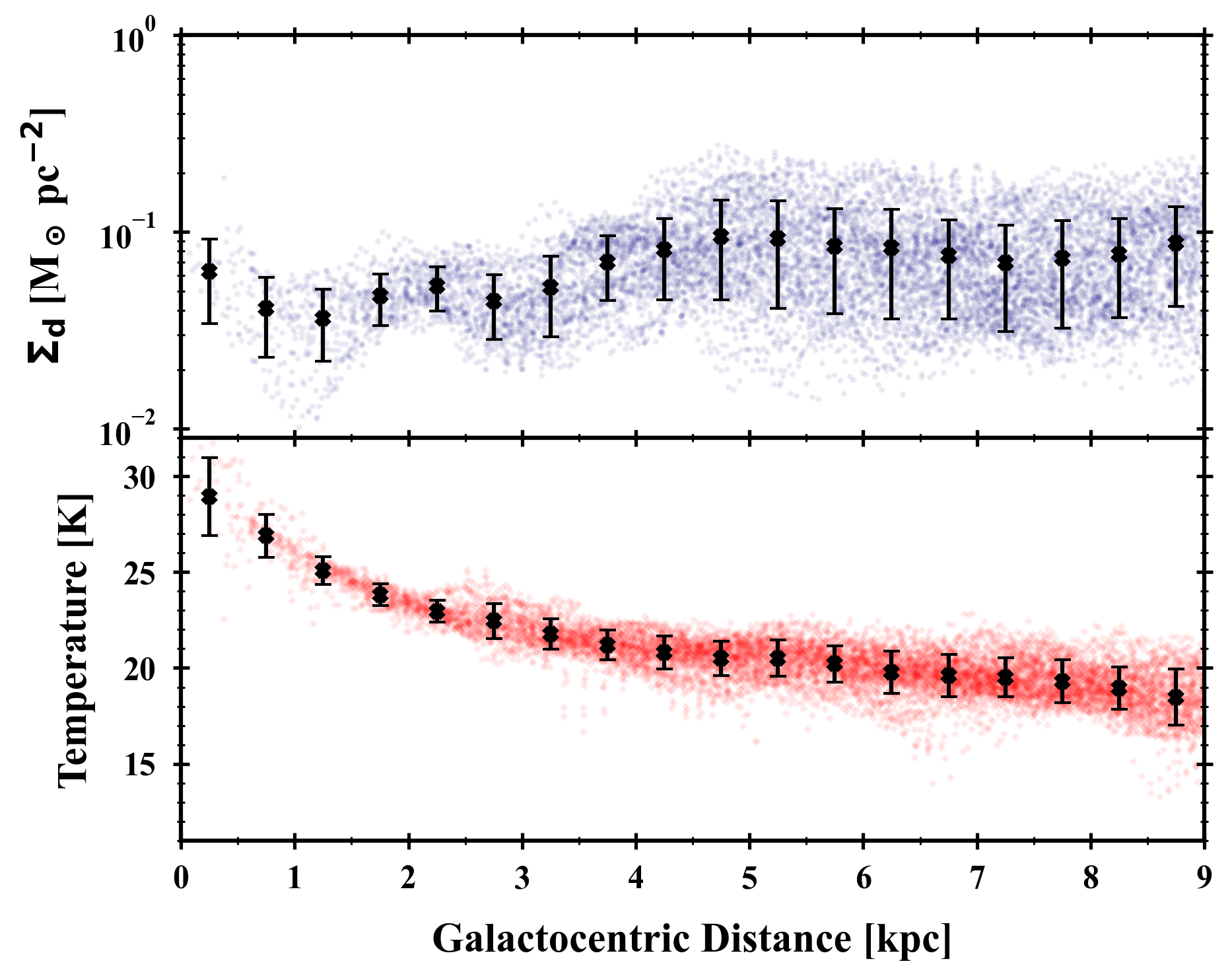}
\caption{Radial profiles of face-on dust mass surface density (top) and dust temperature (bottom) as a function of deprojected galactocentric radius in M\,81 derived from fixed-$\beta$ SMBB fits. The semi-transparent coloured points show the best-fitting values for individual unmasked pixels across the disc. The black points show the radial mean profiles, obtained by averaging pixel values within radial bins of width $\Delta$R\,=\,0.5\,kpc (i.e. [0, 0.5), [0.5, 1.0), ...). The error bars on the black points indicate the $1\sigma$ scatter of pixel values within each radial bin; individual pixel error bars are omitted for clarity. The profiles are shown only out to the first radial bin in which fewer than 50\% of the pixels assigned to that bin have defined values.}
\label{fig:MCMC_AllResults}
\end{figure}

The dust surface density maps from the fixed-$\beta$ SMBB and PPMAP analyses are shown in Figure\,\ref{fig:MCMC_fb} and Figure\,\ref{fig:PPMAP_fb}. Both methods show that dust in M\,81 is distributed throughout the disc with the highest concentrations along the inner spiral arms and lower surface densities in the inter-arm regions and outer disc. The dust mass surface densities from SMBB modelling range from $8.7$\,$\times$\,$10^{-3}$ to 5.4\,$\times$\,$10^{-1}$\,M$_\odot\,{\rm pc}^{-2}$, with a median of 0.123\,M$_\odot\,{\rm pc}^{-2}$. PPMAP gives a higher surface density, with values from $7.9$\,$\times$\,$10^{-2}$ to $2.0$\,M$_\odot\,{\rm pc}^{-2}$ and a median of 0.195\,M$_\odot\,{\rm pc}^{-2}$. The corresponding integrated dust masses are (1.77\,$\pm$\,0.09)\,$\times$\,$10^{7}$\,M$_\odot$ for the SMBB fit and (2.65\,$\pm$\,0.13)\,$\times$\,$10^{7}$\,M$_\odot$ for PPMAP. These masses assume a dust opacity of $\kappa_{\SI{350}{\um}}$\,=\,1.92\,$\mathrm{cm^2\,g^{-1}}$, and the quoted uncertainties correspond to a conservative 5\% observational uncertainty, largely due to uncertainties in the calibration. The PPMAP mass is therefore larger by a factor of $\sim$1.5.

Our integrated SMBB dust mass is approximately 48\% lower than the global value of (3.4\,$\pm$\,0.5)\,$\times$\,$10^{7}$\,M$_\odot$ obtained by \cite{Bendo2010} using global flux measurements from \textit{Herschel} 160\,--\,\SI{500}{\um}. Pixel-by-pixel fitting would often lead to a higher dust mass estimate due to a reduced bias to warmer dust temperatures. However, our pixels above our signal-to-noise criteria only account for 85.4\% and 79.9\% of the total 250 and \SI{850}{\um} flux-densities, respectively. \cite{Bendo2010} also use a $\beta$\,=\,2.0 compared to our $\beta$\,=\,1.9, and their results were in the first \textit{Herschel} release with an earlier calibration of the data. These may explain the difference to our quoted dust masses. \citet{Nersesian2019, Nersesian2026} obtained (1.05\,$\pm$\,0.16)\,$\times$\,$10^{7}$\,M$_\odot$ with CIGALE and THEMIS, while \citet{Verstocken2020} derived a dust mass of (1.28\,$\pm$\,0.27)\,$\times$\,$10^{7}$\,M$_\odot$ with SKIRT and THEMIS. While the THEMIS model uses a full grain model with a distribution of dust temperatures, rescaling the dust mass obtained by \cite{Nersesian2019} from their dust opacity of $\kappa_{\SI{350}{\um}}$\,=\,3.50\,$\mathrm{cm^2\,g^{-1}}$ to our adopted value of 1.92\,$\mathrm{cm^2\,g^{-1}}$, their dust mass estimates would increase to $\sim$1.91\,$\times$\,10$^7$\,M$_\odot$, in much better agreement with our estimate.

The higher PPMAP mass relative to the SMBB mass is expected when comparing an SMBB fit that assumes a single dust temperature along the line of sight with PPMAP, which allows dust mass to be distributed over multiple temperature components. If there is a mixture of warm and cold dust, the warm dust dominates the emitted flux, so the fitted temperature can be biased toward the warmer dust, which emits more efficiently per unit mass. This can therefore bias the fitted temperature high, leading to an underestimate of the dust mass \citep[e.g.][]{Sommovigo2025}. By contrast, PPMAP does not assume a single effective temperature to each line of sight; instead, it distributes the dust mass over the adopted temperature grid. 

\begin{figure}
\centering
\includegraphics[width=\linewidth]{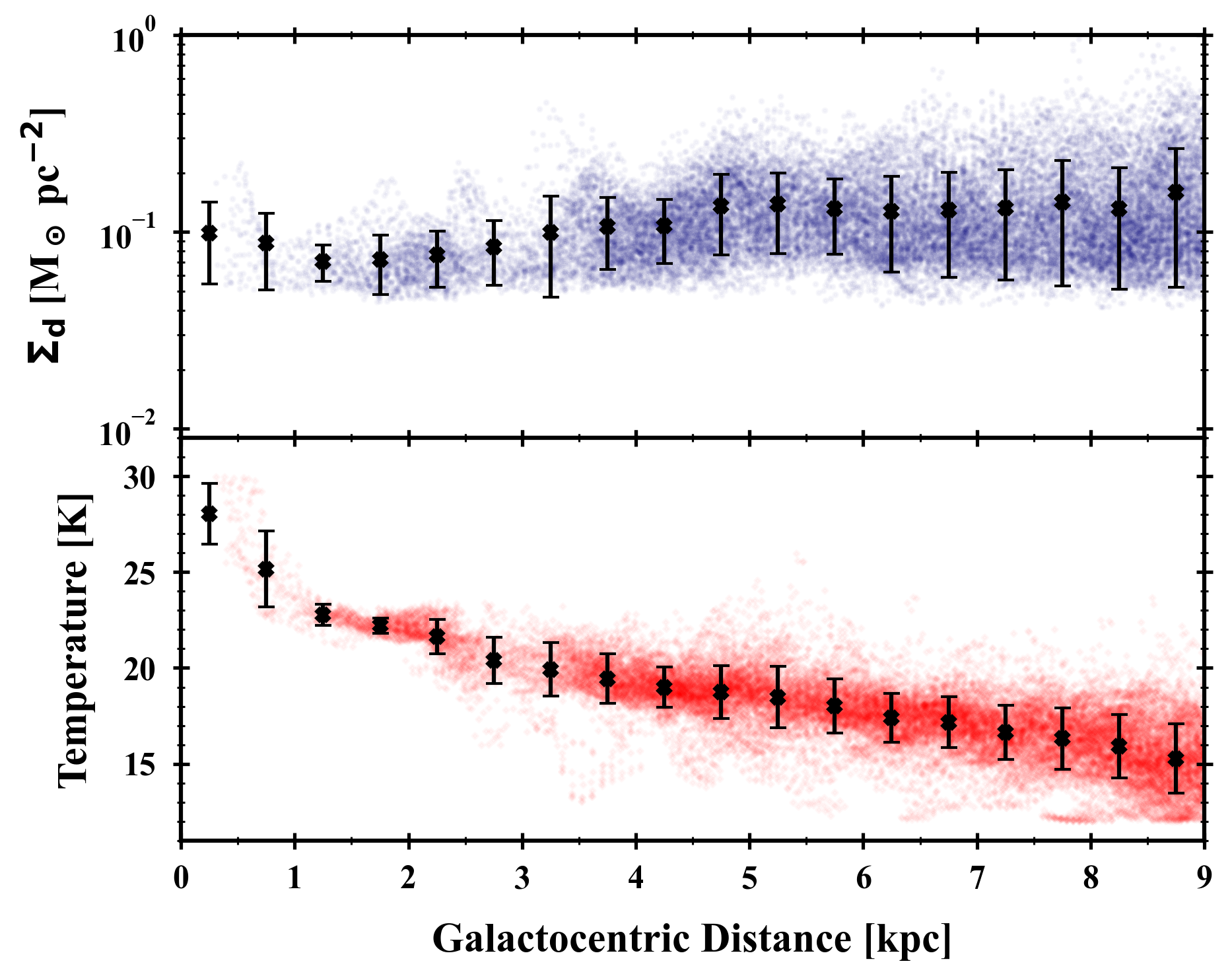}
\caption{Radial profiles of face-on dust mass surface density (top) and surface-density-weighted dust temperature (bottom) as a function of deprojected galactocentric radius in M\,81 derived from fixed-$\beta$ PPMAP fits. The semi-transparent coloured points show the best-fitting values for individual unmasked pixels across the disc. The black points show the radial mean profiles, obtained by averaging pixel values within radial bins of width $\Delta$R\,=\,0.5\,kpc. The error bars on the black points indicate the $1\sigma$ scatter of pixel values within each radial bin; individual pixel error bars are omitted for clarity. The profiles are shown only out to the first radial bin in which fewer than 50\% of the pixels assigned to that bin have defined values.}
\label{fig:PPMAP_AllResults}
\end{figure}

The temperature maps show that the dust in M\,81 is warmest in the central regions and becomes cooler toward the outer disc (Figure\,\ref{fig:MCMC_fb} and Figure\,\ref{fig:PPMAP_fb}). In the fixed-$\beta$ SMBB fit, the fitted temperatures span 10.0\,--\,31.7\,K, with a median of 20.1\,K and a mean of 20.2\,K. The PPMAP temperature map has a median of 16.5\,K and a mean of 16.4\,K, while spanning the adopted temperature grid from 12.0 to 30.0\,K. The exact minimum and maximum PPMAP temperatures therefore reflect the limits of the temperature grid used in the fitting process. In both methods, the warmest dust is associated with the bulge and inner spiral structure, where the ISRF is expected to be strongest, while the cooler dust is found in the outer disc. The resulting negative radial temperature gradient is also seen in other massive spirals; for example, in M\,31 the dust temperature falls from $\sim$30\,K in the centre to $\sim$15\,K in the outer ring \citep{Smith2012}. The M\,81 temperatures are therefore broadly consistent with those typically found in nearby spiral galaxies (15\,--\,30\,K, \citealt{Smith2012}), but are well below the high nuclear temperatures reported for M\,33, where two-component MBB fits yield $T_{\rm d}\sim$58\,K \citep{Tabatabaei2014}.

\begin{figure}
\centering
\includegraphics[width=\linewidth]{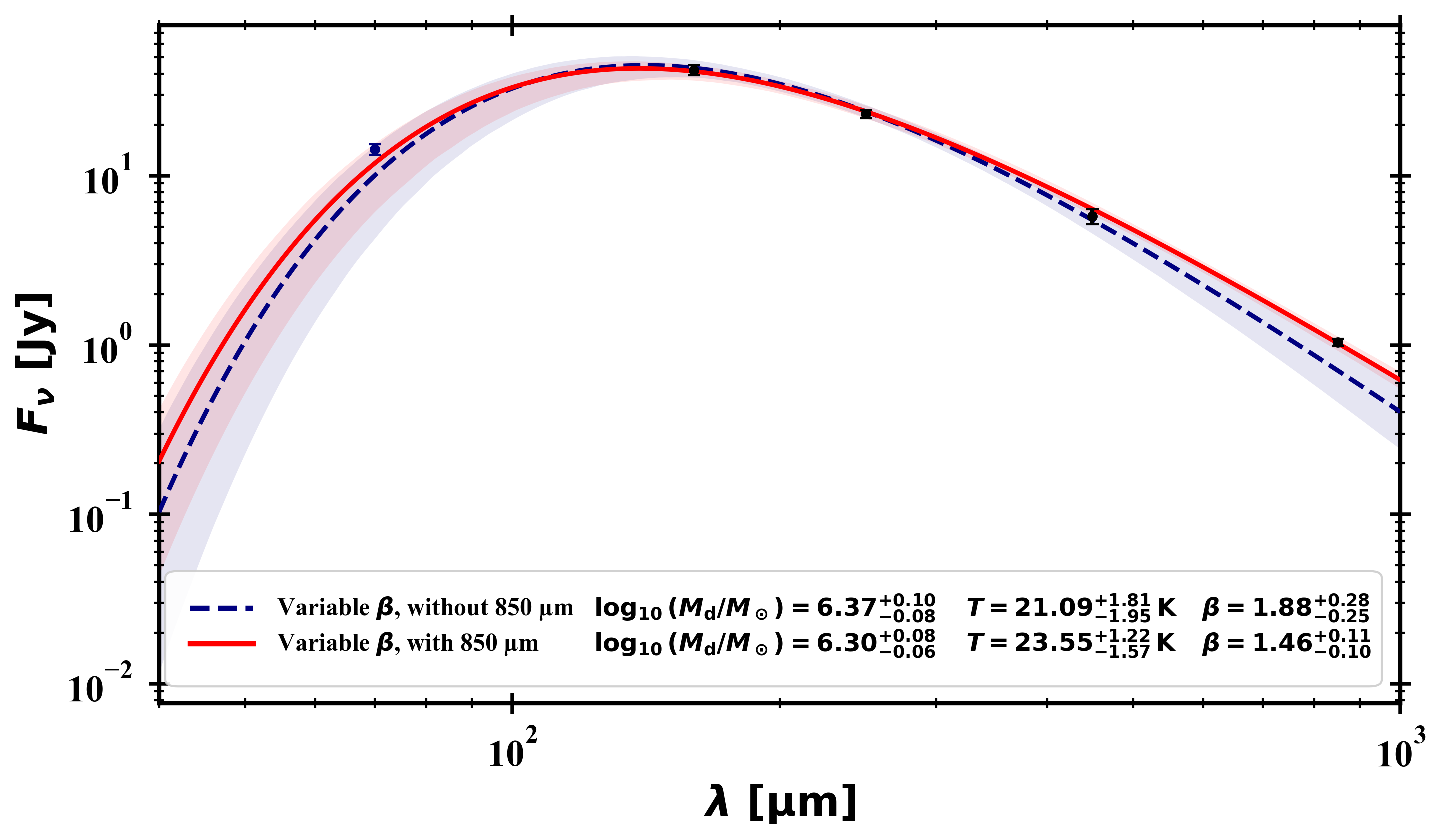}
\caption{Integrated SED of the northern region of M\,81 within a circular aperture of $\sim$\,2\,kpc radius, fitted using variable-$\beta$ SMBB models with (red solid line) and without (navy dashed line) the \SI{850}{\um} observation. The shaded regions represent the 95\% posterior credible intervals. The observed fluxes are shown as points with their uncertainties, with the \SI{70}{\um} observation shown in navy and treated as an upper limit during the fitting.}
\label{fig:excess}
\end{figure}

We investigated if we have any evidence for a sub-mm excess, which has been reported in some studies \citep{RemyRuyer2013, Gordon2014, Chang2026}. We calculated the residuals at \SI{850}{\um} using the fixed-$\beta$\,=\,1.9 SMBB model at \SI{18.2}{\arcsec} resolution. We found positive residuals in the northern part of the galaxy, where the variable-$\beta$ fits indicate lower emissivity indices (see Figure\,\ref{fig:wcs_vb}). To investigate this region, we selected a circular aperture of $\sim$2\,kpc radius and fitted its integrated SED using the variable-$\beta$ model, both including and excluding the \SI{850}{\um} observations (see Figure\,\ref{fig:excess}). The \SI{70}{\um} measurement was treated as an upper limit in both fits. Excluding \SI{850}{\um}, the model predicts a flux of 0.71\,Jy, compared with the observed 1.04\,Jy, corresponding to a positive residual of approximately 46\%. However, accounting for the posterior uncertainty in the predicted flux and the observational uncertainty, the excess has a significance of only $\sim$\,2.06$\sigma$. Including \SI{850}{\um} in the fitting reduces the emissivity index from $\beta$\,=\,$1.88^{+0.28}_{-0.25}$ to $\beta$\,=\,$1.46^{+0.11}_{-0.10}$. Therefore, while there is a suggestion that an excess exists in this one area, it is not statistically significant. A variable-$\beta$ model provides a reasonable fit to the data, with total $\chi^2$ values of $\sim$0.8 and $\sim$1.7 for the fits excluding and including the \SI{850}{\um} observation, respectively, with the \SI{70}{\um} upper limit excluded from both $\chi^2$ calculations. Further observations would be required to differentiate with other models (e.g. the broken emissivity model used by \citealt{Gordon2014}).

\subsection{Radial profiles of dust mass, temperature, and emissivity}

We used radial profiles to investigate the variation of dust properties with galactocentric distance, applying two complementary approaches. In the first approach, we derived profiles directly from the fitted two dimensional dust maps. For both the fixed-$\beta$ SMBB and PPMAP results, the dust surface density and temperature maps were averaged in radial bins of width $\Delta R$\,=\,0.5\,kpc. These profiles are useful for comparing the spatially resolved SMBB and PPMAP maps, but they are affected by the uncertainty cuts applied to the fitted pixels. In particular, the outer disc contains a larger fraction of masked pixels, and the remaining pixels are preferentially those with higher surface brightness. The resulting means therefore represent the well-constrained pixels in each radial bin, rather than all pixels in the annulus. This selection effect can make the outer radial profiles appear artificially flat.

\begin{figure}
\centering
\includegraphics[width=\linewidth]{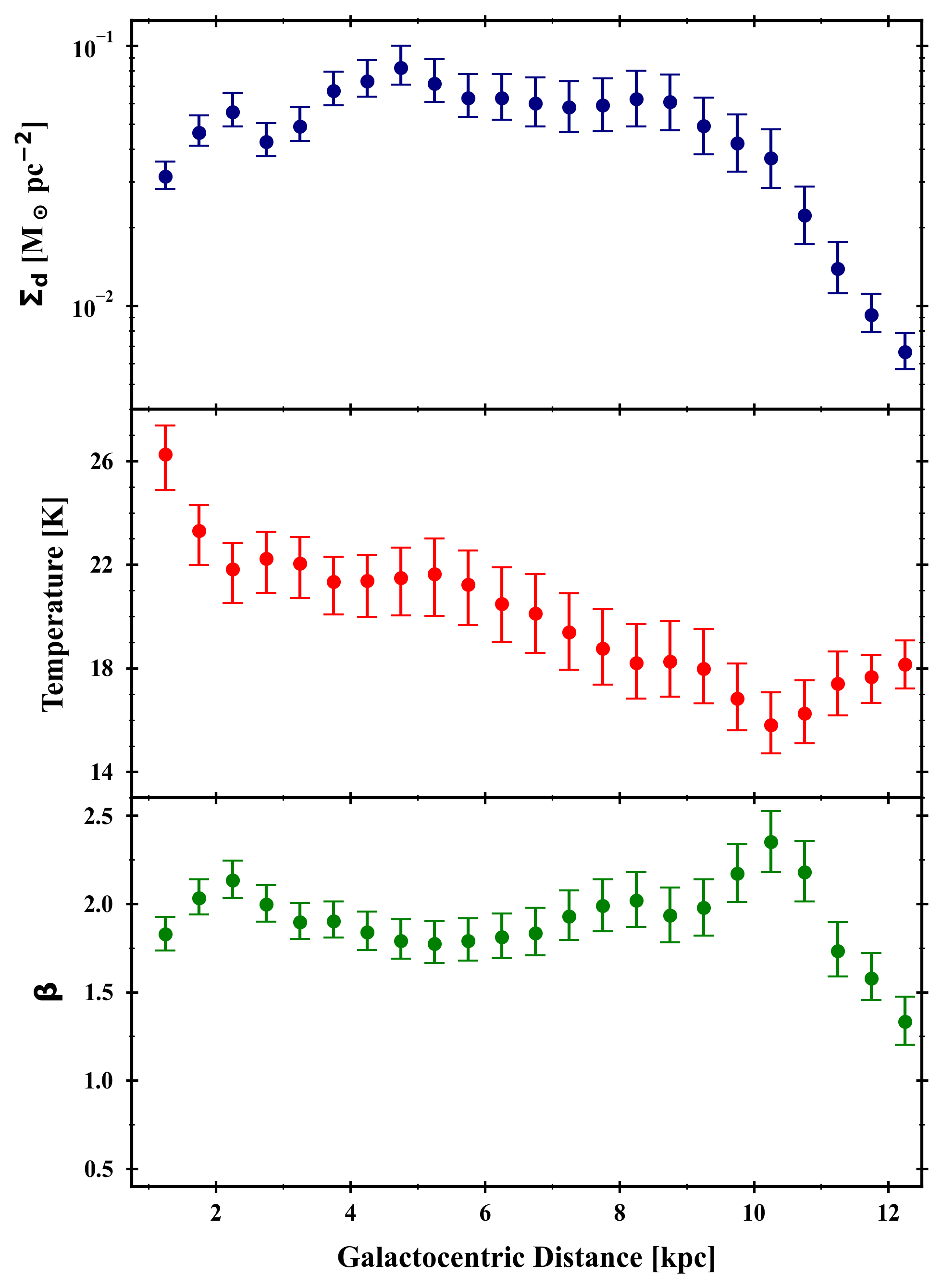}
\caption{Radial profiles of M\,81 derived from variable-$\beta$ SED fits to the mean flux densities measured in elliptical annuli of width 0.5\,kpc. The fits use the 70, 160, 250, 450, and \SI{850}{\um} maps. The navy, red, and green points show the posterior median values in each annulus for the face-on dust surface density, dust temperature, and dust emissivity index, respectively, with error bars indicating the corresponding posterior uncertainties. The x-axis shows the deprojected galactocentric distance from the centre of M\,81.}
\label{fig:RadialProfile_bySEDs}
\end{figure}

\begin{figure}
\centering
\includegraphics[width=\linewidth]{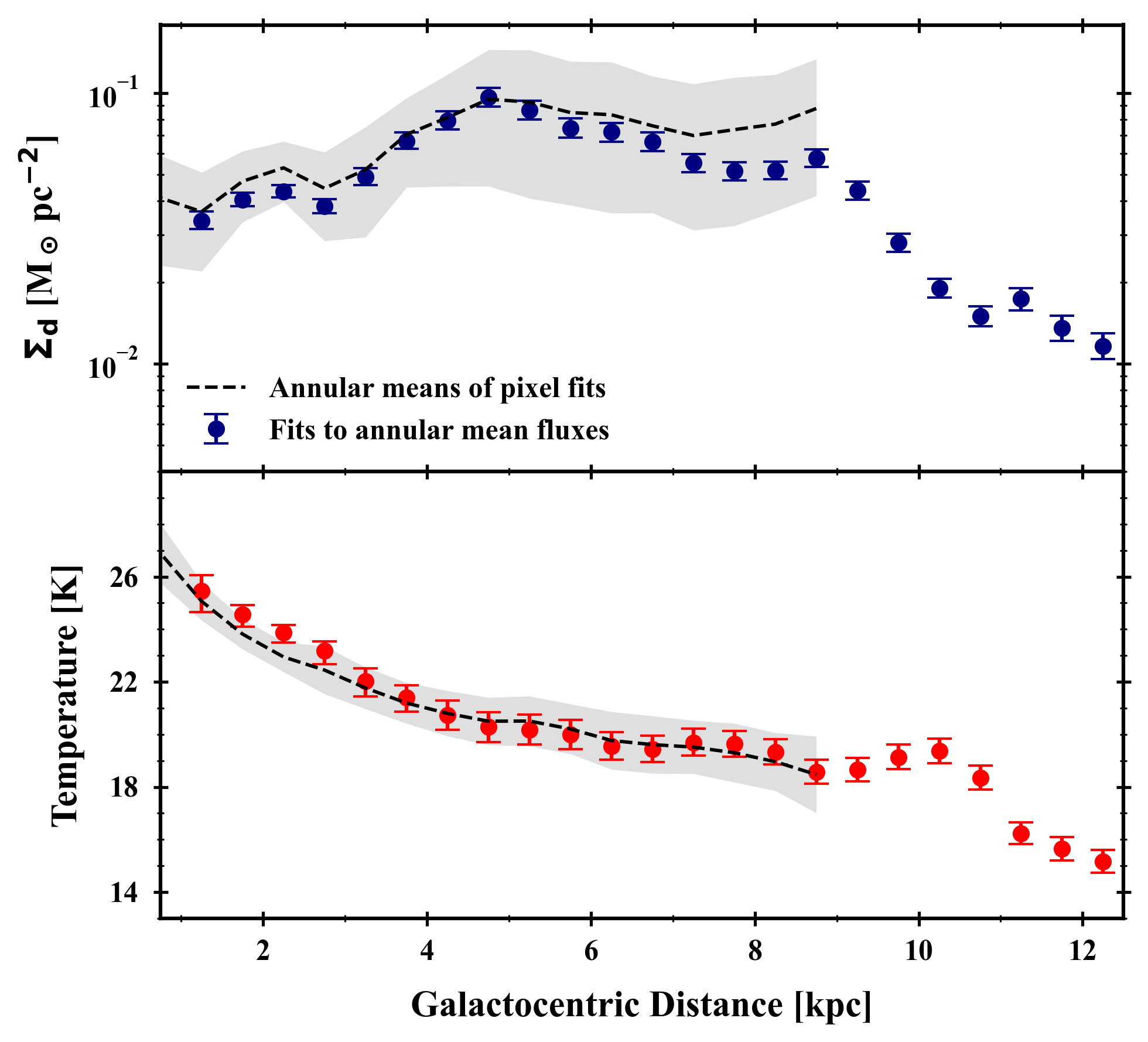}
\caption{Radial profiles of M\,81 derived from fixed-$\beta$ SED fits, with $\beta$\,=\,1.9, to the mean flux densities measured in elliptical annuli of width 0.5\,kpc. The fits use the 70, 160, 250, 450, and \SI{850}{\um} maps. The navy and red points show the posterior median values in each annulus for the face-on dust surface density and dust temperature, respectively, with error bars indicating the corresponding posterior uncertainties. The black dashed lines show the radial profiles obtained by averaging the fitted SMBB maps within the same annuli. The grey bands show the corresponding 1$\sigma$ scatter of pixel values
within each radial bin. The x-axis shows the deprojected galactocentric distance from the centre of M\,81.}
\label{fig:RadialProfile_bySEDs_FixedB}
\end{figure}

In the second approach, we performed SED fitting on azimuthally averaged flux densities to derive radial profiles that were less sensitive to the pixel-level uncertainty masking applied in the resolved analysis. We measured the mean flux density in concentric elliptical annuli, each with a deprojected radial width of 0.5\,kpc (corresponding to approximately \SI{30}{\arcsec} along the projected major axis), from the galaxy centre out to D$_{25}$. The central 1\,kpc was excluded because this region is most strongly affected by potential synchrotron contamination from the nucleus. A background level for each band was estimated within an elliptical annulus between 1.0 and 1.05\,$\times$\,D$_{25}$ and was subtracted before fitting. The resulting mean flux densities at 70, 160, 250, 450, and \SI{850}{\um} were fitted with an SMBB model using two configurations: one in which the dust surface density, dust temperature, and $\beta$ were allowed to vary, and another in which $\beta$ was fixed at 1.9. This approach reduces the impact of missing individual pixels and increases the effective SNR in the outer disc by averaging many pixels before fitting, making it less sensitive to individual low-SNR pixels in the outer disc. The map-derived SMBB and PPMAP radial profiles are shown in Figures\,\ref{fig:MCMC_AllResults} and \ref{fig:PPMAP_AllResults}, respectively. The radial profiles derived from the azimuthally averaged SEDs are shown in Figure\,\ref{fig:RadialProfile_bySEDs} for the variable-$\beta$ fits and Figure\,\ref{fig:RadialProfile_bySEDs_FixedB} for the fixed-$\beta$ fits.

In the first approach (Figure\,\ref{fig:MCMC_AllResults} and Figure\,\ref{fig:PPMAP_AllResults}), both methods show a clear decline in dust temperature with galactocentric radius. In the fixed-$\beta$ SMBB profile, the radial mean temperature decreases from the highest value of 28.9\,K in the central bin to the lowest value of 18.5\,K at the largest radius shown, with an average value of 21.7\,K. The PPMAP profile follows the same overall trend but reaches lower temperatures in the outer disc, decreasing from the highest temperature of 28.0\,K in the centre to the lowest temperature of 15.3\,K at the largest fitted radius, with an average value of 19.6\,K. This difference is consistent with the lower temperatures recovered by PPMAP, which can distribute dust mass across the adopted temperature grid rather than assigning a single effective temperature to each line of sight.

The dust surface density profiles show weaker radial variation than the temperature profiles. In the SMBB case, the radial mean dust surface density spans 0.04\,--\,0.10\,M$_\odot\,{\rm pc}^{-2}$, with an average radial-bin value of 0.07\,M$_\odot\,{\rm pc}^{-2}$. The highest radial means occur at R\,=\,4.75\,kpc, corresponding to the spiral-arm region. PPMAP gives systematically higher radial surface densities, with radial-bin means spanning 0.07\,--\,0.16\,M$_\odot\,{\rm pc}^{-2}$ and an average value of 0.11\,M$_\odot\,{\rm pc}^{-2}$, peaking at R\,=\,8.75\,kpc. These surface density profiles should be interpreted with caution because the averages are computed only from pixels that survive the uncertainty cuts. They therefore represent the mean properties of the well-constrained pixels in each radial bin, rather than averages over all pixels in the full annulus. This selection effect is most important in the outer disc, where faint pixels are more likely to be masked. As a result, the outer surface density profiles may appear flatter than the true azimuthally averaged dust distribution. 

Figure\,\ref{fig:RadialProfile_bySEDs} and Figure\,\ref{fig:RadialProfile_bySEDs_FixedB} show the results of the annular SED fits, which reveal a non-monotonic radial distribution of dust surface density in M\,81. From the innermost radius of R\,=\,1.25\,kpc, $\Sigma_{\rm d}$ increases from approximately 0.03\,M$_\odot\,{\rm pc}^{-2}$ to a maximum of 0.08\,M$_\odot\,{\rm pc}^{-2}$ at R\,=\,4.75\,kpc. The corresponding maximum in the fixed-$\beta$ fits is 0.10\,M$_\odot\,{\rm pc}^{-2}$ at the same galactocentric distance. The dust surface density remains comparatively high across the main disc before declining beyond R\,$\simeq$\,9.0\,kpc. By R\,=\,12.25\,kpc annulus, $\Sigma_{\rm d}$ has decreased by a factor of approximately 12.4 and 8.3 relative to its peak value in the variable and fixed-$\beta$ fits, respectively. The radial profile is therefore not well described by a single exponential over the full radial extent, but instead consists of an inner rise, a broad dust-rich region associated with the main spiral disc, and an outer decline. The comparison between the two radial profile methods in Figure\,\ref{fig:RadialProfile_bySEDs_FixedB} shows that they agree reasonably well, although averaging the fitted dust maps within annuli yields higher dust surface densities beyond R\,$\simeq$\,6--7\,kpc than fitting the annular mean flux densities. This is consistent with the exclusion of faint pixels by the uncertainty cuts applied after the pixel-by-pixel SED fitting, whereas averaging the flux densities before SED fitting retains their contribution and improves the SNR in the outer disc.

The fixed and variable-$\beta$ fits recover similar large-scale $\Sigma_{\rm d}$ profiles over 1\,<\,R\,<\,12.5\,kpc. The median absolute difference between the two profiles over this radial range is 6.7\,$\times$\,10$^{-3}$\,M$_\odot\,{\rm pc}^{-2}$. Both models identify the same radial location of the dust surface-density maximum and recover a similar outer decline, indicating that the radial dust distribution is relatively insensitive to whether $\beta$ is fixed or allowed to vary.

The effective dust temperature decreases systematically with galactocentric distance. In the fixed-$\beta$ fits, the dust temperature decreases from 25.4\,K at R\,=\,1.25\,kpc to 15.2\,K at R\,=\,12.25\,kpc. The decline is steepest in the inner disc and becomes shallower across the main spiral disc, consistent with a radial decrease in the mean intensity of the interstellar radiation field. In the variable-$\beta$ fits, $T_{\rm d}$ decreases from 26.3\,K at R\,=\,1.25\,kpc to 15.8\,K at R\,=\,10.25\,kpc. Beyond this radius the temperature rises again while the dust surface density becomes very low and the fitted emissivity index drops rapidly, reaching $\beta$\,=\,1.3 at R\,=\,12.25\,kpc. 

Over the radial range of 1.0\,<\,R\,<\,12.5\,kpc, the fitted emissivity index spans approximately $\beta$\,=\,1.3\,--\,2.4, with a median value of 1.9, consistent with the fixed value adopted in the main SMBB analysis. The radial changes in $\beta$ are accompanied by opposing changes in T$_{\rm d}$, where annuli with $\beta$\,<\,1.9 generally yield higher temperatures (average value of 20.7\,K) than the fixed-$\beta$ model, whereas those with $\beta$\,>\,1.9 yield lower temperatures (average value of 19.2\,K). This indicates that at least part of the observed structure may be produced by the T$_{\rm d}$-$\beta$ degeneracy.

\subsection{Gas-to-Dust Ratio}

We derived a gas-to-dust ratio (GDR) map using the dust mass surface density from the fixed-$\beta$ SMBB fits at the SPIRE \SI{250}{\um} resolution. For the atomic gas component, we used the naturally weighted VLA \HI\ moment-0 map from THINGS \citep{Walter2008}. \citet{Walter2008} noted that the M\,81 data are affected by difficulties in continuum subtraction because the \HI\ emission fills almost the entire observed bandpass, together with contamination from strong sidelobes associated with M\,82. Consistent with these reported limitations, the THINGS moment-0 map contains a prominent ring-like feature around the centre of M\,81, reaching integrated intensities of approximately 4\,$\times$\,10$^{4}$\,Jy\,beam$^{-1}$\,m\,s$^{-1}$, which is not reproduced in the independent, lower-resolution \HI\ map of \citet{deBlok2018}. We therefore replaced the affected central region of the THINGS map with the corresponding values from the \citet{deBlok2018} \HI\ map. The replacement was restricted to a small circular aperture enclosing the anomalous feature, corresponding to a maximum deprojected galactocentric distance of $\sim$2.8\,kpc, in order to retain the higher-resolution THINGS data over the remainder of the disc.

The THINGS \HI\ map was convolved to the common angular resolution of \SI{18.2}{\arcsec} and reprojected onto the dust surface-density grid before the central replacement was applied. The \citet{deBlok2018} \HI\ map was reprojected onto the same grid but was not additionally convolved because its native beam is broader than the adopted \SI{18.2}{\arcsec} resolution. The central replacement region therefore retains the lower effective resolution of the \citet{deBlok2018} \HI\ data (\SI{38.1}{\arcsec}\,$\times$\,\SI{30.9}{\arcsec} for natural weighted data), while the remainder of the map has a resolution of \SI{18.2}{\arcsec}. The integrated \HI\ intensity was converted to \HI\ surface density and multiplied by a factor of 1.36 to account for helium. We then calculated the \HI+He-based GDR as GDR\,=\,$\Sigma_{\rm \HI+He}$\,/\,$\Sigma_{\rm d}$.

\begin{figure}
\centering
\includegraphics[width=\linewidth]{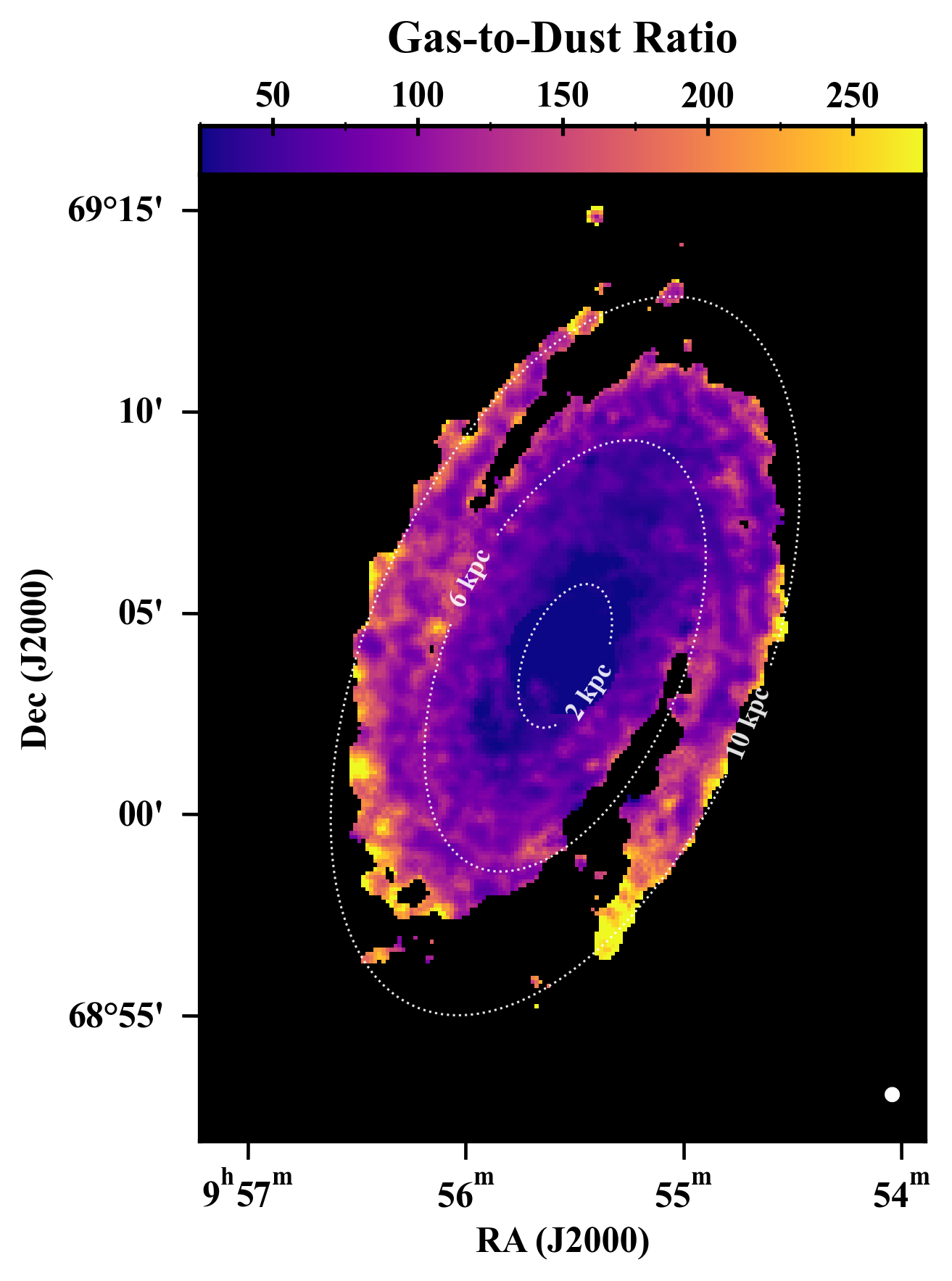}
\caption{Map of the \HI+He-based gas-to-dust ratio, $(1.36\,\Sigma_{\rm \HI})/\Sigma_{\rm d}$, across M\,81 at the common resolution used in the SED analysis (SPIRE \SI{250}{\um} resolution of \SI{18.2}{\arcsec}). Lower values indicate dust-enhanced regions relative to the local atomic gas surface density, while higher values are found in the more extended outer disc where dust emission is faint. The central replacement region at R\,<\,3\,kpc retains the lower effective resolution of the \citet{deBlok2018} \HI\ data.}
\label{fig:gdr}
\end{figure}

We also estimated the possible contribution from molecular gas using the JCMT/HARP CO(\textit{J}\,=\,3\,--\,2) observations of M\,81 obtained as part of the NGLS \citep{Sanchez-Gallego2011, Wilson2012}. The observations cover a rectangular region of approximately \SI{13.2}{\arcmin}\,$\times$\,\SI{7.7}{\arcmin}, roughly corresponding to the inner D$_{25}$/2 region of the galaxy. Significant CO(\textit{J}\,=\,3\,--\,2) emission was identified in only nine localized regions within this footprint (see Figure\,\ref{fig:all_data}), consistent with the weak CO emission previously reported for M\,81. We measured the CO flux in radial annuli by stacking the pixel spectra, assuming that the CO velocity at each pixel follows the local \HI\ velocity.

Both the CO moment-0 map and the cube are calibrated on the T$_{\rm A}^{*}$ temperature scale and were converted to main-beam brightness temperature using a main-beam efficiency of $\eta_{\rm MB}$\,=\,0.6. For the annular GDR analysis, the native-resolution CO spectra were shifted relative to the local THINGS \HI\ velocity and rebinned to 10\,km\,s$^{-1}$. The CO line window was taken to be within $\pm$\,50\,km\,s\,$^{-1}$ of the local \HI\ velocity, while the adjacent intervals between 50 and 150\,km\,s\,$^{-1}$ on either side were used to estimate the local baseline and channel noise. The mean spectra were then calculated in deprojected annuli of width 0.5\,kpc.

\begin{figure}
\centering
\includegraphics[width=\linewidth]{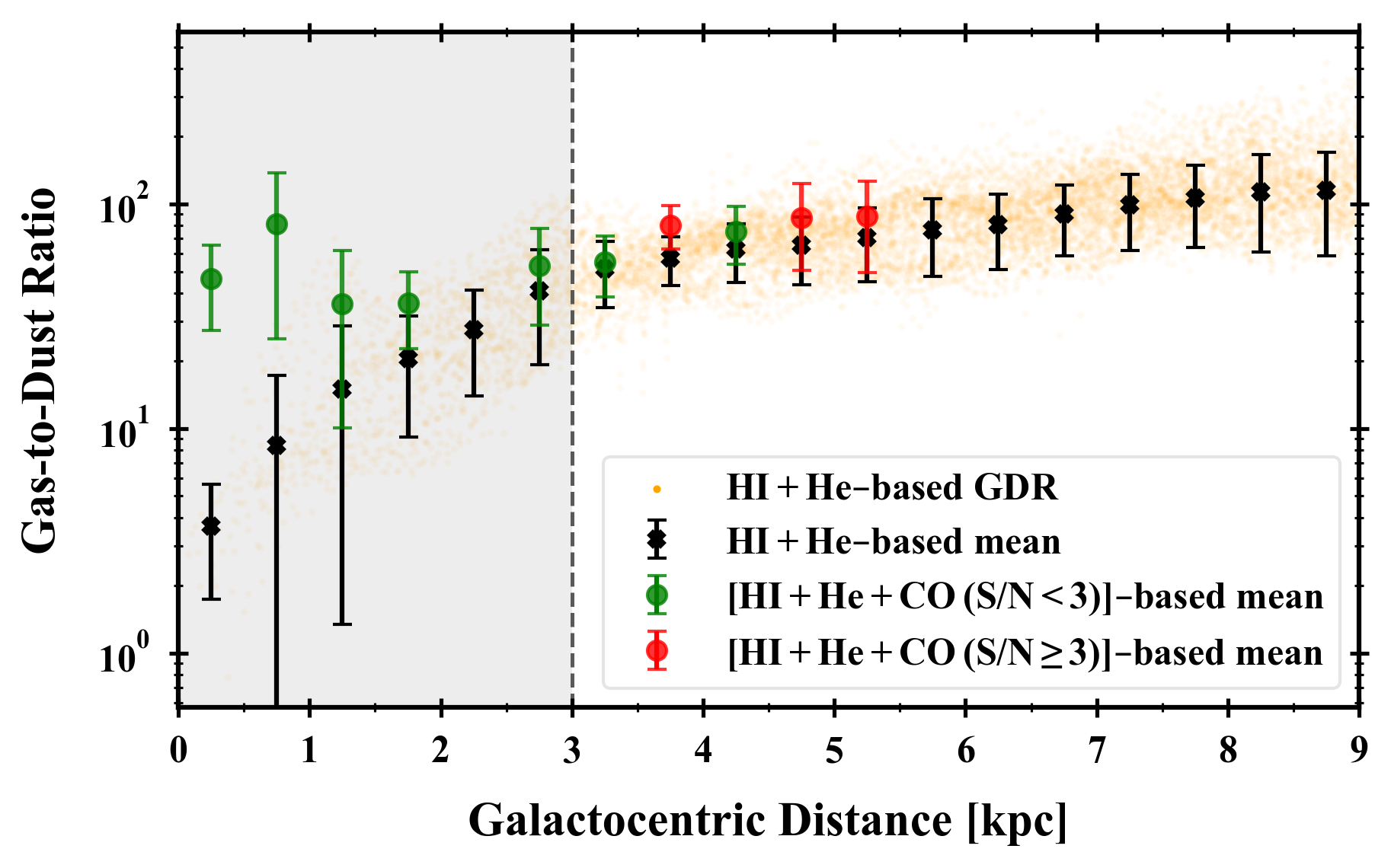}
\caption{Radial profile of the gas-to-dust ratio in M\,81. Orange points show the individual \HI+He-based pixel values from Figure\,\ref{fig:gdr}. Black points show the annular GDR, calculated as the ratio of the summed gas and dust surface densities within radial bins of width $\Delta R$\,=\,0.5\,kpc, and their error bars indicate the standard deviation of the pixel values in each bin. Green and red points show the CO-included GDR estimates for annuli with non-detected (SNR\,<\,3) and detected (SNR\,$\geq$\,3) CO emission, respectively. Annuli with negative measured CO intensities were excluded. The CO-included estimates are plotted only for annuli with complete CO observational coverage, extending to the annulus centred at R\,=\,5.25\,kpc. The shaded region at R\,<\,3\,kpc was excluded from the radial-gradient fit because it is affected by the central \HI\ replacement and is most sensitive to a molecular-gas contribution not included in the \HI+He-based profile.}
\label{fig:gdr_radial}
\end{figure}

The CO(\textit{J}\,=\,3\,--\,2) intensities were converted to equivalent CO(\textit{J}\,=\,1\,--\,0) intensities by adopting a line ratio of 0.43 \citep{Sanchez-Gallego2011}. We then adopted $\alpha_{\rm CO}$\,=\,4.35\,M$_\odot$\,${\rm pc}^{-2}$\,(K\,km\,s\,$^{-1})^{-1}$ including the contribution from helium \citep{Bolatto2013}. Before deriving the molecular-gas surface densities, the CO moment-0 map and the spectral cube were convolved to the SPIRE \SI{250}{\um} resolution and reprojected onto the same grid.

Annuli with CO measurements having SNR\,$\geq$\,3 were classified as detections, while those with SNR\,<\,3 were classified as low-significance measurements. Annuli with negative measured CO intensities were excluded from the CO-included comparison. The CO map extends only to $\sim$5.5\,kpc; however, beyond $\sim$3.0\,kpc the gas surface density is dominated by the atomic component, so omitting the molecular gas beyond the CO coverage is unlikely to significantly affect the results.

The resulting \HI+He-based GDR map is shown in Figure\,\ref{fig:gdr}. GDR remains relatively flat across most of M\,81, and has a median value of $\sim$86.0. The radial median GDR increases from $\sim$3.7 in the innermost annulus at R\,=\,0.25\,kpc to $\sim$27.6 at R\,=\,2.25\,kpc and $\sim$51.2 at R\,=\,3.25\,kpc. It continues to increase more gradually across the disc, reaching $\sim$70.5 at R\,=\,5.25\,kpc, $\sim$90.1 at R\,=\,6.75\,kpc, and $\sim$114.6 in the outermost annulus at R\,=\,8.75\,kpc (Figure\,\ref{fig:gdr_radial}). The systematic outward increase is clearly visible in both the resolved map and the azimuthally averaged profile (Figure\,\ref{fig:gdr_radial}). Spatially, the lowest GDR values occur in the central region, while the highest values are predominantly found toward the outer disc, indicating that the dust surface density decreases more rapidly with radius than the \HI\ surface density. A similar outward increase in GDR has been reported for M\,31 \citep{Smith2012}.

We assessed how molecular gas could modify this result using the measured annular CO intensities within the rectangular NGLS region. The centre of the outermost annulus with complete CO coverage is at R\,=\,5.25\,kpc. Beyond R\,=\,5.25\,kpc, CO observation coverage extend drops to $\sim$0.3\% of total pixels in the annulus centred at 9.25\,kpc.

In the innermost annulus, the positive but low-significance CO measurement increases the estimated GDR from $\sim$3.7 to $\sim$46.4, giving a value approximately 12.6 times higher than the \HI+He-only estimate. %Because this CO measurement has SNR\,<\,3, the magnitude of the correction is poorly constrained.
CO was detected in the annuli centred at R\,=\,3.75, 4.75, and 5.25\,kpc. Including the measured molecular-gas contribution increases the GDR from 57.3 to 80.4, from 65.5 to 86.8, and from 70.5 to 88.1, respectively. These correspond to increases of approximately 40\%, 33\%, and 25\%. The molecular-gas correction is therefore strongest and least constrained in the central region, but remains significant across the CO-covered disc.

The comparison therefore leads to two main conclusions. First, the outward increase in the \HI+He-based GDR across the disc is a robust observational result, showing that the atomic-gas and dust distributions do not decline radially at the same rate. The CO-included GDR broadly preserve this radial increase, although the low-significance measurements in the central region have large spatial scatter. Second, the molecular correction is moderate beyond R\,=\,3\,kpc, where the measured corrections are approximately 8\,--\,40 percent between 3.0 to 5.5\,kpc. Lastly, the exceptionally low central \HI+He-based GDR cannot be interpreted securely as the total-gas GDR, because the molecular contribution can increase the central value from approximately 3.7 to as much as 46.4. However, this central CO-included value remains poorly constrained.

We investigated whether the radial variation in the GDR is consistent with that expected from the metallicity gradient. We fitted $\log_{10}({\rm GDR})$ over 3\,<\,R\,<\,9\,kpc, excluding the central region because it is most sensitive to the molecular-gas contribution and is affected by the replacement of the THINGS \HI\ data. The fit gives an \HI+He-based dust-to-gas gradient of ${\rm d}\log_{10}({\rm GDR})/{\rm d}R$\,=\,0.065\,$\pm$\,0.002\,dex\,kpc$^{-1}$. This interval also overlaps the main-disc region sampled by the blue-supergiant measurements used by \citet{Kudritzki2012} to derive a metallicity gradient of ${\rm d}\log_{10}(Z)/{\rm d}R$\,=\,-0.033\,$\pm$\,0.009\,dex\,kpc$^{-1}$ at their adopted distance of 3.47\,Mpc. Rescaling this gradient to our adopted distance of 3.63\,Mpc gives -0.0316\,$\pm$\,0.0086\,dex\,kpc$^{-1}$. Combining this with our measured GDR gradient gives ${\rm d}\log_{10}(D/Z)/{\rm d}R$\,=\,-0.034\,$\pm$\,0.009\,dex\,kpc$^{-1}$.

The GDR relation is close to linear over this interval, with $R^2$\,=\,0.992 and a residual rms of 0.011\,dex. Taken at face value, this result suggests that the fraction of metals locked into dust decreases with galactocentric radius, similar to the behaviour reported for M\,31 by \citet{Mattsson2014HELGA}. This could be consistent with more efficient grain growth in the denser and more metal-rich inner disc. However, unlike \citet{Mattsson2014HELGA}, we do not fit a galactic dust-evolution model to the dust-to-metals profile.

The inferred dust-to-metals gradient is highly sensitive to the adopted metallicity gradient. Using the H{\sc ii}-region oxygen-abundance gradient of -0.088\,$\pm$\,0.013\,dex\,kpc$^{-1}$ from \citet{Stanghellini2014} gives a weak positive dust-to-metals gradient of 0.023\,$\pm$\,0.013\,dex\,kpc$^{-1}$. Restricting our GDR fit to 5\,<\,R\,<\,9\,kpc, which more closely matches the radial range of these H{\sc ii}-region measurements, gives a very similar gradient of 0.023\,$\pm$\,0.014\,dex\,kpc$^{-1}$. Both are therefore consistent with a radially constant dust-to-metals ratio. In contrast, adopting the shallower H{\sc ii}-region oxygen-abundance gradient of -0.013\,$\pm$\,0.006\,dex\,kpc$^{-1}$ from \citet{Patterson2012} gives a significantly negative dust-to-metals gradient of -0.052\,$\pm$\,0.006\,dex\,kpc$^{-1}$. 

The outward increase in the GDR is directly measured from our gas and dust maps. However, determining the radial variation in the fraction of metals locked into dust also requires an adopted metallicity gradient. The different published metallicity gradients imply either a decreasing, approximately constant, or weakly increasing dust-to-metals ratio with radius. The comparison with the blue-supergiant metallicities is consistent with a larger fraction of metals being locked into dust in the inner disc, potentially due to more efficient grain growth, but this interpretation is not supported by all published metallicity measurements.

\subsection{Star Formation Rate}
\label{sec:sfr}

We constructed a star formation rate (SFR) map of M\,81 using combined UV and IR maps. However evolved stellar populations can contribute to the UV emission and also heat dust, particularly in regions of low SFR. This can cause hybrid prescriptions with a spatially constant infrared coefficient to overestimate the SFR. This effect may be important in M\,81 because of its prominent bulge and relatively quiescent inner disc.

We therefore adopted the resolved calibration of \citet{Belfiore2023}, in which the coefficient multiplying the infrared emission varies with the FUV-to-WISE W1 luminosity ratio. This ratio acts as an empirical proxy for the relative contributions of young and evolved stellar populations and reduces the infrared contribution in regions where diffuse dust heating is expected to be important.

We retrieved the \textit{GALEX} FUV and NUV, \textit{WISE} W1 \SI{3.4}{\um}, and \textit{Spitzer} MIPS \SI{24}{\um} maps from the DustPedia archive \citep{Clark2018}. The NUV map was used only to identify foreground-star contamination. A median background was measured independently in each band within an elliptical annulus outside D$_{25}$ and was subtracted before the maps were combined. Milky Way foreground extinction was corrected using
\begin{equation}
    I_{\nu,\mathrm{corr}}
    =
    I_{\nu,\mathrm{obs}}\,
    10^{0.4R_{\lambda}E(B-V)},
\end{equation}
with E(B-V)\,=\,0.08 \citep{Makarova2026}. We adopted $R_{\rm FUV}$\,=\,8.01, $R_{\rm NUV}$\,=\,6.79 \citep{Wall2019}, and $R_{\rm W1}$\,=\,0.18 \citep{Yuan2013}. No foreground-extinction correction was applied at \SI{24}{\um}, where its effect is negligible. Each map was convolved from its native PSF to the SPIRE \SI{250}{\um} resolution of \SI{18.2}{\arcsec} using the corresponding convolution kernel. The convolved maps were then reprojected onto the SPIRE \SI{250}{\um} grid, which has a pixel size of \SI{6}{\arcsec}.

Foreground stars can increase the W1 intensity and thereby change the FUV/W1-dependent coefficient. Following the criterion used by \citet{Ford2013}, we identified candidate foreground-star pixels by requiring both $I_{\rm NUV}/I_{\rm FUV}$\,>\,15 and $I_{\rm NUV}$\,>\,$5\sigma_{\rm NUV}$. Identified pixels were masked consistently in all four input maps.

The hybrid luminosity calibration is
\begin{equation}
    {\rm SFR}
    =
    C_{\rm FUV}L_{\rm FUV}
    +
    C_{\rm 24}^{\rm FUV}(Q)L_{\rm 24},
    \label{eq:Belfiore_luminosity}
\end{equation}
where all monochromatic luminosities are expressed in energy units,
$L$\,$\equiv$\,$\nu L_\nu$, and
\begin{equation}
    Q
    =
    \frac{L_{\rm FUV}}{L_{\rm W1}}
    =
    \frac{\nu_{\rm FUV}I_{\nu,\rm FUV}}
         {\nu_{\rm W1}I_{\nu,\rm W1}}.
\end{equation}
We adopted $\log_{10}C_{\rm FUV}$\,=\,-43.42, with $C$ in $\mathrm{M_\odot\,yr^{-1}/(erg\,s^{-1})}$.

\begin{figure}
\centering
\includegraphics[width=\linewidth]{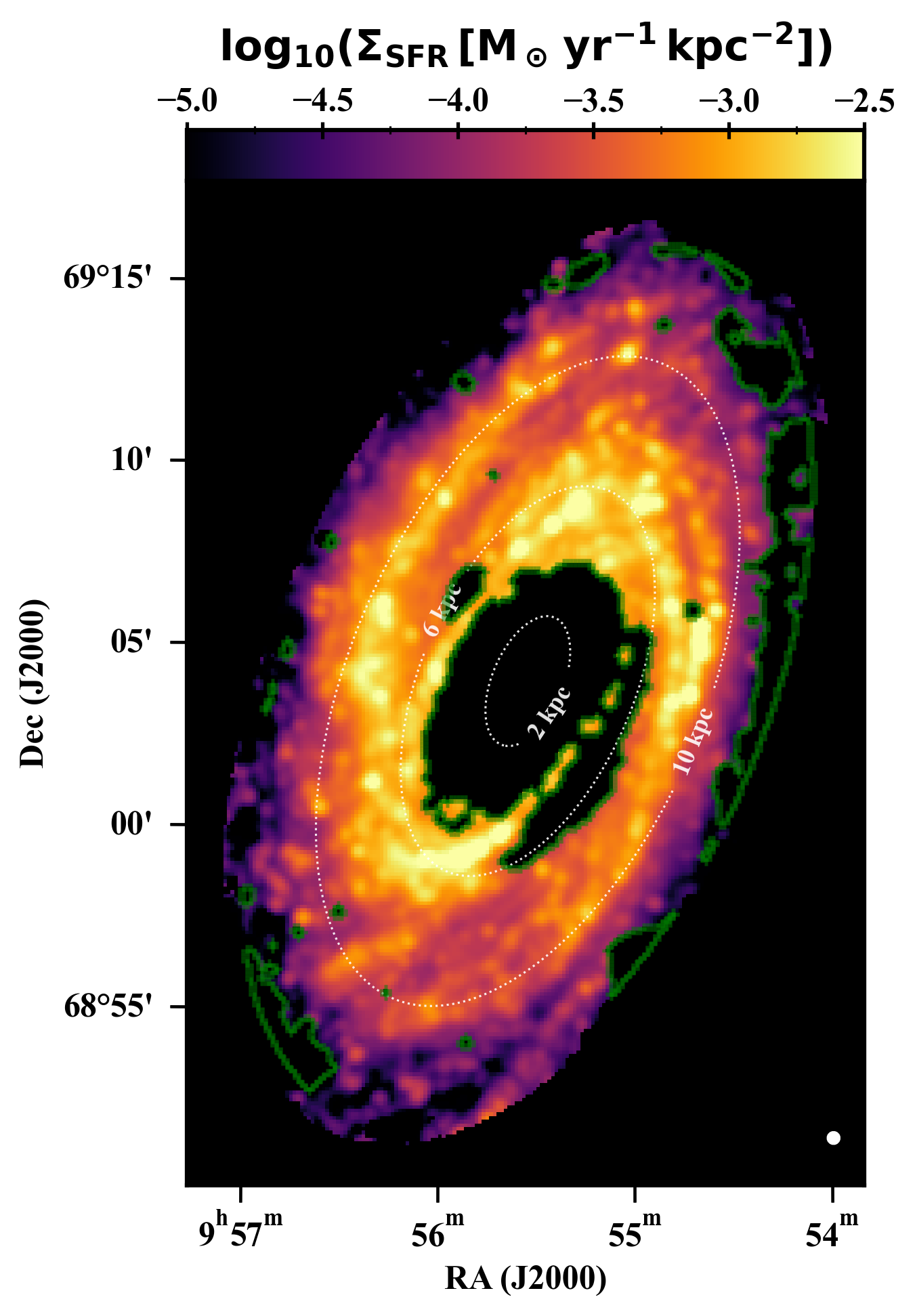}
\caption{Inclination-corrected SFR surface-density map of M\,81, derived from the \textit{GALEX} FUV and \textit{Spitzer} MIPS \SI{24}{\um} maps using the FUV/W1-dependent calibration of \citet{Belfiore2023}. The map is at the SPIRE \SI{250}{\um} resolution of \SI{18.2}{\arcsec}. The green contour encloses pixels with $\log_{10}(L_{\rm FUV}/L_{\rm W1})$\,<\,-0.64, which lie outside the calibrated range and were excluded. The dotted ellipses indicate deprojected galactocentric radii of 2, 6, and 10\,kpc, and the white circle shows the common PSF.}
\label{fig:SFR_maps}
\end{figure}

\begin{figure}
\centering
\includegraphics[width=\linewidth]{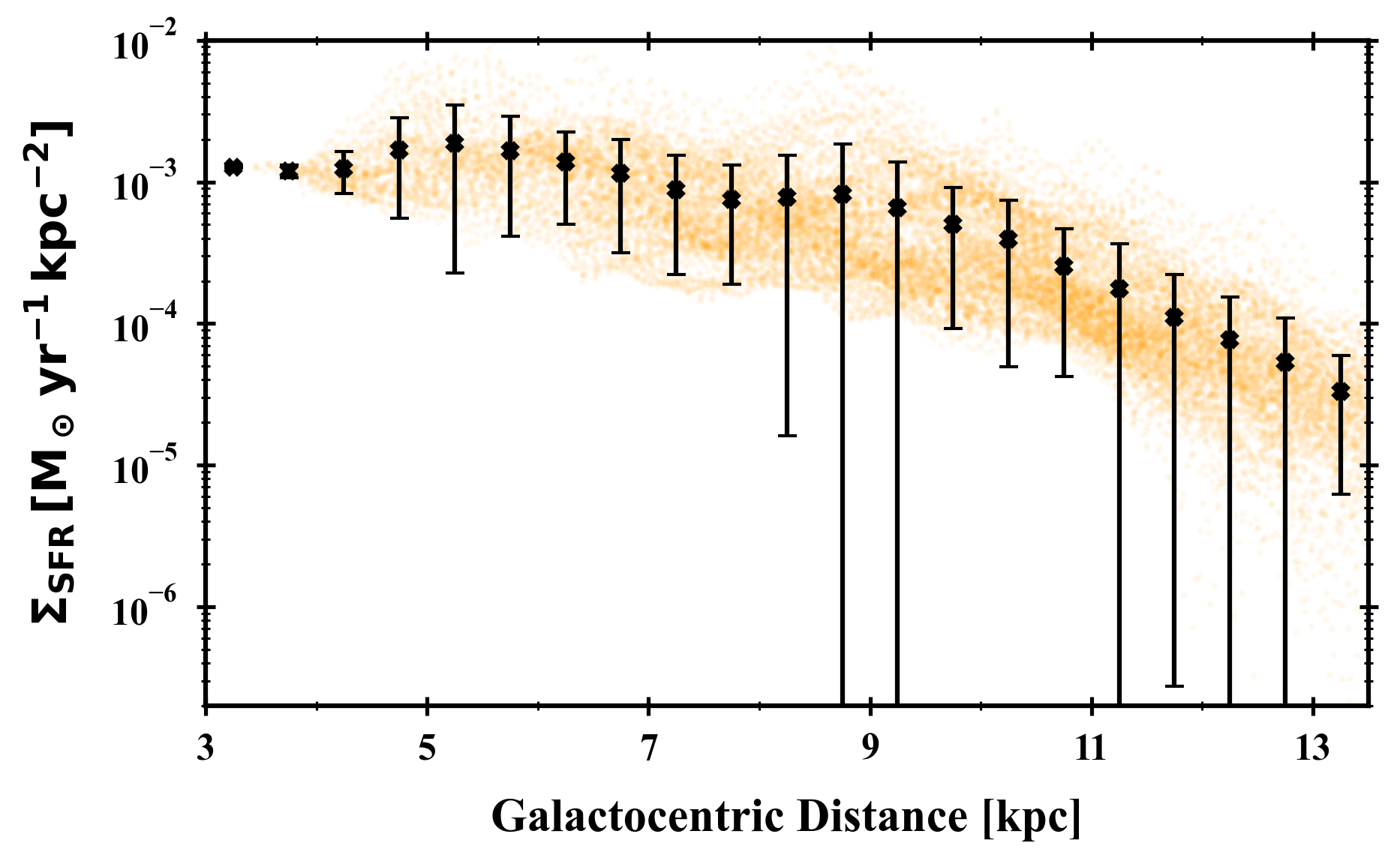}
\caption{Deprojected radial profiles of the inclination-corrected SFR surface density. Orange points show individual valid pixels. The black points show the radial mean profiles, calculated within radial bins of width $\Delta R$\,=\,0.5\,kpc. The error bars indicate the standard deviation of the pixel values within each annulus. The absence of a Belfiore profile inside approximately 3\,kpc results from the lower FUV/W1 calibration limit, $\log_{10}Q_{\rm min}$\,=\,-0.64. The profiles are shown only out to the first radial bin in which fewer than 50\% of the pixels assigned to that bin have defined values.}
\label{fig:SFR_radial}
\end{figure}

The calibration of \citet{Belfiore2023} was derived using WISE W4 \SI{22}{\um} emission. They find $\log_{10}L_{24}$\,=\,$\log_{10}L_{\rm W4}$\,-0.08 for matched MIPS \SI{24}{\um} and W4 measurements. We therefore increased the logarithmic infrared coefficient by 0.08\,dex when applying the prescription to the MIPS \SI{24}{\um} map. Defining $q\equiv\log_{10}Q$, the W4 coefficient is
\begin{equation}
\log_{10}C_{\rm 24}^{\rm FUV}
=
\log_{10}C_{\rm max,24}
+a_1\min(q-q_{\rm max},0),
\qquad q\geq q_{\rm min},
\end{equation}
where $q_{\rm max}$\,=\,0.60, $q_{\rm min}$\,=\,-0.64, $a_1$\,=\,0.23, and $\log_{10}C_{\rm max}$\,=\,-42.73\,+\,0.08\,=\,-42.65 \citep[Table\,3 in][]{Belfiore2023}. 

The lowest FUV/W1 ratio directly sampled by the PHANGS calibration is $\log_{10}Q_{\rm min}$\,=\,-0.64, corresponding to the fifth percentile of its calibration sample. This boundary is a limit of the empirical model rather than a physical threshold in star formation, and \citet{Belfiore2023} advise against extrapolating below it. We therefore excluded pixels with $q$\,<$q_{\rm min}$. Consequently, the absence of an SFR measurement across much of the centre of M\,81 indicates that its FUV/W1 colour lies outside the calibrated domain, largely because the W1 emission is dominated by the evolved stellar population in the prominent bulge, rather than demonstrating an absence of central star formation.

The SFR distribution follows the spiral-arm structure more closely than a simple monotonic radial trend. The radial mean $\Sigma_{\rm SFR}$ peaks at R\,=\,5.25\,kpc, reaching 1.86\,$\times$\,10$^{-3}$\,$\mathrm{M_\odot\,yr^{-1}\,kpc^{-2}}$, and declines progressively towards the outer disc. Integrating unmasked pixels within the D$_{25}$ gives SFR\,=\,0.27\,$\pm$\,0.03\,$\mathrm{M_\odot\,yr^{-1}}$. The quoted uncertainty combines the photometric calibration uncertainties of the input maps (see Table\,1 of \citealt{Clark2018}), the uncertainty in the IR coefficient normalisation, and the scatter in the conversion between MIPS \SI{24}{\um} and \textit{WISE} W4 (see Table\,3 and Section\,2.2 of \citealt{Belfiore2023}, respectively). Our SFR is slightly lower than the value of 0.345\,$\pm$\,0.130\,M$_\odot$\,yr$^{-1}$ obtained by \citet{Nersesian2019} from integrated CIGALE SED fitting assuming a Salpeter IMF \citep{Salpeter1955}, but falls within their reported uncertainty.

\subsection{Star Formation Efficiency}

\begin{figure}
\centering
\includegraphics[width=\linewidth]{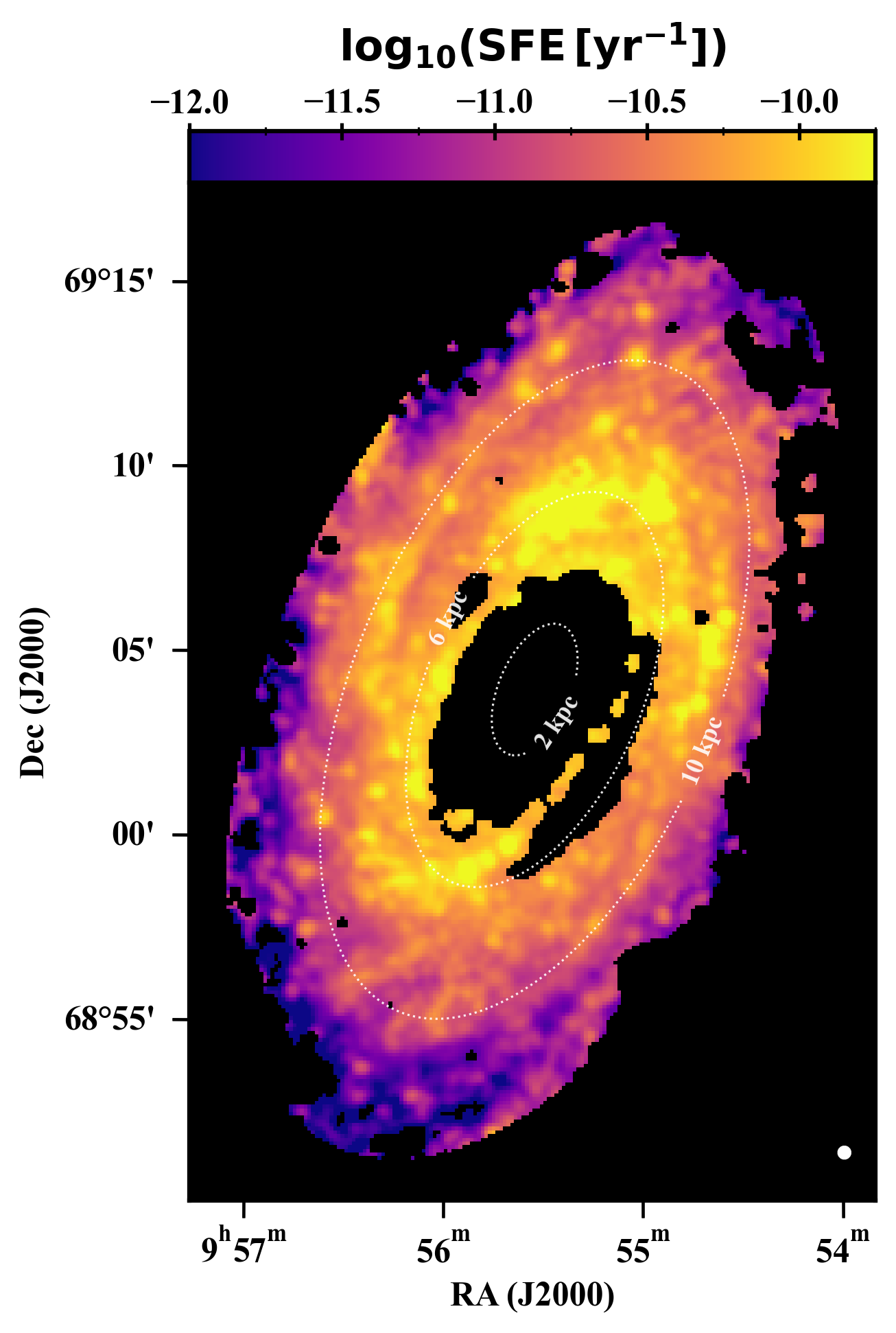}
\caption{\HI+He-based star formation efficiency in M\,81, calculated as ${\rm SFE}$\,=$\Sigma_{\rm SFR}$\,/\,$\Sigma_{\rm \HI+He}$ using the \citet{Belfiore2023} SFR map. The analysis was performed at the common SPIRE \SI{250}{\micron} resolution of \SI{18.2}{\arcsec}. The white dotted ellipses indicate deprojected galactocentric radii of 2, 6, and 10\,kpc. The absence of central values results primarily from the adopted lower FUV/W1 calibration boundary in the SFR calculation.}
\label{fig:SFE_maps}
\end{figure}

To investigate how recent star formation varies relative to the atomic-gas reservoir in M\,81, we constructed a spatially resolved star formation efficiency (SFE) map by dividing the SFR surface density by the atomic-gas surface density ($\Sigma_{\rm \HI+He}$). Figure\,\ref{fig:SFE_maps} shows that SFE is generally higher across the main spiral disc than in the outer \HI-dominated regions. It has a median value of $\log_{10}({\rm SFE}/{\rm yr}^{-1})$\,=\,-10.59, corresponding to atomic-gas depletion times of approximately 38.9\,Gyr.

\begin{figure}
\centering
\includegraphics[width=\linewidth]{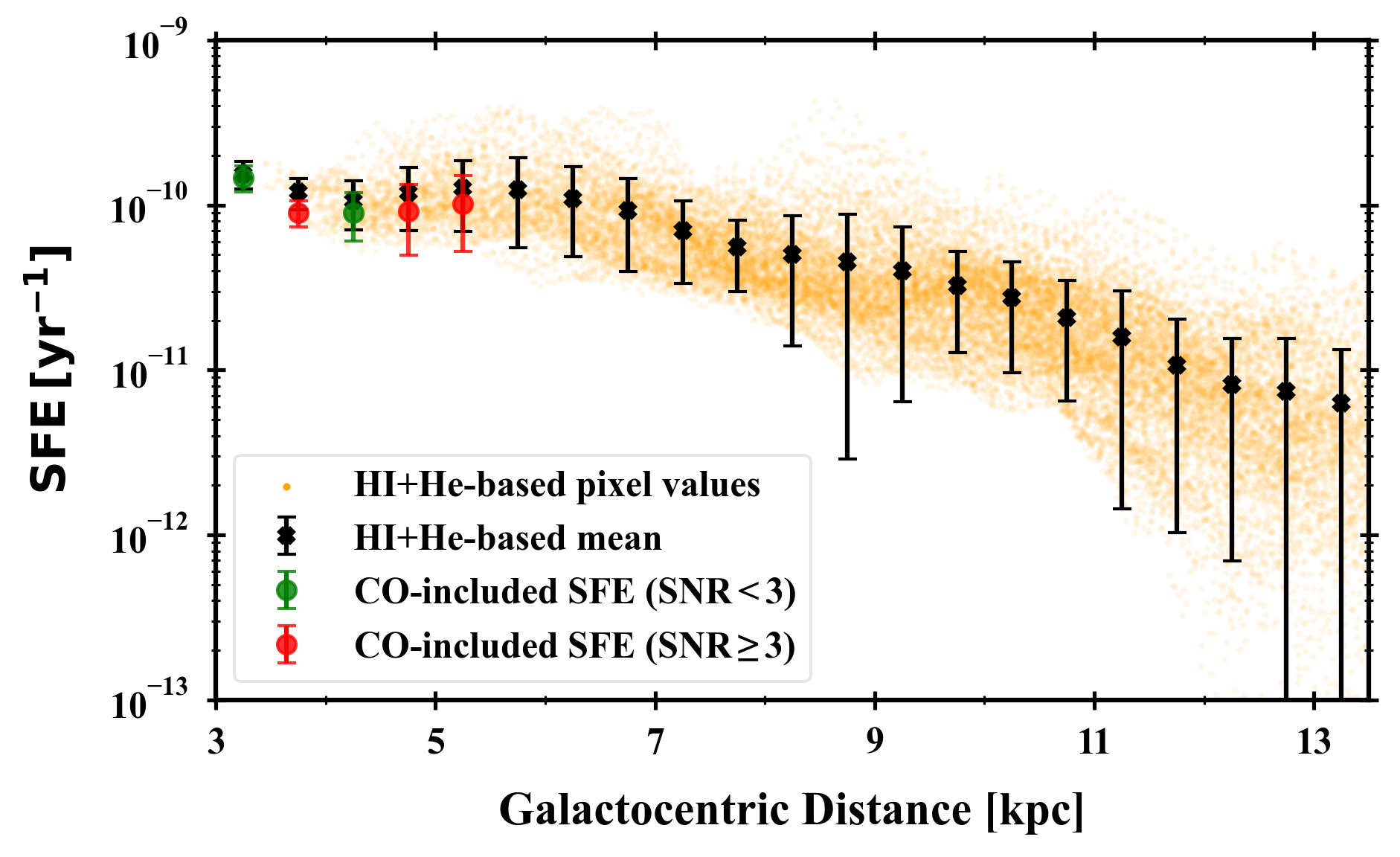}
\caption{Radial profile of SFE calculated using the SFR prescriptions of \citet{Belfiore2023}. Orange points show individual \HI+He-based pixel values. Black points show the radial mean profiles, calculated within radial bins of width $\Delta R$\,=\,0.5\,kpc, while the error bars indicate the standard deviation of the pixel values in each bin. Green and red points show the CO-included SFE estimates for annuli with non-detected (SNR\,<\,3) and detected (SNR\,$\geq$\,3) CO emission, respectively. CO-included radial mean points are shown only for annuli with complete CO observational coverage, up to R\,=\,5.25\,kpc. The profiles are shown only out to the first radial bin in which fewer than 50\% of the pixels assigned to that bin have defined values.}
\label{fig:SFE_radial}
\end{figure}

We calculated the radial mean SFE in deprojected annuli of width $\Delta R$\,=\,0.5\,kpc (Figure\,\ref{fig:SFE_radial}). Including the CO-traced molecular component produces only a negligible change relative to the \HI+He-based profile. The radial profile is approximately flat between 3.25 and 6.25\,kpc, where the mean ${\rm SFE}_{\rm atom}$ remains within (1.1\,--\,1.5)\,$\times$\,10$^{-10}$\,yr$^{-1}$, before declining systematically towards larger radii, reaching $\sim$\,2.8\,$\times$\,10$^{-11}$\,yr$^{-1}$ at 10.25\,kpc.

The mean CO-inclusive efficiencies range from 0.9 to 1.5\,$\times$10$^{-10}$\,yr$^{-1}$ between 3.25 and 5.25\,kpc. Including molecular gas reduces the mean SFE by 4.8\,--\,24.6\%, equivalent to 0.02\,--\,0.12\,dex. For the CO detections at 3.75, 4.75, and 5.25\,kpc, the reductions are 24.6, 23.5, and 20.3\%, respectively. The non-detected measurements at 3.25 and 4.25\,kpc produce reductions of 4.8 and 15.0\%, respectively. These latter estimates are more sensitive to the uncertain CO measurements and are not treated as formal limits.

The molecular-gas correction therefore lowers the inferred SFE but does not change its order of magnitude or its approximately flat behaviour across the inner spiral disc. For each annulus included in the CO measurements, the corresponding annular molecular-gas surface density was added to the SFE calculation. Complete CO coverage extends only to 5.25\,kpc, consequently it does not provide any information for the region beyond $\sim$\,6\,kpc where the \HI-based SFE begins to decline. Nevertheless, the subsequent decline in the \HI-based SFE is consistent with an increasingly inefficient conversion of atomic gas into the molecular phase in the outer disc \citep{Leroy2008}.

\subsection{Morphology-Dependent Dust and ISM Properties}
\label{sec:morph_comparison}

\begin{figure}
\centering
\includegraphics[width=\linewidth]{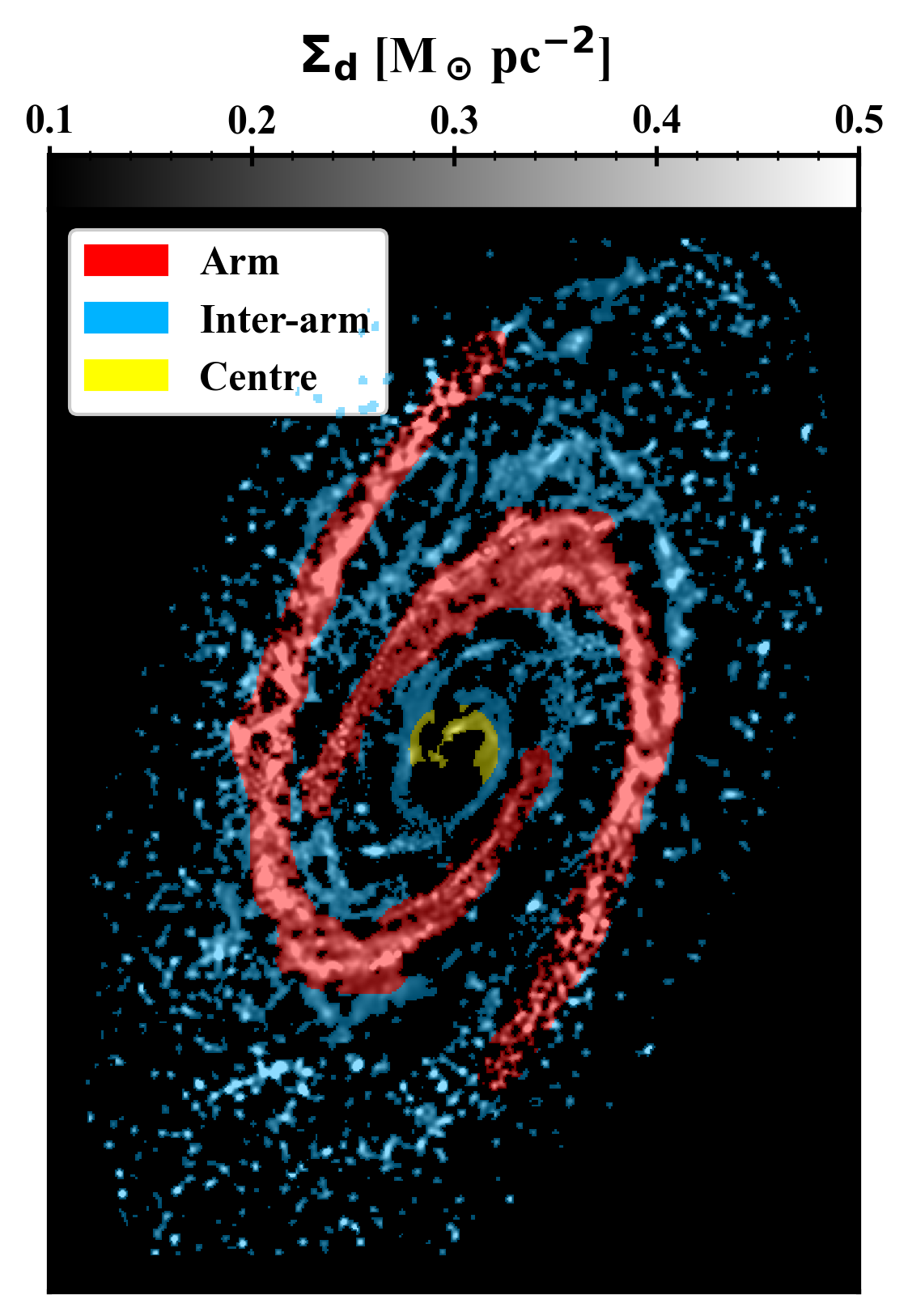}
\caption{Morphological masks used for the regional analysis of M\,81, shown on the dust mass surface density map derived using PPMAP. The spiral arms, central region, and inter-arm disc are shown in red, yellow, and blue, respectively, overlaid on the dust mass surface density map derived from PPMAP. These masks, defined manually in DS9 on the SPIRE \SI{250}{\um} image, are applied consistently to all subsequent maps.}
\label{fig:Selection_over_MBB_Surface_Density}
\end{figure}

To investigate how the dust and ISM properties vary with galactic environment, we separate M\,81 into three morphological components: spiral arms, inter-arm regions, and the central region. We used SAOImage DS9 to define the spiral arms manually on the SPIRE \SI{250}{\um} image, tracing the bright dust lanes with the polygon selection tool. The central region is defined as a circular aperture of radius 1\arcmin\ ($\sim$1.06\,kpc) centred on the galaxy centre. The inter-arm component is then defined as all remaining pixels inside the $D_{25}$ ellipse, outside the central aperture and not assigned to the spiral-arm masks. These region definitions are subsequently applied to all maps in this work. Figure\,\ref{fig:Selection_over_MBB_Surface_Density} illustrates the resulting masks overlaid on the PPMAP's dust mass surface density map (see left panel in Figure\,\ref{fig:PPMAP_fb}).

For the dust surface density, both the SMBB and PPMAP results show clear environmental differences between the spiral arms and inter-arm regions (Table\,\ref{tab:morph_stats}). In the SMBB maps, the median dust surface density in the spiral arms is $\Sigma_{\rm d}$\,=\,0.19\,M$_\odot\,{\rm pc}^{-2}$, with a 16th\,--\,84th percentile range of 0.12\,--\,0.29\,M$_\odot\,{\rm pc}^{-2}$. This is approximately a factor of two higher than in the inter-arm regions, where the median is 0.09\,M$_\odot\,{\rm pc}^{-2}$, with a 16th\,--\,84th percentile range of 0.06\,--\,0.14\,M$_\odot\,{\rm pc}^{-2}$. However, such an enhancement is expected to some extent because the spiral-arm regions were defined using the bright dust-emission structure. The central region has a lower median surface density, $\Sigma_{\rm d}$\,=\,0.08\,M$_\odot\,{\rm pc}^{-2}$, closer to the inter-arm values. The Mann--Whitney U and Kolmogorov--Smirnov tests indicate a clear difference between the arm and inter-arm SMBB dust surface density distributions ($D_{\rm KS}$\,=\,0.599). The large positive Cliff's delta, $\delta$\,=\,0.76, means that a randomly selected arm pixel has a higher dust surface density than a randomly selected inter-arm pixel in approximately 88\% of pairwise comparisons.

\begin{table*}
\centering
\caption{Median dust and ISM properties in the centre, spiral arms, and inter-arm regions of M\,81. Values in brackets give the 16th\,--\,84th percentile range. The Mann--Whitney U and KS tests compare the arm and inter-arm pixel distributions, which are statistically significant for all quantities (p\,<\,4\,$\times$\,10$^{-4}$).}
\label{tab:morph_stats}
\begin{tabular}{lccccc}
\hline
Quantity & Centre & Arm & Inter-arm & $D_{\rm KS}$ & $\delta_{\rm Cliff}$ \\
\hline
SMBB $\Sigma_{\rm d}$
& $0.08\,[0.05,0.12]$
& $0.19\,[0.12,0.29]$
& $0.09\,[0.06,0.14]$
& $0.60$
& $0.76$ \\

SMBB $T_{\rm d}$
& $25.78\,[24.15,27.64]$
& $19.62\,[18.11,21.04]$
& $20.35\,[18.58,22.20]$
& $0.20$
& $-0.26$ \\

PPMAP $\Sigma_{\rm d}$
& $0.15\,[0.11,0.20]$
& $0.26\,[0.17,0.44]$
& $0.16\,[0.11,0.26]$
& $0.40$
& $0.55$ \\

PPMAP $T_{\rm d}$
& $22.70\,[21.89,25.22]$
& $17.64\,[15.19,19.33]$
& $15.17\,[12.32,18.64]$
& $0.35$
& $0.42$ \\

$\log_{10}\!\left(\Sigma_{\rm SFR}/[{\rm M_\odot\,yr^{-1}\,kpc^{-2}}]\right)$
& $--$
& $-2.90\,[-3.14,-2.65]$
& $-3.78\,[-4.52,-3.24]$
& $0.76$
& $0.89$ \\

\HI+He-based GDR
& $8.01\,[5.54,14.45]$
& $88.25\,[56.17,135.92]$
& $86.61\,[42.92,152.69]$
& $0.13$
& $0.04$ \\
\hline
\end{tabular}

\begin{flushleft}
\footnotesize
Notes: Dust surface densities are in M$_\odot\,{\rm pc}^{-2}$ and temperatures are in K. The PPMAP temperature is the mass-weighted temperature. No central $\Sigma_{\rm SFR}$ value is reported because the final SFR map contains no valid pixels within the central aperture after masking. The GDR is calculated using \HI\ and helium rather than the total gas surface density.
\end{flushleft}
\end{table*}

The PPMAP dust surface density map shows the same qualitative behaviour but with higher absolute surface densities. The median PPMAP dust surface density is 0.26\,M$_\odot\,{\rm pc}^{-2}$ in the spiral arms and 0.16\,M$_\odot\,{\rm pc}^{-2}$ in the inter-arm regions, corresponding to an arm/inter-arm contrast of $\sim$1.6. The corresponding 16th\,--\,84th percentile ranges are 0.17\,--\,0.44\,M$_\odot\,{\rm pc}^{-2}$ and 0.11\,--\,0.26\,M$_\odot\,{\rm pc}^{-2}$, respectively. The central region again lies closer to the inter-arm distribution, with a median of 0.15\,M$_\odot\,{\rm pc}^{-2}$. The arm and inter-arm PPMAP distributions are also strongly separated ($D_{\rm KS}$\,=\,0.399; $\delta$\,=\,0.55). Thus, both fitting methods show that the spiral arms have systematically higher dust columns than the surrounding inter-arm disc, although the contrast is stronger in the SMBB map than in the PPMAP reconstruction.

The dust temperature shows a strong central enhancement and a method-dependent arm/inter-arm contrast (Table\,\ref{tab:morph_stats}). In the central region, both methods give substantially warmer dust than in the disc. The median central temperature is $T_{\rm d}$\,=\,25.78\,K in the SMBB map, with a 16th\,--\,84th percentile range of 24.15\,--\,27.63\,K, and $T_{\rm d}$\,=\,22.70\,K in the PPMAP mass-weighted temperature map, with a 16th\,--\,84th percentile range of 21.89\,--\,25.22\,K. This central enhancement is consistent with a stronger ISRF in the bulge and nuclear environment.

Outside the centre, the arm/inter-arm temperature contrast depends on the fitting method. In the SMBB map, the inter-arm regions are slightly warmer than the spiral arms, with median temperatures of 20.35\,K and 19.62\,K, respectively. The median offset is only $\sim$0.7\,K, and the distributions show a modest separation ($D_{\rm KS}$\,=\,0.199; $\delta$\,=\,-0.263). By contrast, PPMAP gives higher mass-weighted temperatures in the spiral arms than in the inter-arm regions, with medians of 17.64\,K and 15.17\,K, respectively. The PPMAP arm and inter-arm temperature distributions show a more pronounced separation ($D_{\rm KS}$\,=\,0.346; $\delta$\,=\,0.418).

This difference likely reflects the different treatment of temperature structure in the two methods. The SMBB fit returns a single effective temperature in each pixel, which can be biased by line-of-sight mixtures of warm and cold dust. PPMAP instead distributes the dust mass over a temperature components. The individual temperature slices (Figure\,\ref{fig:PPMAP_perT}) reveal a spatial variation in the temperature distribution: the colder components extend into the outer spiral structure, the 19\,K component traces the spiral arms most prominently, and the warmer components become progressively more concentrated toward the inner disc and nucleus. The two methods are therefore not expected to yield identical arm/inter-arm temperature contrasts. Nevertheless, both agree on the principal result that the nuclear region is substantially warmer than the disc, while the magnitude of the arm/inter-arm temperature contrast depend on the adopted fitting method.

The origin of this central enhancement is likely connected to the stronger ISRF in the inner disc and bulge, where both young and evolved stellar populations may contribute to dust heating. Previous studies of M\,81 have argued that the total stellar population plays an important role in heating dust emitting at wavelengths longer than \SI{160}{\um} \citep{Bendo2012}. Similarly, radiative-transfer modelling of M\,31 found that evolved stellar populations dominate the dust heating, accounting for 91\% of the absorbed luminosity, with the bulge producing a strong radiation field \citep{Viaene2017}. In addition, the SKIRT model of M\,81 showed that only 10.4\% of the heating within the central 3\,kpc is caused by young stars \citep{Verstocken2020}. This supports our interpretation that the temperature enhancement in the central region is primarily associated with the evolved stellar populations. We also note that M\,81 hosts a well-known low-luminosity AGN \citep{Ho1996}, so additional non-stellar heating may contribute locally in the innermost region. Overall, we interpret the central temperature enhancement primarily as a signature of stellar/bulge dust heating, consistent with previous FIR/sub-mm studies.

The morphological separation is also apparent in the SFR and GDR maps. The median star formation rate surface density in the spiral arms is $\log_{10}(\Sigma_{\rm SFR}/[M_\odot\,{\rm yr}^{-1}\,{\rm kpc}^{-2}])$\,=\,-2.90, with a 16th\,--\,84th percentile range of -3.14 to -2.65. In the inter-arm regions, the median is lower, $\log_{10}(\Sigma_{\rm SFR}/[M_\odot\,{\rm yr}^{-1}\,{\rm kpc}^{-2}])$\,=\,-3.78, with a 16th\,--\,84th percentile range of -4.52 to -3.24. This corresponds to an arm enhancement of $\Delta\log_{10}\Sigma_{\rm SFR}$\,$\simeq$\,0.88\,dex, or a factor of $\sim$7.6 at the median. The Mann--Whitney U and Kolmogorov--Smirnov tests show that the arm and inter-arm $\Sigma_{\rm SFR}$ distributions are strongly separated ($D_{\rm KS}$\,=\,0.76). The large  Cliff's delta $\delta$\,=\,0.89, indicates that an arm pixel has a higher $\Sigma_{\rm SFR}$ than an inter-arm pixel in approximately 94\% of pairwise comparisons, consistent with the spiral arms hosting most of the recent star formation traced in Figure\,\ref{fig:SFR_maps}. No central value is reported because the adopted SFR map contains no valid pixels within the central aperture after the final masking.

In contrast to $\Sigma_{\rm SFR}$, the \HI-based GDR shows no substantial shift between the spiral arms and inter-arm regions. The median GDR is 88.25 in the spiral arms and 86.61 in the inter-arm regions, with broad 16th\,--\,84th percentile ranges of 56.17\,--\,135.92 and 42.92\,--\,152.69, respectively. Although the formal two-sample tests indicate differences between the distributions ($D_{\rm KS}$\,=\,0.126), the median difference is only approximately 2\%, and Cliff's delta is close to zero ($\delta$\,=\,0.04). The small Cliff's delta indicates that the typical arm and inter-arm GDR values are essentially indistinguishable.

The central region has a much lower median \HI-based GDR of 8.01, with a 16th\,--\,84th percentile range of 5.54\,--\,14.45. This is most naturally associated with the low \HI\ surface density relative to the dust surface density in the central region. Overall, the region-based statistics show that the spiral arms host enhanced dust surface density and strongly enhanced $\Sigma_{\rm SFR}$, whereas the nuclear region is characterised by warmer dust and a substantially reduced \HI-based GDR. The typical \HI-based GDR of the spiral arms, however, is not meaningfully different from that of the inter-arm disc.

\section{Conclusion}
\label{sec:Conclusion}

We have presented a resolved FIR/sub-mm study of dust in M\,81 using \textit{Herschel} PACS/SPIRE and JCMT SCUBA-2 observations from 70 to \SI{850}{\um}. We fitted the dust SEDs with an SMBB model using conventional pixel-by-pixel SED fitting and PPMAP that recovers dust surface density over a grid of temperatures. The fixed-$\beta$ SMBB analysis with $\beta$\,=\,$1.9$ gives an integrated dust mass of (1.77\,$\pm$\,0.09)\,$\times$\,10$^{7}\,M_\odot$, while PPMAP gives a higher dust mass of (2.65\,$\pm$\,0.13)\,$\times$\,10$^{7}\,M_\odot$. This mass difference is consistent with the fact that the SMBB fit returns a single effective temperature for each line of sight, which can favour the warmer dust component and therefore lead to a lower dust mass than recovered by PPMAP. Both methods show that the dust surface density is enhanced along the spiral arms, while the dust temperature is highest in the central regions and decreases with galactocentric radius. The SMBB temperature map has a median temperature of 20.1\,K, while the PPMAP mass-weighted temperature map has a lower median value of 16.5\,K.

The variable-$\beta$ fits show that the emissivity index is difficult to constrain over the full disc at high resolution. At \SI{250}{\um} resolution, reliable $\beta$ measurements are mainly restricted to the bright spiral arms after applying an uncertainty cut, whereas the \SI{500}{\um}-resolution fit provides more continuous spatial coverage because it includes the additional SPIRE \SI{350}{\um} and \SI{500}{\um} bands. However, because $\beta$ is only well constrained in the brightest regions and remains strongly degenerate with dust temperature, we adopt a fixed emissivity index for the main dust mass and temperature analysis.

Combining the dust maps with the \HI\ data, we find that the \HI+He-based GDR increases systematically with radius, from $\sim$4 in the innermost annulus at R\,=\,0.25\,kpc to $\sim$115 by R\,=\,8.75\,kpc. These values should be regarded as lower bounds on the total-gas GDR because the molecular-gas contribution is not included. A linear fit to the radial profile over 3\,<\,R\,<\,9\,kpc gives ${\rm d}\log_{10}({\rm GDR})/{\rm d}R$\,=\,0.065\,$\pm$\,0.002\,dex\,kpc$^{-1}$. Comparison with the blue-supergiant metallicity gradient gives a dust-to-metals gradient of -0.034\,$\pm$\,0.009\,dex\,kpc$^{-1}$, suggesting that the fraction of metals locked into dust may decrease with galactocentric radius. Following the slope-based argument used in the HELGA analysis of M\,31, this could be consistent with more efficient grain growth in the denser, metal-rich inner disc. However, we interpret this comparison as indicative because the inferred dust-to-metals gradient depends strongly on the adopted metallicity gradient, while the gas map is \HI-dominated and the molecular contribution is least certain in the centre of this CO-poor galaxy.

Finally, $\Sigma_{\rm SFR}$ is strongly concentrated along the spiral arms, giving a global SFR of 0.27\,$\pm$\,0.03\,M$_\odot\,{\rm yr}^{-1}$ from the valid pixels in the Belfiore SFR map. A morphological comparison of the spiral arms, inter-arm regions, and centre shows that the spiral arms host enhanced dust surface densities and strongly enhanced $\Sigma_{\rm SFR}$, whereas the nuclear region is characterised by warmer dust and a substantially reduced \HI-based gas-to-dust ratio. Including the CO-traced molecular-gas contribution raises the inferred central GDR and therefore reduces the contrast with the disc. The typical \HI-based GDR in the spiral arms is nevertheless similar to that in the inter-arm disc. Together, these results show that M\,81 is a spiral galaxy with modest star formation, in which the dust and recent star formation remain strongly structured by the spiral arms, while the central region is dominated by warmer dust and an enhanced radiation field.

\section*{Data Availability}

SPIRE and PACS data used in this article are available in NASA/IPAC Infrared Science Archive (IRSA)\footnote{\url{https://irsa.ipac.caltech.edu/data/Herschel/DustPedia/overview.html}}. SCUBA-2 raw data can be downloaded from Canadian Astronomy Data Centre with proposal ID of M20AL011.

\section*{Acknowledgements}

This work is supported by the UKRI AIMLAC CDT and the Science and Technology Facilities Council, funded by grant EP/S023992/1. MWLS and SE are grateful to the Science and Technology Facilities Council (consolidated grant ST/K000926/1 and UKRI5123) for financial support. 
% I. De Looze
IDL acknowledges funding from the European Research Council (ERC) under the European Union's Horizon 2020 research and innovation program DustOrigin (ERC-2019-StG-851622) and funding from the Flemish Fund for Scientific Research (FWO-Vlaanderen) through the research project 3G023821.
% Lapo Fanciullo
LF acknowledges support from the National Science and Technology Council of Taiwan under grant No. 114-2811-M-032-005.
% Luis Ho
LCH was supported by the National Science Foundation of China (12233001), the National Key R\&D Program of China (2022YFF0503401), and the China Manned Space Program (CMS-CSST-2025-A09).
% F. Kemper
FK acknowledges support from the Spanish Ministry of Science, Innovation and Universities, under project PID2023-149918NB-I00, financed by MCIU/AEI/10.13039/501100011033 / FEDER, EU. FK was also partly supported by the Spanish program Unidad de Excelencia María de Maeztu CEX2020-001058-M, financed by MCIN/AEI/10.13039/501100011033.
% Kate Pattle
KP is a Royal Society University Research Fellow, supported by grant no. URF/R1/211322.
% Relano
MR acknowledges support from project PID2023-150178NB-I00 financed by MCIU/AEI/10.13039/501100011033, and by FEDER, UE.
% S. Van Der Giessen
SAvdG acknowledge support by the French National Research Agency under the contract REDEEMING (ANR-24-CE31-2530).
% Thomas G. Williams
TGW gratefully acknowledges support from the UK ALMA Regional Centre (ARC) Node, which is supported by the Science and Technology Facilities Council grant number ST/Y004108/1.
% Christine Wilson
CDW acknowledges support form the Natural Science and Engineering Research Council of Canada (NSERC) and the Canada Research Chairs program (CRC-2022-00184).
% JCMT
These observations were obtained by the James Clerk Maxwell Telescope, operated by the East Asian Observatory on behalf of The National Astronomical Observatory of Japan; Academia Sinica Institute of Astronomy and Astrophysics; the Korea Astronomy and Space Science Institute; the National Astronomical Research Institute of Thailand; Center for Astronomical Mega-Science (as well as the National Key R\&D Program of China with No. 2017YFA0402700). Additional funding support is provided by the Science and Technology Facilities Council of the United Kingdom and participating universities and organizations in the United Kingdom and Canada.
% SCUBA-2
Additional funds for the construction of SCUBA-2 were provided by the Canada Foundation for Innovation.
% Maunakea
The authors wish to recognize and acknowledge the very significant cultural role and reverence that the summit of Maunakea has always had within the indigenous Hawaiian community. We are most fortunate to have the opportunity to conduct observations from this mountain.
% Astropy
This research made use of Astropy, a community-developed core Python package for Astronomy (Astropy Collaboration, 2013).

%%%%%%%%%%%%%%%%%%%%%%%%%%%%%%%%%%%%%%%%%%%%%%%%%%%%%%%%%%%%%%%%%%
% References
\bibliographystyle{mnras}
\bibliography{refs}

@ARTICLE{Pilbratt2010,
       author = {{Pilbratt}, G.~L. and {Riedinger}, J.~R. and {Passvogel}, T. and {Crone}, G. and {Doyle}, D. and {Gageur}, U. and {Heras}, A.~M. and {Jewell}, C. and {Metcalfe}, L. and {Ott}, S. and {Schmidt}, M.},
        title = "{Herschel Space Observatory. An ESA facility for far-infrared and submillimetre astronomy}",
      journal = {\aap},
         year = 2010,
        month = jul,
       volume = {518},
          eid = {L1},
        pages = {L1},
          doi = {10.1051/0004-6361/201014759},
archivePrefix = {arXiv},
       eprint = {1005.5331},
 primaryClass = {astro-ph.IM},
       adsurl = {https://ui.adsabs.harvard.edu/abs/2010A&A...518L...1P}
}

@ARTICLE{Griffin2010,
       author = {{Griffin}, M.~J. and {Abergel}, A. and {Abreu}, A. and {Ade}, P.~A.~R. and {Andr{\'e}}, P. and {Augueres}, J. -L. and {Babbedge}, T. and {Bae}, Y. and {Baillie}, T. and {Baluteau}, J. -P. and {Barlow}, M.~J. and {Bendo}, G. and {Benielli}, D. and {Bock}, J.~J. and {Bonhomme}, P. and {Brisbin}, D. and {Brockley-Blatt}, C. and {Caldwell}, M. and {Cara}, C. and {Castro-Rodriguez}, N. and {Cerulli}, R. and {Chanial}, P. and {Chen}, S. and {Clark}, E. and {Clements}, D.~L. and {Clerc}, L. and {Coker}, J. and {Communal}, D. and {Conversi}, L. and {Cox}, P. and {Crumb}, D. and {Cunningham}, C. and {Daly}, F. and {Davis}, G.~R. and {de Antoni}, P. and {Delderfield}, J. and {Devin}, N. and {di Giorgio}, A. and {Didschuns}, I. and {Dohlen}, K. and {Donati}, M. and {Dowell}, A. and {Dowell}, C.~D. and {Duband}, L. and {Dumaye}, L. and {Emery}, R.~J. and {Ferlet}, M. and {Ferrand}, D. and {Fontignie}, J. and {Fox}, M. and {Franceschini}, A. and {Frerking}, M. and {Fulton}, T. and {Garcia}, J. and {Gastaud}, R. and {Gear}, W.~K. and {Glenn}, J. and {Goizel}, A. and {Griffin}, D.~K. and {Grundy}, T. and {Guest}, S. and {Guillemet}, L. and {Hargrave}, P.~C. and {Harwit}, M. and {Hastings}, P. and {Hatziminaoglou}, E. and {Herman}, M. and {Hinde}, B. and {Hristov}, V. and {Huang}, M. and {Imhof}, P. and {Isaak}, K.~J. and {Israelsson}, U. and {Ivison}, R.~J. and {Jennings}, D. and {Kiernan}, B. and {King}, K.~J. and {Lange}, A.~E. and {Latter}, W. and {Laurent}, G. and {Laurent}, P. and {Leeks}, S.~J. and {Lellouch}, E. and {Levenson}, L. and {Li}, B. and {Li}, J. and {Lilienthal}, J. and {Lim}, T. and {Liu}, S.~J. and {Lu}, N. and {Madden}, S. and {Mainetti}, G. and {Marliani}, P. and {McKay}, D. and {Mercier}, K. and {Molinari}, S. and {Morris}, H. and {Moseley}, H. and {Mulder}, J. and {Mur}, M. and {Naylor}, D.~A. and {Nguyen}, H. and {O'Halloran}, B. and {Oliver}, S. and {Olofsson}, G. and {Olofsson}, H. -G. and {Orfei}, R. and {Page}, M.~J. and {Pain}, I. and {Panuzzo}, P. and {Papageorgiou}, A. and {Parks}, G. and {Parr-Burman}, P. and {Pearce}, A. and {Pearson}, C. and {P{\'e}rez-Fournon}, I. and {Pinsard}, F. and {Pisano}, G. and {Podosek}, J. and {Pohlen}, M. and {Polehampton}, E.~T. and {Pouliquen}, D. and {Rigopoulou}, D. and {Rizzo}, D. and {Roseboom}, I.~G. and {Roussel}, H. and {Rowan-Robinson}, M. and {Rownd}, B. and {Saraceno}, P. and {Sauvage}, M. and {Savage}, R. and {Savini}, G. and {Sawyer}, E. and {Scharmberg}, C. and {Schmitt}, D. and {Schneider}, N. and {Schulz}, B. and {Schwartz}, A. and {Shafer}, R. and {Shupe}, D.~L. and {Sibthorpe}, B. and {Sidher}, S. and {Smith}, A. and {Smith}, A.~J. and {Smith}, D. and {Spencer}, L. and {Stobie}, B. and {Sudiwala}, R. and {Sukhatme}, K. and {Surace}, C. and {Stevens}, J.~A. and {Swinyard}, B.~M. and {Trichas}, M. and {Tourette}, T. and {Triou}, H. and {Tseng}, S. and {Tucker}, C. and {Turner}, A. and {Vaccari}, M. and {Valtchanov}, I. and {Vigroux}, L. and {Virique}, E. and {Voellmer}, G. and {Walker}, H. and {Ward}, R. and {Waskett}, T. and {Weilert}, M. and {Wesson}, R. and {White}, G.~J. and {Whitehouse}, N. and {Wilson}, C.~D. and {Winter}, B. and {Woodcraft}, A.~L. and {Wright}, G.~S. and {Xu}, C.~K. and {Zavagno}, A. and {Zemcov}, M. and {Zhang}, L. and {Zonca}, E.},
        title = "{The Herschel-SPIRE instrument and its in-flight performance}",
      journal = {\aap},
         year = 2010,
        month = jul,
       volume = {518},
          eid = {L3},
        pages = {L3},
          doi = {10.1051/0004-6361/201014519},
archivePrefix = {arXiv},
       eprint = {1005.5123},
 primaryClass = {astro-ph.IM},
       adsurl = {https://ui.adsabs.harvard.edu/abs/2010A&A...518L...3G}
}

@ARTICLE{Poglitsch2010,
       author = {{Poglitsch}, A. and {Waelkens}, C. and {Geis}, N. and {Feuchtgruber}, H. and {Vandenbussche}, B. and {Rodriguez}, L. and {Krause}, O. and {Renotte}, E. and {van Hoof}, C. and {Saraceno}, P. and {Cepa}, J. and {Kerschbaum}, F. and {Agn{\`e}se}, P. and {Ali}, B. and {Altieri}, B. and {Andreani}, P. and {Augueres}, J. -L. and {Balog}, Z. and {Barl}, L. and {Bauer}, O.~H. and {Belbachir}, N. and {Benedettini}, M. and {Billot}, N. and {Boulade}, O. and {Bischof}, H. and {Blommaert}, J. and {Callut}, E. and {Cara}, C. and {Cerulli}, R. and {Cesarsky}, D. and {Contursi}, A. and {Creten}, Y. and {De Meester}, W. and {Doublier}, V. and {Doumayrou}, E. and {Duband}, L. and {Exter}, K. and {Genzel}, R. and {Gillis}, J. -M. and {Gr{\"o}zinger}, U. and {Henning}, T. and {Herreros}, J. and {Huygen}, R. and {Inguscio}, M. and {Jakob}, G. and {Jamar}, C. and {Jean}, C. and {de Jong}, J. and {Katterloher}, R. and {Kiss}, C. and {Klaas}, U. and {Lemke}, D. and {Lutz}, D. and {Madden}, S. and {Marquet}, B. and {Martignac}, J. and {Mazy}, A. and {Merken}, P. and {Montfort}, F. and {Morbidelli}, L. and {M{\"u}ller}, T. and {Nielbock}, M. and {Okumura}, K. and {Orfei}, R. and {Ottensamer}, R. and {Pezzuto}, S. and {Popesso}, P. and {Putzeys}, J. and {Regibo}, S. and {Reveret}, V. and {Royer}, P. and {Sauvage}, M. and {Schreiber}, J. and {Stegmaier}, J. and {Schmitt}, D. and {Schubert}, J. and {Sturm}, E. and {Thiel}, M. and {Tofani}, G. and {Vavrek}, R. and {Wetzstein}, M. and {Wieprecht}, E. and {Wiezorrek}, E.},
        title = "{The Photodetector Array Camera and Spectrometer (PACS) on the Herschel Space Observatory}",
      journal = {\aap},
         year = 2010,
        month = jul,
       volume = {518},
          eid = {L2},
        pages = {L2},
          doi = {10.1051/0004-6361/201014535},
archivePrefix = {arXiv},
       eprint = {1005.1487},
 primaryClass = {astro-ph.IM},
       adsurl = {https://ui.adsabs.harvard.edu/abs/2010A&A...518L...2P}
}

@ARTICLE{Holland2013,
       author = {{Holland}, W.~S. and {Bintley}, D. and {Chapin}, E.~L. and {Chrysostomou}, A. and {Davis}, G.~R. and {Dempsey}, J.~T. and {Duncan}, W.~D. and {Fich}, M. and {Friberg}, P. and {Halpern}, M. and {Irwin}, K.~D. and {Jenness}, T. and {Kelly}, B.~D. and {MacIntosh}, M.~J. and {Robson}, E.~I. and {Scott}, D. and {Ade}, P.~A.~R. and {Atad-Ettedgui}, E. and {Berry}, D.~S. and {Craig}, S.~C. and {Gao}, X. and {Gibb}, A.~G. and {Hilton}, G.~C. and {Hollister}, M.~I. and {Kycia}, J.~B. and {Lunney}, D.~W. and {McGregor}, H. and {Montgomery}, D. and {Parkes}, W. and {Tilanus}, R.~P.~J. and {Ullom}, J.~N. and {Walther}, C.~A. and {Walton}, A.~J. and {Woodcraft}, A.~L. and {Amiri}, M. and {Atkinson}, D. and {Burger}, B. and {Chuter}, T. and {Coulson}, I.~M. and {Doriese}, W.~B. and {Dunare}, C. and {Economou}, F. and {Niemack}, M.~D. and {Parsons}, H.~A.~L. and {Reintsema}, C.~D. and {Sibthorpe}, B. and {Smail}, I. and {Sudiwala}, R. and {Thomas}, H.~S.},
        title = "{SCUBA-2: the 10 000 pixel bolometer camera on the James Clerk Maxwell Telescope}",
      journal = {\mnras},
         year = 2013,
        month = apr,
       volume = {430},
       number = {4},
        pages = {2513-2533},
          doi = {10.1093/mnras/sts612},
archivePrefix = {arXiv},
       eprint = {1301.3650},
 primaryClass = {astro-ph.IM},
       adsurl = {https://ui.adsabs.harvard.edu/abs/2013MNRAS.430.2513H}
}

@ARTICLE{Thompson1980,
       author = {{Thompson}, A.~R. and {Clark}, B.~G. and {Wade}, C.~M. and {Napier}, P.~J.},
        title = "{The Very Large Array.}",
      journal = {\apjs},
         year = 1980,
        month = oct,
       volume = {44},
        pages = {151-167},
          doi = {10.1086/190688},
       adsurl = {https://ui.adsabs.harvard.edu/abs/1980ApJS...44..151T}
}

@ARTICLE{Galex,
       author = {{Martin}, D. Christopher and {Fanson}, James and {Schiminovich}, David and {Morrissey}, Patrick and {Friedman}, Peter G. and {Barlow}, Tom A. and {Conrow}, Tim and {Grange}, Robert and {Jelinsky}, Patrick N. and {Milliard}, Bruno and {Siegmund}, Oswald H.~W. and {Bianchi}, Luciana and {Byun}, Yong-Ik and {Donas}, Jose and {Forster}, Karl and {Heckman}, Timothy M. and {Lee}, Young-Wook and {Madore}, Barry F. and {Malina}, Roger F. and {Neff}, Susan G. and {Rich}, R. Michael and {Small}, Todd and {Surber}, Frank and {Szalay}, Alex S. and {Welsh}, Barry and {Wyder}, Ted K.},
        title = "{The Galaxy Evolution Explorer: A Space Ultraviolet Survey Mission}",
      journal = {\apjl},
         year = 2005,
        month = jan,
       volume = {619},
       number = {1},
        pages = {L1-L6},
          doi = {10.1086/426387},
archivePrefix = {arXiv},
       eprint = {astro-ph/0411302},
 primaryClass = {astro-ph},
       adsurl = {https://ui.adsabs.harvard.edu/abs/2005ApJ...619L...1M}
}

@ARTICLE{Spitzer-MIPS,
       author = {{Rieke}, G.~H. and {Young}, E.~T. and {Engelbracht}, C.~W. and {Kelly}, D.~M. and {Low}, F.~J. and {Haller}, E.~E. and {Beeman}, J.~W. and {Gordon}, K.~D. and {Stansberry}, J.~A. and {Misselt}, K.~A. and {Cadien}, J. and {Morrison}, J.~E. and {Rivlis}, G. and {Latter}, W.~B. and {Noriega-Crespo}, A. and {Padgett}, D.~L. and {Stapelfeldt}, K.~R. and {Hines}, D.~C. and {Egami}, E. and {Muzerolle}, J. and {Alonso-Herrero}, A. and {Blaylock}, M. and {Dole}, H. and {Hinz}, J.~L. and {Le Floc'h}, E. and {Papovich}, C. and {P{\'e}rez-Gonz{\'a}lez}, P.~G. and {Smith}, P.~S. and {Su}, K.~Y.~L. and {Bennett}, L. and {Frayer}, D.~T. and {Henderson}, D. and {Lu}, N. and {Masci}, F. and {Pesenson}, M. and {Rebull}, L. and {Rho}, J. and {Keene}, J. and {Stolovy}, S. and {Wachter}, S. and {Wheaton}, W. and {Werner}, M.~W. and {Richards}, P.~L.},
        title = "{The Multiband Imaging Photometer for Spitzer (MIPS)}",
      journal = {\apjs},
         year = 2004,
        month = sep,
       volume = {154},
       number = {1},
        pages = {25-29},
          doi = {10.1086/422717},
       adsurl = {https://ui.adsabs.harvard.edu/abs/2004ApJS..154...25R}
}

@ARTICLE{Spitzer-IRAC,
       author = {{Fazio}, G.~G. and {Hora}, J.~L. and {Allen}, L.~E. and {Ashby}, M.~L.~N. and {Barmby}, P. and {Deutsch}, L.~K. and {Huang}, J.-S. and {Kleiner}, S. and {Marengo}, M. and {Megeath}, S.~T. and {Melnick}, G.~J. and {Pahre}, M.~A. and {Patten}, B.~M. and {Polizotti}, J. and {Smith}, H.~A. and {Taylor}, R.~S. and {Wang}, Z. and {Willner}, S.~P. and {Hoffmann}, W.~F. and {Pipher}, J.~L. and {Forrest}, W.~J. and {McMurty}, C.~W. and {McCreight}, C.~R. and {McKelvey}, M.~E. and {McMurray}, R.~E. and {Koch}, D.~G. and {Moseley}, S.~H. and {Arendt}, R.~G. and {Mentzell}, J.~E. and {Marx}, C.~T. and {Losch}, P. and {Mayman}, P. and {Eichhorn}, W. and {Krebs}, D. and {Jhabvala}, M. and {Gezari}, D.~Y. and {Fixsen}, D.~J. and {Flores}, J. and {Shakoorzadeh}, K. and {Jungo}, R. and {Hakun}, C. and {Workman}, L. and {Karpati}, G. and {Kichak}, R. and {Whitley}, R. and {Mann}, S. and {Tollestrup}, E.~V. and {Eisenhardt}, P. and {Stern}, D. and {Gorjian}, V. and {Bhattacharya}, B. and {Carey}, S. and {Nelson}, B.~O. and {Glaccum}, W.~J. and {Lacy}, M. and {Lowrance}, P.~J. and {Laine}, S. and {Reach}, W.~T. and {Stauffer}, J.~A. and {Surace}, J.~A. and {Wilson}, G. and {Wright}, E.~L. and {Hoffman}, A. and {Domingo}, G. and {Cohen}, M.},
        title = "{The Infrared Array Camera (IRAC) for the Spitzer Space Telescope}",
      journal = {\apjs},
         year = 2004,
        month = sep,
       volume = {154},
       number = {1},
        pages = {10-17},
          doi = {10.1086/422843},
archivePrefix = {arXiv},
       eprint = {astro-ph/0405616},
 primaryClass = {astro-ph},
       adsurl = {https://ui.adsabs.harvard.edu/abs/2004ApJS..154...10F}
}

@ARTICLE{Gould&Salpeter1963,
       author = {{Gould}, Robert J. and {Salpeter}, Edwin E.},
        title = "{The Interstellar Abundance of the Hydrogen Molecule. I. Basic Processes.}",
      journal = {\apj},
         year = 1963,
        month = aug,
       volume = {138},
        pages = {393},
          doi = {10.1086/147654},
       adsurl = {https://ui.adsabs.harvard.edu/abs/1963ApJ...138..393G}
}

@ARTICLE{Draine1978,
       author = {{Draine}, B.~T.},
        title = "{Photoelectric heating of interstellar gas.}",
      journal = {\apjs},
         year = 1978,
        month = apr,
       volume = {36},
        pages = {595-619},
          doi = {10.1086/190513},
       adsurl = {https://ui.adsabs.harvard.edu/abs/1978ApJS...36..595D}
}

@ARTICLE{Dempsey2013,
       author = {{Dempsey}, J.~T. and {Friberg}, P. and {Jenness}, T. and {Tilanus}, R.~P.~J. and {Thomas}, H.~S. and {Holland}, W.~S. and {Bintley}, D. and {Berry}, D.~S. and {Chapin}, E.~L. and {Chrysostomou}, A. and {Davis}, G.~R. and {Gibb}, A.~G. and {Parsons}, H. and {Robson}, E.~I.},
        title = "{SCUBA-2: on-sky calibration using submillimetre standard sources}",
      journal = {\mnras},
         year = 2013,
        month = apr,
       volume = {430},
       number = {4},
        pages = {2534-2544},
          doi = {10.1093/mnras/stt090},
archivePrefix = {arXiv},
       eprint = {1301.3773},
 primaryClass = {astro-ph.IM},
       adsurl = {https://ui.adsabs.harvard.edu/abs/2013MNRAS.430.2534D}
}

@ARTICLE{Smith2016,
       author = {{Smith}, Matthew W.~L. and {Eales}, Stephen A. and {De Looze}, Ilse and {Baes}, Maarten and {Bendo}, George J. and {Bianchi}, Simone and {Boquien}, M{\'e}d{\'e}ric and {Boselli}, Alessandro and {Buat}, Veronique and {Ciesla}, Laure and {Clemens}, Marcel and {Clements}, David L. and {Cooray}, Asantha R. and {Cortese}, Luca and {Davies}, Jonathan I. and {Fritz}, Jacopo and {Gomez}, Haley L. and {Hughes}, Thomas M. and {Karczewski}, Oskar {\L}. and {Lu}, Nanyao and {Oliver}, Seb J. and {Remy-Ruyer}, Aur{\'e}lie and {Spinoglio}, Luigi and {Viaene}, Sebastien},
        title = "{Far-reaching dust distribution in galaxy discs}",
      journal = {\mnras},
         year = 2016,
        month = oct,
       volume = {462},
       number = {1},
        pages = {331-344},
          doi = {10.1093/mnras/stw1611},
archivePrefix = {arXiv},
       eprint = {1607.01020},
 primaryClass = {astro-ph.GA},
       adsurl = {https://ui.adsabs.harvard.edu/abs/2016MNRAS.462..331S}
}

@ARTICLE{Draine2003,
       author = {{Draine}, B.~T.},
        title = "{Interstellar Dust Grains}",
      journal = {\araa},
         year = 2003,
        month = jan,
       volume = {41},
        pages = {241-289},
          doi = {10.1146/annurev.astro.41.011802.094840},
archivePrefix = {arXiv},
       eprint = {astro-ph/0304489},
 primaryClass = {astro-ph},
       adsurl = {https://ui.adsabs.harvard.edu/abs/2003ARA&A..41..241D}
}

@ARTICLE{Smith2012,
       author = {{Smith}, M.~W.~L. and {Eales}, S.~A. and {Gomez}, H.~L. and {Roman-Duval}, J. and {Fritz}, J. and {Braun}, R. and {Baes}, M. and {Bendo}, G.~J. and {Blommaert}, J.~A.~D.~L. and {Boquien}, M. and {Boselli}, A. and {Clements}, D.~L. and {Cooray}, A.~R. and {Cortese}, L. and {De Looze}, I. and {Ford}, G.~P. and {Gear}, W.~K. and {Gentile}, G. and {Gordon}, K.~D. and {Kirk}, J. and {Lebouteiller}, V. and {Madden}, S. and {Mentuch}, E. and {O'Halloran}, B. and {Page}, M.~J. and {Schulz}, B. and {Spinoglio}, L. and {Verstappen}, J. and {Wilson}, C.~D. and {Thilker}, D.~A.},
        title = "{The Herschel Exploitation of Local Galaxy Andromeda (HELGA). II. Dust and Gas in Andromeda}",
      journal = {\apj},
         year = 2012,
        month = sep,
       volume = {756},
       number = {1},
          eid = {40},
        pages = {40},
          doi = {10.1088/0004-637X/756/1/40},
archivePrefix = {arXiv},
       eprint = {1204.0785},
 primaryClass = {astro-ph.CO},
       adsurl = {https://ui.adsabs.harvard.edu/abs/2012ApJ...756...40S}
}

@ARTICLE{Bendo2012,
       author = {{Bendo}, G.~J. and {Boselli}, A. and {Dariush}, A. and {Pohlen}, M. and {Roussel}, H. and {Sauvage}, M. and {Smith}, M.~W.~L. and {Wilson}, C.~D. and {Baes}, M. and {Cooray}, A. and {Clements}, D.~L. and {Cortese}, L. and {Foyle}, K. and {Galametz}, M. and {Gomez}, H.~L. and {Lebouteiller}, V. and {Lu}, N. and {Madden}, S.~C. and {Mentuch}, E. and {O'Halloran}, B. and {Page}, M.~J. and {Remy}, A. and {Schulz}, B. and {Spinoglio}, L.},
        title = "{Investigations of dust heating in M81, M83 and NGC 2403 with the Herschel Space Observatory}",
      journal = {\mnras},
         year = 2012,
        month = jan,
       volume = {419},
       number = {3},
        pages = {1833-1859},
          doi = {10.1111/j.1365-2966.2011.19735.x},
archivePrefix = {arXiv},
       eprint = {1109.0237},
 primaryClass = {astro-ph.CO},
       adsurl = {https://ui.adsabs.harvard.edu/abs/2012MNRAS.419.1833B}
}

@ARTICLE{deBlok2018,
       author = {{de Blok}, W.~J.~G. and {Walter}, Fabian and {Ferguson}, Annette M.~N. and {Bernard}, Edouard J. and {van der Hulst}, J.~M. and {Neeleman}, Marcel and {Leroy}, Adam K. and {Ott}, J{\"u}rgen and {Zschaechner}, Laura K. and {Zwaan}, Martin A. and {Yun}, Min S. and {Langston}, Glen and {Keating}, Katie M.},
        title = "{A High-resolution Mosaic of the Neutral Hydrogen in the M81 Triplet}",
      journal = {\apj},
         year = 2018,
        month = sep,
       volume = {865},
       number = {1},
          eid = {26},
        pages = {26},
          doi = {10.3847/1538-4357/aad557},
archivePrefix = {arXiv},
       eprint = {1808.02840},
 primaryClass = {astro-ph.GA},
       adsurl = {https://ui.adsabs.harvard.edu/abs/2018ApJ...865...26D}
}

@ARTICLE{Lagache2005,
       author = {{Lagache}, Guilaine and {Puget}, Jean-Loup and {Dole}, Herv{\'e}},
        title = "{Dusty Infrared Galaxies: Sources of the Cosmic Infrared Background}",
      journal = {\araa},
         year = 2005,
        month = sep,
       volume = {43},
       number = {1},
        pages = {727-768},
          doi = {10.1146/annurev.astro.43.072103.150606},
archivePrefix = {arXiv},
       eprint = {astro-ph/0507298},
 primaryClass = {astro-ph},
       adsurl = {https://ui.adsabs.harvard.edu/abs/2005ARA&A..43..727L}
}

@ARTICLE{Marsh2015,
       author = {{Marsh}, K.~A. and {Whitworth}, A.~P. and {Lomax}, O.},
        title = "{Temperature as a third dimension in column-density mapping of dusty astrophysical structures associated with star formation}",
      journal = {\mnras},
         year = 2015,
        month = dec,
       volume = {454},
       number = {4},
        pages = {4282-4292},
          doi = {10.1093/mnras/stv2248},
archivePrefix = {arXiv},
       eprint = {1509.08699},
 primaryClass = {astro-ph.IM},
       adsurl = {https://ui.adsabs.harvard.edu/abs/2015MNRAS.454.4282M}
}

@ARTICLE{Galametz2016,
       author = {{Galametz}, M. and {Hony}, S. and {Albrecht}, M. and {Galliano}, F. and {Cormier}, D. and {Lebouteiller}, V. and {Lee}, M.~Y. and {Madden}, S.~C. and {Bolatto}, A. and {Bot}, C. and {Hughes}, A. and {Israel}, F. and {Meixner}, M. and {Oliviera}, J.~M. and {Paradis}, D. and {Pellegrini}, E. and {Roman-Duval}, J. and {Rubio}, M. and {Sewi{\l}o}, M. and {Fukui}, Y. and {Kawamura}, A. and {Onishi}, T.},
        title = "{The dust properties and physical conditions of the interstellar medium in the LMC massive star-forming complex N11}",
      journal = {\mnras},
         year = 2016,
        month = feb,
       volume = {456},
       number = {2},
        pages = {1767-1790},
          doi = {10.1093/mnras/stv2773},
archivePrefix = {arXiv},
       eprint = {1511.07457},
 primaryClass = {astro-ph.GA},
       adsurl = {https://ui.adsabs.harvard.edu/abs/2016MNRAS.456.1767G}
}

@ARTICLE{Tabatabaei2014,
       author = {{Tabatabaei}, F.~S. and {Braine}, J. and {Xilouris}, E.~M. and {Kramer}, C. and {Boquien}, M. and {Combes}, F. and {Henkel}, C. and {Relano}, M. and {Verley}, S. and {Gratier}, P. and {Israel}, F. and {Wiedner}, M.~C. and {R{\"o}llig}, M. and {Schuster}, K.~F. and {van der Werf}, P.},
        title = "{Variation in the dust emissivity index across M 33 with Herschel and Spitzer (HerM 33es)}",
      journal = {\aap},
         year = 2014,
        month = jan,
       volume = {561},
          eid = {A95},
        pages = {A95},
          doi = {10.1051/0004-6361/201321441},
archivePrefix = {arXiv},
       eprint = {1310.4155},
 primaryClass = {astro-ph.CO},
       adsurl = {https://ui.adsabs.harvard.edu/abs/2014A&A...561A..95T}
}

@ARTICLE{Kirkpatrick2014,
       author = {{Kirkpatrick}, Allison and {Calzetti}, Daniela and {Kennicutt}, Robert and {Galametz}, Maud and {Gordon}, Karl and {Groves}, Brent and {Hunt}, Leslie and {Dale}, Daniel and {Hinz}, Joannah and {Tabatabaei}, Fatemeh},
        title = "{Untangling the Nature of Spatial Variations of Cold Dust Properties in Star Forming Galaxies}",
      journal = {\apj},
         year = 2014,
        month = jul,
       volume = {789},
       number = {2},
          eid = {130},
        pages = {130},
          doi = {10.1088/0004-637X/789/2/130},
archivePrefix = {arXiv},
       eprint = {1405.4872},
 primaryClass = {astro-ph.GA},
       adsurl = {https://ui.adsabs.harvard.edu/abs/2014ApJ...789..130K}
}

@ARTICLE{Whitworth2019,
       author = {{Whitworth}, A.~P. and {Marsh}, K.~A. and {Cigan}, P.~J. and {Dalcanton}, J.~J. and {Smith}, M.~W.~L. and {Gomez}, H.~L. and {Lomax}, O. and {Griffin}, M.~J. and {Eales}, S.~A.},
        title = "{The dust in M31}",
      journal = {\mnras},
         year = 2019,
        month = nov,
       volume = {489},
       number = {4},
        pages = {5436-5452},
          doi = {10.1093/mnras/stz2166},
archivePrefix = {arXiv},
       eprint = {1908.03458},
 primaryClass = {astro-ph.GA},
       adsurl = {https://ui.adsabs.harvard.edu/abs/2019MNRAS.489.5436W}
}

@ARTICLE{KINGFISH,
       author = {{Kennicutt}, R.~C. and {Calzetti}, D. and {Aniano}, G. and {Appleton}, P. and {Armus}, L. and {Beir{\~a}o}, P. and {Bolatto}, A.~D. and {Brandl}, B. and {Crocker}, A. and {Croxall}, K. and {Dale}, D.~A. and {Donovan Meyer}, J. and {Draine}, B.~T. and {Engelbracht}, C.~W. and {Galametz}, M. and {Gordon}, K.~D. and {Groves}, B. and {Hao}, C. -N. and {Helou}, G. and {Hinz}, J. and {Hunt}, L.~K. and {Johnson}, B. and {Koda}, J. and {Krause}, O. and {Leroy}, A.~K. and {Li}, Y. and {Meidt}, S. and {Montiel}, E. and {Murphy}, E.~J. and {Rahman}, N. and {Rix}, H. -W. and {Roussel}, H. and {Sandstrom}, K. and {Sauvage}, M. and {Schinnerer}, E. and {Skibba}, R. and {Smith}, J.~D.~T. and {Srinivasan}, S. and {Vigroux}, L. and {Walter}, F. and {Wilson}, C.~D. and {Wolfire}, M. and {Zibetti}, S.},
        title = "{KINGFISH{\textemdash}Key Insights on Nearby Galaxies: A Far-Infrared Survey with Herschel: Survey Description and Image Atlas}",
      journal = {\pasp},
         year = 2011,
        month = dec,
       volume = {123},
       number = {910},
        pages = {1347},
          doi = {10.1086/663818},
archivePrefix = {arXiv},
       eprint = {1111.4438},
 primaryClass = {astro-ph.CO},
       adsurl = {https://ui.adsabs.harvard.edu/abs/2011PASP..123.1347K}
}

@ARTICLE{Dustpedia,
       author = {{Davies}, J.~I. and {Baes}, M. and {Bianchi}, S. and {Jones}, A. and {Madden}, S. and {Xilouris}, M. and {Bocchio}, M. and {Casasola}, V. and {Cassara}, L. and {Clark}, C. and {De Looze}, I. and {Evans}, R. and {Fritz}, J. and {Galametz}, M. and {Galliano}, F. and {Lianou}, S. and {Mosenkov}, A.~V. and {Smith}, M. and {Verstocken}, S. and {Viaene}, S. and {Vika}, M. and {Wagle}, G. and {Ysard}, N.},
        title = "{DustPedia: A Definitive Study of Cosmic Dust in the Local Universe}",
      journal = {\pasp},
         year = 2017,
        month = apr,
       volume = {129},
       number = {974},
        pages = {044102},
          doi = {10.1088/1538-3873/129/974/044102},
archivePrefix = {arXiv},
       eprint = {1609.06138},
 primaryClass = {astro-ph.GA},
       adsurl = {https://ui.adsabs.harvard.edu/abs/2017PASP..129d4102D}
}

@ARTICLE{Aniano2011,
       author = {{Aniano}, G. and {Draine}, B.~T. and {Gordon}, K.~D. and {Sandstrom}, K.},
        title = "{Common-Resolution Convolution Kernels for Space- and Ground-Based Telescopes}",
      journal = {\pasp},
         year = 2011,
        month = oct,
       volume = {123},
       number = {908},
        pages = {1218},
          doi = {10.1086/662219},
archivePrefix = {arXiv},
       eprint = {1106.5065},
 primaryClass = {astro-ph.IM},
       adsurl = {https://ui.adsabs.harvard.edu/abs/2011PASP..123.1218A}
}

@ARTICLE{Bajaja1979,
       author = {{Bajaja}, E. and {van Albada}, G.~D.},
        title = "{Complementing aperture synthesis radio data by short spacing components from single dish observations.}",
      journal = {\aap},
         year = 1979,
        month = may,
       volume = {75},
       number = {1-2},
        pages = {251-254},
       adsurl = {https://ui.adsabs.harvard.edu/abs/1979A&A....75..251B}
}

@ARTICLE{Smith2021_HASHTAG,
       author = {{Smith}, Matthew W.~L. and {Eales}, Stephen A. and {Williams}, Thomas G. and {Lee}, Bumhyun and {Li}, Zongnan and {Barmby}, Pauline and {Bureau}, Martin and {Chapman}, Scott and {Cho}, Brian S. and {Chung}, Aeree and {Chung}, Eun Jung and {Chung}, Hui-Hsuan and {Clark}, Christopher J.~R. and {Clements}, David L. and {Davis}, Timothy A. and {De Looze}, Ilse and {Eden}, David J. and {Athikkat-Eknath}, Gayathri and {Ford}, George P. and {Gao}, Yu and {Gear}, Walter and {Gomez}, Haley L. and {de Grijs}, Richard and {He}, Jinhua and {Ho}, Luis C. and {Hughes}, Thomas M. and {Jiao}, Sihan and {Li}, Zhiyuan and {Kemper}, Francisca and {Kirchschlager}, Florian and {Koch}, Eric W. and {Kong}, Albert K.~H. and {Lee}, Chien-Hsiu and {Lin}, En-Tzu and {Mairs}, Steve and {Micha{\l}owski}, Micha{\l} J. and {Pattle}, Kate and {Peng}, Yingjie and {Ragan}, Sarah E. and {Rawlings}, Mark G. and {Rigopoulou}, Dimitra and {Saintonge}, Amelie and {Schruba}, Andreas and {Tang}, Xindi and {Wang}, Junfeng and {Whitworth}, Anthony P. and {Wilson}, Christine D. and {Yim}, Kijeong and {Zhu}, Ming},
        title = "{The HASHTAG Project: The First Submillimeter Images of the Andromeda Galaxy from the Ground}",
      journal = {\apjs},
         year = 2021,
        month = dec,
       volume = {257},
       number = {2},
          eid = {52},
        pages = {52},
          doi = {10.3847/1538-4365/ac23d0},
archivePrefix = {arXiv},
       eprint = {2110.00011},
 primaryClass = {astro-ph.GA},
       adsurl = {https://ui.adsabs.harvard.edu/abs/2021ApJS..257...52S}
}

@ARTICLE{Bendo2010,
       author = {{Bendo}, G.~J. and {Wilson}, C.~D. and {Pohlen}, M. and {Sauvage}, M. and {Auld}, R. and {Baes}, M. and {Barlow}, M.~J. and {Bock}, J.~J. and {Boselli}, A. and {Bradford}, M. and {Buat}, V. and {Castro-Rodriguez}, N. and {Chanial}, P. and {Charlot}, S. and {Ciesla}, L. and {Clements}, D.~L. and {Cooray}, A. and {Cormier}, D. and {Cortese}, L. and {Davies}, J.~I. and {Dwek}, E. and {Eales}, S.~A. and {Elbaz}, D. and {Galametz}, M. and {Galliano}, F. and {Gear}, W.~K. and {Glenn}, J. and {Gomez}, H.~L. and {Griffin}, M. and {Hony}, S. and {Isaak}, K.~G. and {Levenson}, L.~R. and {Lu}, N. and {Madden}, S. and {O'Halloran}, B. and {Okumura}, K. and {Oliver}, S. and {Page}, M.~J. and {Panuzzo}, P. and {Papageorgiou}, A. and {Parkin}, T.~J. and {Perez-Fournon}, I. and {Rangwala}, N. and {Rigby}, E.~E. and {Roussel}, H. and {Rykala}, A. and {Sacchi}, N. and {Schulz}, B. and {Schirm}, M.~R.~P. and {Smith}, M.~W.~L. and {Spinoglio}, L. and {Stevens}, J.~A. and {Sundar}, S. and {Symeonidis}, M. and {Trichas}, M. and {Vaccari}, M. and {Vigroux}, L. and {Wozniak}, H. and {Wright}, G.~S. and {Zeilinger}, W.~W.},
        title = "{The Herschel Space Observatory view of dust in M81}",
      journal = {\aap},
         year = 2010,
        month = jul,
       volume = {518},
          eid = {L65},
        pages = {L65},
          doi = {10.1051/0004-6361/201014568},
archivePrefix = {arXiv},
       eprint = {1005.1889},
 primaryClass = {astro-ph.CO},
       adsurl = {https://ui.adsabs.harvard.edu/abs/2010A&A...518L..65B}
}

@MISC{Ciesla2013,
       author = {{Ciesla}, L. and {Boselli}, A. and {Smith}, M.~W.~L. and {Bendo}, G.~J. and {Cortese}, L. and {Eales}, S. and {Bianchi}, S. and {Boquien}, M. and {Buat}, V. and {Davies}, J. and {Pohlen}, M. and {Zibetti}, S. and {Baes}, M. and {Cooray}, A. and {De Looze}, I. and {di Serego Alighieri}, S. and {Galametz}, M. and {Gomez}, H.~L. and {Lebouteiller}, V. and {Madden}, S.~C. and {Pappalardo}, C. and {Remy}, A. and {Spinoglio}, L. and {Vaccari}, M. and {Auld}, R. and {Clements}, D.~L.},
        title = "{Submillimetre photometry of 323 nearby galaxies from the Herschel Reference Survey (Corrigendum)}",
 howpublished = {Astronomy \& Astrophysics, Volume 550, id.C1, 82 pp.},
         year = 2013,
        month = feb,
          eid = {C1},
        pages = {C1},
          doi = {10.1051/0004-6361/201219216e},
       adsurl = {https://ui.adsabs.harvard.edu/abs/2013A&A...550C...1C}
}

@ARTICLE{2019Williams,
       author = {{Williams}, Thomas G. and {Gear}, Walter K. and {Smith}, Matthew W.~L.},
        title = "{A high-resolution, dust-selected molecular cloud catalogue of M33, the Triangulum Galaxy}",
      journal = {\mnras},
         year = 2019,
        month = mar,
       volume = {483},
       number = {4},
        pages = {5135-5149},
          doi = {10.1093/mnras/sty3437},
archivePrefix = {arXiv},
       eprint = {1812.06103},
 primaryClass = {astro-ph.GA},
       adsurl = {https://ui.adsabs.harvard.edu/abs/2019MNRAS.483.5135W}
}

@ARTICLE{HASHTAG1_2020Li,
       author = {{Li}, Zongnan and {Li}, Zhiyuan and {Smith}, Matthew W.~L. and {Wilson}, Christine D. and {Gao}, Yu and {Eales}, Stephen A. and {Ao}, Yiping and {Bureau}, Martin and {Chung}, Aeree and {Davis}, Timothy A. and {de Grijs}, Richard and {Eden}, David J. and {He}, Jinhua and {Hughes}, Tom M. and {Jiang}, Xuejian and {Kemper}, Francisca and {Lamperti}, Isabella and {Lee}, Bumhyun and {Lee}, Chien-Hsiu and {Micha{\l}owski}, Micha{\l} J. and {Parsons}, Harriet and {Ragan}, Sarah and {Scicluna}, Peter and {Shi}, Yong and {Tang}, Xindi and {Tomi{\v{c}}i{\'c}}, Neven and {Viaene}, Sebastien and {Williams}, Thomas G. and {Zhu}, Ming},
        title = "{The HASHTAG project I. A survey of CO(3-2) emission from the star forming disc of M31}",
      journal = {\mnras},
         year = 2020,
        month = feb,
       volume = {492},
       number = {1},
        pages = {195-209},
          doi = {10.1093/mnras/stz3409},
archivePrefix = {arXiv},
       eprint = {1912.02403},
 primaryClass = {astro-ph.GA},
       adsurl = {https://ui.adsabs.harvard.edu/abs/2020MNRAS.492..195L}
}

@ARTICLE{Wilson2012,
       author = {{Wilson}, C.~D. and {Warren}, B.~E. and {Israel}, F.~P. and {Serjeant}, S. and {Attewell}, D. and {Bendo}, G.~J. and {Butner}, H.~M. and {Chanial}, P. and {Clements}, D.~L. and {Golding}, J. and {Heesen}, V. and {Irwin}, J. and {Leech}, J. and {Matthews}, H.~E. and {M{\"u}hle}, S. and {Mortier}, A.~M.~J. and {Petitpas}, G. and {S{\'a}nchez-Gallego}, J.~R. and {Sinukoff}, E. and {Shorten}, K. and {Tan}, B.~K. and {Tilanus}, R.~P.~J. and {Usero}, A. and {Vaccari}, M. and {Wiegert}, T. and {Zhu}, M. and {Alexander}, D.~M. and {Alexander}, P. and {Azimlu}, M. and {Barmby}, P. and {Brar}, R. and {Bridge}, C. and {Brinks}, E. and {Brooks}, S. and {Coppin}, K. and {C{\^o}t{\'e}}, S. and {C{\^o}t{\'e}}, P. and {Courteau}, S. and {Davies}, J. and {Eales}, S. and {Fich}, M. and {Hudson}, M. and {Hughes}, D.~H. and {Ivison}, R.~J. and {Knapen}, J.~H. and {Page}, M. and {Parkin}, T.~J. and {Rigopoulou}, D. and {Rosolowsky}, E. and {Seaquist}, E.~R. and {Spekkens}, K. and {Tanvir}, N. and {van der Hulst}, J.~M. and {van der Werf}, P. and {Vlahakis}, C. and {Webb}, T.~M. and {Weferling}, B. and {White}, G.~J.},
        title = "{The JCMT Nearby Galaxies Legacy Survey {\textemdash} VIII. CO data and the L$_{CO(3-2)}$-L$_{FIR}$ correlation in the SINGS sample}",
      journal = {\mnras},
         year = 2012,
        month = aug,
       volume = {424},
       number = {4},
        pages = {3050-3080},
          doi = {10.1111/j.1365-2966.2012.21453.x},
archivePrefix = {arXiv},
       eprint = {1206.1629},
 primaryClass = {astro-ph.CO},
       adsurl = {https://ui.adsabs.harvard.edu/abs/2012MNRAS.424.3050W}
}

@ARTICLE{Hildebrand1983,
       author = {{Hildebrand}, R.~H.},
        title = "{The determination of cloud masses and dust characteristics from submillimetre thermal emission.}",
      journal = {\qjras},
         year = 1983,
        month = sep,
       volume = {24},
        pages = {267-282},
       adsurl = {https://ui.adsabs.harvard.edu/abs/1983QJRAS..24..267H}
}

@ARTICLE{Alton2004,
       author = {{Alton}, P.~B. and {Xilouris}, E.~M. and {Misiriotis}, A. and {Dasyra}, K.~M. and {Dumke}, M.},
        title = "{The emissivity of dust grains in spiral galaxies}",
      journal = {\aap},
         year = 2004,
        month = oct,
       volume = {425},
        pages = {109-120},
          doi = {10.1051/0004-6361:20040438},
archivePrefix = {arXiv},
       eprint = {astro-ph/0406389},
 primaryClass = {astro-ph},
       adsurl = {https://ui.adsabs.harvard.edu/abs/2004A&A...425..109A}
}

@ARTICLE{Clark2019,
       author = {{Clark}, C.~J.~R. and {De Vis}, P. and {Baes}, M. and {Bianchi}, S. and {Casasola}, V. and {Cassar{\`a}}, L.~P. and {Davies}, J.~I. and {Dobbels}, W. and {Lianou}, S. and {De Looze}, I. and {Evans}, R. and {Galametz}, M. and {Galliano}, F. and {Jones}, A.~P. and {Madden}, S.~C. and {Mosenkov}, A.~V. and {Verstocken}, S. and {Viaene}, S. and {Xilouris}, E.~M. and {Ysard}, N.},
        title = "{The first maps of {\ensuremath{\kappa}}$_{d}$ - the dust mass absorption coefficient - in nearby galaxies, with DustPedia}",
      journal = {\mnras},
         year = 2019,
        month = nov,
       volume = {489},
       number = {4},
        pages = {5256-5283},
          doi = {10.1093/mnras/stz2257},
archivePrefix = {arXiv},
       eprint = {1908.04318},
 primaryClass = {astro-ph.GA},
       adsurl = {https://ui.adsabs.harvard.edu/abs/2019MNRAS.489.5256C}
}

@INPROCEEDINGS{Mutschke2013,
       author = {{Mutschke}, H.},
        title = "{Optical Properties of Interstellar Dust from Cosmic Dust Analogs Studied in the Lab}",
    booktitle = {Proceedings of The Life Cycle of Dust in the Universe: Observations},
         year = 2013,
       editor = {{Kemper}, Ciska and {Andersen}, Anja and {Baes}, Maarten and {Gomez}, Haley and {Watson}, Darach},
        month = jan,
          eid = {42},
        pages = {42},
          doi = {10.22323/1.207.0042},
       adsurl = {https://ui.adsabs.harvard.edu/abs/2013lcdu.confE..42M}
}

@ARTICLE{Jones2013,
       author = {{Jones}, A.~P. and {Fanciullo}, L. and {K{\"o}hler}, M. and {Verstraete}, L. and {Guillet}, V. and {Bocchio}, M. and {Ysard}, N.},
        title = "{The evolution of amorphous hydrocarbons in the ISM: dust modelling from a new vantage point}",
      journal = {\aap},
         year = 2013,
        month = oct,
       volume = {558},
          eid = {A62},
        pages = {A62},
          doi = {10.1051/0004-6361/201321686},
archivePrefix = {arXiv},
       eprint = {1411.6293},
 primaryClass = {astro-ph.GA},
       adsurl = {https://ui.adsabs.harvard.edu/abs/2013A&A...558A..62J}
}

@ARTICLE{Kim1994,
       author = {{Kim}, Sang-Hee and {Martin}, P.~G. and {Hendry}, Paul D.},
        title = "{The Size Distribution of Interstellar Dust Particles as Determined from Extinction}",
      journal = {\apj},
         year = 1994,
        month = feb,
       volume = {422},
        pages = {164},
          doi = {10.1086/173714},
       adsurl = {https://ui.adsabs.harvard.edu/abs/1994ApJ...422..164K}
}

@ARTICLE{Sanchez-Gallego2011,
       author = {{S{\'a}nchez-Gallego}, J.~R. and {Knapen}, J.~H. and {Heiner}, J.~S. and {Wilson}, C.~D. and {Warren}, B.~E. and {Allen}, R.~J. and {Azimlu}, M. and {Barmby}, P. and {Bendo}, G.~J. and {Comer{\'o}n}, S. and {Israel}, F.~P. and {Serjeant}, S. and {Tilanus}, R.~P.~J. and {Vlahakis}, C. and {van der Werf}, P.},
        title = "{The JCMT Nearby Galaxies Legacy Survey. VI. The distribution of gas and star formation in M 81}",
      journal = {\aap},
         year = 2011,
        month = mar,
       volume = {527},
          eid = {A16},
        pages = {A16},
          doi = {10.1051/0004-6361/201014930},
archivePrefix = {arXiv},
       eprint = {1011.5116},
 primaryClass = {astro-ph.GA},
       adsurl = {https://ui.adsabs.harvard.edu/abs/2011A&A...527A..16S}
}

@ARTICLE{Clark2018,
       author = {{Clark}, C.~J.~R. and {Verstocken}, S. and {Bianchi}, S. and {Fritz}, J. and {Viaene}, S. and {Smith}, M.~W.~L. and {Baes}, M. and {Casasola}, V. and {Cassara}, L.~P. and {Davies}, J.~I. and {De Looze}, I. and {De Vis}, P. and {Evans}, R. and {Galametz}, M. and {Jones}, A.~P. and {Lianou}, S. and {Madden}, S. and {Mosenkov}, A.~V. and {Xilouris}, M.},
        title = "{DustPedia: Multiwavelength photometry and imagery of 875 nearby galaxies in 42 ultraviolet-microwave bands}",
      journal = {\aap},
         year = 2018,
        month = jan,
       volume = {609},
          eid = {A37},
        pages = {A37},
          doi = {10.1051/0004-6361/201731419},
archivePrefix = {arXiv},
       eprint = {1708.05335},
 primaryClass = {astro-ph.GA},
       adsurl = {https://ui.adsabs.harvard.edu/abs/2018A&A...609A..37C}
}

@ARTICLE{Kudritzki2012,
       author = {{Kudritzki}, Rolf-Peter and {Urbaneja}, Miguel A. and {Gazak}, Zachary and {Bresolin}, Fabio and {Przybilla}, Norbert and {Gieren}, Wolfgang and {Pietrzy{\'n}ski}, Grzegorz},
        title = "{Quantitative Spectroscopy of Blue Supergiant Stars in the Disk of M81: Metallicity, Metallicity Gradient, and Distance}",
      journal = {\apj},
         year = 2012,
        month = mar,
       volume = {747},
       number = {1},
          eid = {15},
        pages = {15},
          doi = {10.1088/0004-637X/747/1/15},
archivePrefix = {arXiv},
       eprint = {1112.3643},
 primaryClass = {astro-ph.CO},
       adsurl = {https://ui.adsabs.harvard.edu/abs/2012ApJ...747...15K}
}

@ARTICLE{Freedman1994,
       author = {{Freedman}, Wendy L. and {Hughes}, Shaun M. and {Madore}, Barry F. and {Mould}, Jeremy R. and {Lee}, Myung Gyoon and {Stetson}, Peter and {Kennicutt}, Robert C. and {Turner}, Anne and {Ferrarese}, Laura and {Ford}, Holland and {Graham}, John A. and {Hill}, Robert and {Hoessel}, John G. and {Huchra}, John and {Illingworth}, Garth D.},
        title = "{The Hubble Space Telescope Extragalactic Distance Scale Key Project. I. The Discovery of Cepheids and a New Distance to M81}",
      journal = {\apj},
         year = 1994,
        month = jun,
       volume = {427},
        pages = {628},
          doi = {10.1086/174172},
       adsurl = {https://ui.adsabs.harvard.edu/abs/1994ApJ...427..628F}
}

@ARTICLE{Freedman2001,
       author = {{Freedman}, Wendy L. and {Madore}, Barry F. and {Gibson}, Brad K. and {Ferrarese}, Laura and {Kelson}, Daniel D. and {Sakai}, Shoko and {Mould}, Jeremy R. and {Kennicutt}, Jr., Robert C. and {Ford}, Holland C. and {Graham}, John A. and {Huchra}, John P. and {Hughes}, Shaun M.~G. and {Illingworth}, Garth D. and {Macri}, Lucas M. and {Stetson}, Peter B.},
        title = "{Final Results from the Hubble Space Telescope Key Project to Measure the Hubble Constant}",
      journal = {\apj},
         year = 2001,
        month = may,
       volume = {553},
       number = {1},
        pages = {47-72},
          doi = {10.1086/320638},
archivePrefix = {arXiv},
       eprint = {astro-ph/0012376},
 primaryClass = {astro-ph},
       adsurl = {https://ui.adsabs.harvard.edu/abs/2001ApJ...553...47F}
}

@ARTICLE{Parsons2018,
       author = {{Parsons}, H. and {Dempsey}, J.~T. and {Thomas}, H.~S. and {Berry}, D. and {Currie}, M.~J. and {Friberg}, P. and {Wouterloot}, J.~G.~A. and {Chrysostomou}, A. and {Graves}, S. and {Tilanus}, R.~P.~J. and {Bell}, G.~S. and {Rawlings}, M.~G.},
        title = "{The Dusty Galactic Center as Seen by SCUBA-2}",
      journal = {\apjs},
         year = 2018,
        month = feb,
       volume = {234},
       number = {2},
          eid = {22},
        pages = {22},
          doi = {10.3847/1538-4365/aa989c},
       adsurl = {https://ui.adsabs.harvard.edu/abs/2018ApJS..234...22P}
}

@ARTICLE{Doi2005,
       author = {{Doi}, A. and {Kameno}, S. and {Kohno}, K. and {Nakanishi}, K. and {Inoue}, M.},
        title = "{A high-frequency radio survey of low-luminosity active galactic nuclei}",
      journal = {\mnras},
         year = 2005,
        month = oct,
       volume = {363},
       number = {2},
        pages = {692-704},
          doi = {10.1111/j.1365-2966.2005.09471.x},
archivePrefix = {arXiv},
       eprint = {astro-ph/0508598},
 primaryClass = {astro-ph},
       adsurl = {https://ui.adsabs.harvard.edu/abs/2005MNRAS.363..692D}
}

@ARTICLE{Ford2013,
       author = {{Ford}, George P. and {Gear}, Walter K. and {Smith}, Matthew W.~L. and {Eales}, Steve A. and {Baes}, Maarten and {Bendo}, George J. and {Boquien}, M{\'e}d{\'e}ric and {Boselli}, Alessandro and {Cooray}, Asantha R. and {De Looze}, Ilse and {Fritz}, Jacopo and {Gentile}, Gianfranco and {Gomez}, Haley L. and {Gordon}, Karl D. and {Kirk}, Jason and {Lebouteiller}, Vianney and {O'Halloran}, Brian and {Spinoglio}, Luigi and {Verstappen}, Joris and {Wilson}, Christine D.},
        title = "{Herschel Exploitation of Local Galaxy Andromeda (HELGA). III. The Star Formation Law in M31}",
      journal = {\apj},
         year = 2013,
        month = may,
       volume = {769},
       number = {1},
          eid = {55},
        pages = {55},
          doi = {10.1088/0004-637X/769/1/55},
archivePrefix = {arXiv},
       eprint = {1303.6284},
 primaryClass = {astro-ph.CO},
       adsurl = {https://ui.adsabs.harvard.edu/abs/2013ApJ...769...55F}
}

@ARTICLE{Leroy2008,
       author = {{Leroy}, Adam K. and {Walter}, Fabian and {Brinks}, Elias and {Bigiel}, Frank and {de Blok}, W.~J.~G. and {Madore}, Barry and {Thornley}, M.~D.},
        title = "{The Star Formation Efficiency in Nearby Galaxies: Measuring Where Gas Forms Stars Effectively}",
      journal = {\aj},
         year = 2008,
        month = dec,
       volume = {136},
       number = {6},
        pages = {2782-2845},
          doi = {10.1088/0004-6256/136/6/2782},
archivePrefix = {arXiv},
       eprint = {0810.2556},
 primaryClass = {astro-ph},
       adsurl = {https://ui.adsabs.harvard.edu/abs/2008AJ....136.2782L}
}

@ARTICLE{Smith2010,
       author = {{Smith}, M.~W.~L. and {Vlahakis}, C. and {Baes}, M. and {Bendo}, G.~J. and {Bianchi}, S. and {Bomans}, D.~J. and {Boselli}, A. and {Clemens}, M. and {Corbelli}, E. and {Cortese}, L. and {Dariush}, A. and {Davies}, J.~I. and {De Looze}, I. and {di Serego Alighieri}, S. and {Fadda}, D. and {Fritz}, J. and {Garcia-Appadoo}, D.~A. and {Gavazzi}, G. and {Giovanardi}, C. and {Grossi}, M. and {Hughes}, T.~M. and {Hunt}, L.~K. and {Jones}, A.~P. and {Madden}, S. and {Pierini}, D. and {Pohlen}, M. and {Sabatini}, S. and {Verstappen}, J. and {Xilouris}, E.~M. and {Zibetti}, S.},
        title = "{The Herschel Virgo Cluster Survey. IV. Resolved dust analysis of spiral galaxies}",
      journal = {\aap},
         year = 2010,
        month = jul,
       volume = {518},
          eid = {L51},
        pages = {L51},
          doi = {10.1051/0004-6361/201014584},
archivePrefix = {arXiv},
       eprint = {1005.3057},
 primaryClass = {astro-ph.CO},
       adsurl = {https://ui.adsabs.harvard.edu/abs/2010A&A...518L..51S}
}

@ARTICLE{Hunt2015,
       author = {{Hunt}, L.~K. and {Draine}, B.~T. and {Bianchi}, S. and {Gordon}, K.~D. and {Aniano}, G. and {Calzetti}, D. and {Dale}, D.~A. and {Helou}, G. and {Hinz}, J.~L. and {Kennicutt}, R.~C. and {Roussel}, H. and {Wilson}, C.~D. and {Bolatto}, A. and {Boquien}, M. and {Croxall}, K.~V. and {Galametz}, M. and {Gil de Paz}, A. and {Koda}, J. and {Mu{\~n}oz-Mateos}, J.~C. and {Sandstrom}, K.~M. and {Sauvage}, M. and {Vigroux}, L. and {Zibetti}, S.},
        title = "{Cool dust heating and temperature mixing in nearby star-forming galaxies}",
      journal = {\aap},
         year = 2015,
        month = apr,
       volume = {576},
          eid = {A33},
        pages = {A33},
          doi = {10.1051/0004-6361/201424734},
archivePrefix = {arXiv},
       eprint = {1409.5916},
 primaryClass = {astro-ph.GA},
       adsurl = {https://ui.adsabs.harvard.edu/abs/2015A&A...576A..33H}
}

@ARTICLE{Planck2011,
       author = {{Planck Collaboration} and {Ade}, P.~A.~R. and {Aghanim}, N. and {Arnaud}, M. and {Ashdown}, M. and {Aumont}, J. and {Baccigalupi}, C. and {Balbi}, A. and {Banday}, A.~J. and {Barreiro}, R.~B. and {Bartlett}, J.~G. and {Battaner}, E. and {Benabed}, K. and {Beno{\^\i}t}, A. and {Bernard}, J.-P. and {Bersanelli}, M. and {Bhatia}, R. and {Bock}, J.~J. and {Bonaldi}, A. and {Bond}, J.~R. and {Borrill}, J. and {Bouchet}, F.~R. and {Boulanger}, F. and {Bucher}, M. and {Burigana}, C. and {Cabella}, P. and {Cardoso}, J.-F. and {Catalano}, A. and {Cay{\'o}n}, L. and {Challinor}, A. and {Chamballu}, A. and {Chiang}, L.-Y. and {Chiang}, C. and {Christensen}, P.~R. and {Clements}, D.~L. and {Colombi}, S. and {Couchot}, F. and {Coulais}, A. and {Crill}, B.~P. and {Cuttaia}, F. and {Dame}, T.~M. and {Danese}, L. and {Davies}, R.~D. and {Davis}, R.~J. and {de Bernardis}, P. and {de Gasperis}, G. and {de Rosa}, A. and {de Zotti}, G. and {Delabrouille}, J. and {Delouis}, J.-M. and {D{\'e}sert}, F.-X. and {Dickinson}, C. and {Dobashi}, K. and {Donzelli}, S. and {Dor{\'e}}, O. and {D{\"o}rl}, U. and {Douspis}, M. and {Dupac}, X. and {Efstathiou}, G. and {En{\ss}lin}, T.~A. and {Eriksen}, H.~K. and {Falgarone}, E. and {Finelli}, F. and {Forni}, O. and {Fosalba}, P. and {Frailis}, M. and {Franceschi}, E. and {Fukui}, Y. and {Galeotta}, S. and {Ganga}, K. and {Giard}, M. and {Giardino}, G. and {Giraud-H{\'e}raud}, Y. and {Gonz{\'a}lez-Nuevo}, J. and {G{\'o}rski}, K.~M. and {Gratton}, S. and {Gregorio}, A. and {Grenier}, I.~A. and {Gruppuso}, A. and {Hansen}, F.~K. and {Harrison}, D. and {Helou}, G. and {Henrot-Versill{\'e}}, S. and {Herranz}, D. and {Hildebrandt}, S.~R. and {Hivon}, E. and {Hobson}, M. and {Holmes}, W.~A. and {Hovest}, W. and {Hoyland}, R.~J. and {Huffenberger}, K.~M. and {Jaffe}, A.~H. and {Jones}, W.~C. and {Juvela}, M. and {Kawamura}, A. and {Keih{\"a}nen}, E. and {Keskitalo}, R. and {Kisner}, T.~S. and {Kneissl}, R. and {Knox}, L. and {Kurki-Suonio}, H. and {Lagache}, G. and {Lamarre}, J.-M. and {Lasenby}, A. and {Laureijs}, R.~J. and {Lawrence}, C.~R. and {Leach}, S. and {Leonardi}, R. and {Leroy}, C. and {Lilje}, P.~B. and {Linden-V{\o}rnle}, M. and {L{\'o}pez-Caniego}, M. and {Lubin}, P.~M. and {Mac{\'\i}as-P{\'e}rez}, J.~F. and {MacTavish}, C.~J. and {Maffei}, B. and {Maino}, D. and {Mandolesi}, N. and {Mann}, R. and {Maris}, M. and {Martin}, P. and {Mart{\'\i}nez-Gonz{\'a}lez}, E. and {Masi}, S. and {Matarrese}, S. and {Matthai}, F. and {Mazzotta}, P. and {McGehee}, P. and {Meinhold}, P.~R. and {Melchiorri}, A. and {Mendes}, L. and {Mennella}, A. and {Miville-Desch{\^e}nes}, M.-A. and {Moneti}, A. and {Montier}, L. and {Morgante}, G. and {Mortlock}, D. and {Munshi}, D. and {Murphy}, A. and {Naselsky}, P. and {Natoli}, P. and {Netterfield}, C.~B. and {N{\o}rgaard-Nielsen}, H.~U. and {Noviello}, F. and {Novikov}, D. and {Novikov}, I. and {O'Dwyer}, I.~J. and {Onishi}, T. and {Osborne}, S. and {Pajot}, F. and {Paladini}, R. and {Paradis}, D. and {Pasian}, F. and {Patanchon}, G. and {Perdereau}, O. and {Perotto}, L. and {Perrotta}, F. and {Piacentini}, F. and {Piat}, M. and {Plaszczynski}, S. and {Pointecouteau}, E. and {Polenta}, G. and {Ponthieu}, N. and {Poutanen}, T. and {Pr{\'e}zeau}, G. and {Prunet}, S. and {Puget}, J.-L. and {Reach}, W.~T. and {Reinecke}, M. and {Renault}, C. and {Ricciardi}, S. and {Riller}, T. and {Ristorcelli}, I. and {Rocha}, G. and {Rosset}, C. and {Rowan-Robinson}, M. and {Rubi{\~n}o-Mart{\'\i}n}, J.~A. and {Rusholme}, B. and {Sandri}, M. and {Santos}, D. and {Savini}, G. and {Scott}, D. and {Seiffert}, M.~D. and {Shellard}, P. and {Smoot}, G.~F. and {Starck}, J.-L. and {Stivoli}, F. and {Stolyarov}, V. and {Stompor}, R. and {Sudiwala}, R. and {Sygnet}, J.-F. and {Tauber}, J.~A. and {Terenzi}, L. and {Toffolatti}, L. and {Tomasi}, M. and {Torre}, J.-P. and {Tristram}, M. and {Tuovinen}, J. and {Umana}, G. and {Valenziano}, L. and {Vielva}, P.},
        title = "{Planck early results. XIX. All-sky temperature and dust optical depth from Planck and IRAS. Constraints on the ``dark gas'' in our Galaxy}",
      journal = {\aap},
         year = 2011,
        month = dec,
       volume = {536},
          eid = {A19},
        pages = {A19},
          doi = {10.1051/0004-6361/201116479},
archivePrefix = {arXiv},
       eprint = {1101.2029},
 primaryClass = {astro-ph.GA},
       adsurl = {https://ui.adsabs.harvard.edu/abs/2011A&A...536A..19P}
}

@ARTICLE{Langer2014,
       author = {{Langer}, W.~D. and {Velusamy}, T. and {Pineda}, J.~L. and {Willacy}, K. and {Goldsmith}, P.~F.},
        title = "{A Herschel [C ii] Galactic plane survey. II. CO-dark H$_{2}$ in clouds}",
      journal = {\aap},
         year = 2014,
        month = jan,
       volume = {561},
          eid = {A122},
        pages = {A122},
          doi = {10.1051/0004-6361/201322406},
archivePrefix = {arXiv},
       eprint = {1312.3320},
 primaryClass = {astro-ph.GA},
       adsurl = {https://ui.adsabs.harvard.edu/abs/2014A&A...561A.122L}
}

@ARTICLE{Grenier2005,
       author = {{Grenier}, Isabelle A. and {Casandjian}, Jean-Marc and {Terrier}, R{\'e}gis},
        title = "{Unveiling Extensive Clouds of Dark Gas in the Solar Neighborhood}",
      journal = {Science},
         year = 2005,
        month = feb,
       volume = {307},
       number = {5713},
        pages = {1292-1295},
          doi = {10.1126/science.1106924},
       adsurl = {https://ui.adsabs.harvard.edu/abs/2005Sci...307.1292G}
}

@ARTICLE{Wolfire2010,
       author = {{Wolfire}, Mark G. and {Hollenbach}, David and {McKee}, Christopher F.},
        title = "{The Dark Molecular Gas}",
      journal = {\apj},
         year = 2010,
        month = jun,
       volume = {716},
       number = {2},
        pages = {1191-1207},
          doi = {10.1088/0004-637X/716/2/1191},
archivePrefix = {arXiv},
       eprint = {1004.5401},
 primaryClass = {astro-ph.GA},
       adsurl = {https://ui.adsabs.harvard.edu/abs/2010ApJ...716.1191W}
}

@ARTICLE{Desert2008,
       author = {{D{\'e}sert}, F.-X. and {Mac{\'\i}as-P{\'e}rez}, J.~F. and {Mayet}, F. and {Giardino}, G. and {Renault}, C. and {Aumont}, J. and {Beno{\^\i}t}, A. and {Bernard}, J.-Ph. and {Ponthieu}, N. and {Tristram}, M.},
        title = "{Submillimetre point sources from the Archeops experiment: very cold clumps in the Galactic plane}",
      journal = {\aap},
         year = 2008,
        month = apr,
       volume = {481},
       number = {2},
        pages = {411-421},
          doi = {10.1051/0004-6361:20078701},
archivePrefix = {arXiv},
       eprint = {0801.4502},
 primaryClass = {astro-ph},
       adsurl = {https://ui.adsabs.harvard.edu/abs/2008A&A...481..411D}
}

@ARTICLE{Paradis2010,
       author = {{Paradis}, D. and {Veneziani}, M. and {Noriega-Crespo}, A. and {Paladini}, R. and {Piacentini}, F. and {Bernard}, J.~P. and {de Bernardis}, P. and {Calzoletti}, L. and {Faustini}, F. and {Martin}, P. and {Masi}, S. and {Montier}, L. and {Natoli}, P. and {Ristorcelli}, I. and {Thompson}, M.~A. and {Traficante}, A. and {Molinari}, S.},
        title = "{Variations of the spectral index of dust emissivity from Hi-GAL observations of the Galactic plane}",
      journal = {\aap},
         year = 2010,
        month = sep,
       volume = {520},
          eid = {L8},
        pages = {L8},
          doi = {10.1051/0004-6361/201015301},
archivePrefix = {arXiv},
       eprint = {1009.2779},
 primaryClass = {astro-ph.GA},
       adsurl = {https://ui.adsabs.harvard.edu/abs/2010A&A...520L...8P}
}

@ARTICLE{Bracco2011,
       author = {{Bracco}, A. and {Cooray}, A. and {Veneziani}, M. and {Amblard}, A. and {Serra}, P. and {Wardlow}, J. and {Thompson}, M.~A. and {White}, G. and {Auld}, R. and {Baes}, M. and {Bertoldi}, F. and {Buttiglione}, S. and {Cava}, A. and {Clements}, D.~L. and {Dariush}, A. and {de Zotti}, G. and {Dunne}, L. and {Dye}, S. and {Eales}, S. and {Fritz}, J. and {Gomez}, H. and {Hopwood}, R. and {Ibar}, I. and {Ivison}, R.~J. and {Jarvis}, M. and {Lagache}, G. and {Lee}, M.~G. and {Leeuw}, L. and {Maddox}, S. and {Micha{\l}owski}, M. and {Pearson}, C. and {Pohlen}, M. and {Rigby}, E. and {Rodighiero}, G. and {Smith}, D.~J.~B. and {Temi}, P. and {Vaccari}, M. and {van der Werf}, P.},
        title = "{Herschel-ATLAS: statistical properties of Galactic cirrus in the GAMA-9 Hour Science Demonstration Phase Field}",
      journal = {\mnras},
         year = 2011,
        month = apr,
       volume = {412},
       number = {2},
        pages = {1151-1161},
          doi = {10.1111/j.1365-2966.2010.17971.x},
archivePrefix = {arXiv},
       eprint = {1011.0725},
 primaryClass = {astro-ph.CO},
       adsurl = {https://ui.adsabs.harvard.edu/abs/2011MNRAS.412.1151B}
}

@ARTICLE{DraineLi2001_StochasticHeating,
       author = {{Draine}, B.~T. and {Li}, Aigen},
        title = "{Infrared Emission from Interstellar Dust. I. Stochastic Heating of Small Grains}",
      journal = {\apj},
         year = 2001,
        month = apr,
       volume = {551},
       number = {2},
        pages = {807-824},
          doi = {10.1086/320227},
archivePrefix = {arXiv},
       eprint = {astro-ph/0011318},
 primaryClass = {astro-ph},
       adsurl = {https://ui.adsabs.harvard.edu/abs/2001ApJ...551..807D}
}

@ARTICLE{Drabek2012,
       author = {{Drabek}, E. and {Hatchell}, J. and {Friberg}, P. and {Richer}, J. and {Graves}, S. and {Buckle}, J.~V. and {Nutter}, D. and {Johnstone}, D. and {Di Francesco}, J.},
        title = "{Molecular line contamination in the SCUBA-2 450 and 850 {\ensuremath{\mu}}m continuum data}",
      journal = {\mnras},
         year = 2012,
        month = oct,
       volume = {426},
       number = {1},
        pages = {23-39},
          doi = {10.1111/j.1365-2966.2012.21140.x},
archivePrefix = {arXiv},
       eprint = {1204.6180},
 primaryClass = {astro-ph.GA},
       adsurl = {https://ui.adsabs.harvard.edu/abs/2012MNRAS.426...23D}
}

@ARTICLE{Mattsson2014HELGA,
       author = {{Mattsson}, L. and {Gomez}, H.~L. and {Andersen}, A.~C. and {Smith}, M.~W.~L. and {De Looze}, I. and {Baes}, M. and {Viaene}, S. and {Gentile}, G. and {Fritz}, J. and {Spinoglio}, L.},
        title = "{The Herschel exploitation of local galaxy Andromeda (HELGA) - V. Strengthening the case for substantial interstellar grain growth}",
      journal = {\mnras},
         year = 2014,
        month = oct,
       volume = {444},
       number = {1},
        pages = {797-807},
          doi = {10.1093/mnras/stu1228},
archivePrefix = {arXiv},
       eprint = {1406.7649},
 primaryClass = {astro-ph.GA},
       adsurl = {https://ui.adsabs.harvard.edu/abs/2014MNRAS.444..797M}
}

@ARTICLE{Katsioli2023,
       author = {{Katsioli}, S. and {Xilouris}, E.~M. and {Kramer}, C. and {Adam}, R. and {Ade}, P. and {Ajeddig}, H. and {Andr{\'e}}, P. and {Artis}, E. and {Aussel}, H. and {Baes}, M. and {Beelen}, A. and {Beno{\^\i}t}, A. and {Berta}, S. and {Bing}, L. and {Bourrion}, O. and {Calvo}, M. and {Catalano}, A. and {Clark}, C.~J.~R. and {De Looze}, I. and {De Petris}, M. and {D{\'e}sert}, F.-X. and {Doyle}, S. and {Driessen}, E.~F.~C. and {Ejlali}, G. and {Galametz}, M. and {Galliano}, F. and {Gomez}, A. and {Goupy}, J. and {Hanser}, C. and {Hughes}, A. and {Jones}, A.~P. and {K{\'e}ruzor{\'e}}, F. and {Ladjelate}, B. and {Lagache}, G. and {Leclercq}, S. and {Lestrade}, J.-F. and {Mac{\'\i}as-P{\'e}rez}, J.-F. and {Madden}, S.~C. and {Maury}, A. and {Mauskopf}, P. and {Mayet}, F. and {Monfardini}, A. and {Mu{\~n}oz-Echeverr{\'\i}a}, M. and {Nersesian}, A. and {Pantoni}, L. and {Paradis}, D. and {Perotto}, L. and {Pisano}, G. and {Ponthieu}, N. and {Rev{\'e}ret}, V. and {Rigby}, A.~J. and {Ritacco}, A. and {Romero}, C. and {Roussel}, H. and {Ruppin}, F. and {Schuster}, K. and {Sievers}, A. and {Smith}, M.~W.~L. and {Tedros}, J. and {Tabatabaei}, F. and {Tucker}, C. and {Ysard}, N. and {Zylka}, R.},
        title = "{The stratification of ISM properties in the edge-on galaxy NGC 891 revealed by NIKA2}",
      journal = {\aap},
         year = 2023,
        month = nov,
       volume = {679},
          eid = {A7},
        pages = {A7},
          doi = {10.1051/0004-6361/202347020},
archivePrefix = {arXiv},
       eprint = {2309.08403},
 primaryClass = {astro-ph.GA},
       adsurl = {https://ui.adsabs.harvard.edu/abs/2023A&A...679A...7K}
}

@ARTICLE{Lamperti2019,
       author = {{Lamperti}, Isabella and {Saintonge}, Am{\'e}lie and {De Looze}, Ilse and {Accurso}, Gioacchino and {Clark}, Christopher J.~R. and {Smith}, Matthew W.~L. and {Wilson}, Christine D. and {Brinks}, Elias and {Brown}, Toby and {Bureau}, Martin and {Clements}, David L. and {Eales}, Stephen and {Glass}, David H.~W. and {Hwang}, Ho Seong and {Lee}, Jong Chul and {Lin}, Lihwai and {Michalowski}, Michal J. and {Sargent}, Mark and {Williams}, Thomas G. and {Xiao}, Ting and {Yang}, Chentao},
        title = "{JINGLE - V. Dust properties of nearby galaxies derived from hierarchical Bayesian SED fitting}",
      journal = {\mnras},
         year = 2019,
        month = nov,
       volume = {489},
       number = {3},
        pages = {4389-4417},
          doi = {10.1093/mnras/stz2311},
archivePrefix = {arXiv},
       eprint = {1909.05266},
 primaryClass = {astro-ph.GA},
       adsurl = {https://ui.adsabs.harvard.edu/abs/2019MNRAS.489.4389L}
}

@ARTICLE{Pantoni2026,
       author = {{Pantoni}, L. and {Galliano}, F. and {Madden}, S.~C. and {Adam}, R. and {Ade}, P. and {Ajeddig}, H. and {Andr{\'e}}, P. and {Artis}, E. and {Aussel}, H. and {Baes}, M. and {Beelen}, A. and {Beno{\^\i}t}, A. and {Berta}, S. and {Bing}, L. and {Bourrion}, O. and {Calvo}, M. and {Casasola}, V. and {Catalano}, A. and {Chang}, I.-D. and {De Looze}, I. and {De Petris}, M. and {D{\'e}sert}, F.-X. and {Doyle}, S. and {Driessen}, E.~F.~C. and {Ejlali}, G. and {Gomez}, A. and {Goupy}, J. and {Jones}, A.~P. and {Hanser}, C. and {Hughes}, A. and {Katsioli}, S. and {K{\'e}ruzor{\'e}}, F. and {Koch}, E.~W. and {Kramer}, C. and {Ladjelate}, B. and {Lagache}, G. and {Leclercq}, S. and {Lestrade}, J.-F. and {Mac{\'\i}as-P{\'e}rez}, J.~F. and {Maury}, A. and {Mauskopf}, P. and {Mayet}, F. and {Monfardini}, A. and {Moyer-Anin}, A. and {Mu{\~n}oz-Echeverr{\'\i}a}, M. and {Nersesian}, A. and {Paradis}, D. and {Perotto}, L. and {Pisano}, G. and {Ponthieu}, N. and {Rev{\'e}ret}, V. and {Rigby}, A.~J. and {Ritacco}, A. and {Romero}, C. and {Roussel}, H. and {Ruppin}, F. and {Schuster}, K. and {Sievers}, A. and {Smith}, M.~W.~L. and {Tabatabaei}, F.~S. and {Tedros}, J. and {Tucker}, C. and {Ysard}, N. and {Xilouris}, E.~M. and {Zylka}, R.},
        title = "{Spatially resolved interstellar dust properties in the face-on spiral galaxy M 99 as observed by NIKA2}",
      journal = {\aap},
         year = 2026,
        month = apr,
       volume = {709},
          eid = {A1},
        pages = {A1},
          doi = {10.1051/0004-6361/202557583},
archivePrefix = {arXiv},
       eprint = {2602.20393},
 primaryClass = {astro-ph.GA},
       adsurl = {https://ui.adsabs.harvard.edu/abs/2026A&A...709A...1P}
}

@ARTICLE{Vilchez2019,
       author = {{V{\'\i}lchez}, J.~M. and {Rela{\~n}o}, M. and {Kennicutt}, R. and {De Looze}, I. and {Moll{\'a}}, M. and {Galametz}, M.},
        title = "{Metals and dust content across the galaxies M 101 and NGC 628}",
      journal = {\mnras},
         year = 2019,
        month = mar,
       volume = {483},
       number = {4},
        pages = {4968-4983},
          doi = {10.1093/mnras/sty3455},
archivePrefix = {arXiv},
       eprint = {1811.10262},
 primaryClass = {astro-ph.GA},
       adsurl = {https://ui.adsabs.harvard.edu/abs/2019MNRAS.483.4968V}
}

@ARTICLE{Relano2018,
       author = {{Rela{\~n}o}, M. and {De Looze}, I. and {Kennicutt}, R.~C. and {Lisenfeld}, U. and {Dariush}, A. and {Verley}, S. and {Braine}, J. and {Tabatabaei}, F. and {Kramer}, C. and {Boquien}, M. and {Xilouris}, M. and {Gratier}, P.},
        title = "{Spatially resolving the dust properties and submillimetre excess in M 33}",
      journal = {\aap},
         year = 2018,
        month = may,
       volume = {613},
          eid = {A43},
        pages = {A43},
          doi = {10.1051/0004-6361/201732347},
archivePrefix = {arXiv},
       eprint = {1801.04806},
 primaryClass = {astro-ph.GA},
       adsurl = {https://ui.adsabs.harvard.edu/abs/2018A&A...613A..43R}
}

@ARTICLE{vanderGiessen2024,
       author = {{van der Giessen}, S.~A. and {Matsumoto}, K. and {Relano}, M. and {De Looze}, I. and {Romano}, L. and {Hirashita}, H. and {Nagamine}, K. and {Baes}, M. and {Palla}, M. and {Hou}, K.~C. and {Faesi}, C.},
        title = "{Radial properties of dust in galaxies: Comparison between observations and isolated galaxy simulations}",
      journal = {\aap},
         year = 2024,
        month = dec,
       volume = {692},
          eid = {A39},
        pages = {A39},
          doi = {10.1051/0004-6361/202451988},
archivePrefix = {arXiv},
       eprint = {2410.21430},
 primaryClass = {astro-ph.GA},
       adsurl = {https://ui.adsabs.harvard.edu/abs/2024A&A...692A..39V}
}

@ARTICLE{Chiang2018,
       author = {{Chiang}, I-Da and {Sandstrom}, Karin M. and {Chastenet}, J{\'e}r{\'e}my and {Johnson}, L. Clifton and {Leroy}, Adam K. and {Utomo}, Dyas},
        title = "{The Spatially Resolved Dust-to-metals Ratio in M101}",
      journal = {\apj},
         year = 2018,
        month = oct,
       volume = {865},
       number = {2},
          eid = {117},
        pages = {117},
          doi = {10.3847/1538-4357/aadc5f},
archivePrefix = {arXiv},
       eprint = {1808.07164},
 primaryClass = {astro-ph.GA},
       adsurl = {https://ui.adsabs.harvard.edu/abs/2018ApJ...865..117C}
}

@ARTICLE{Park2024,
       author = {{Park}, Hye-Jin and {Battisti}, Andrew J. and {Wisnioski}, Emily and {Cortese}, Luca and {Seibert}, Mark and {Grasha}, Kathryn and {Madore}, Barry F. and {Groves}, Brent and {Rich}, Jeff A. and {Beaton}, Rachael L. and {Chen}, Qian-Hui and {Mun}, Marcie and {McClure-Griffiths}, Naomi M. and {de Blok}, W.~J.~G. and {Kewley}, Lisa J.},
        title = "{The spatially resolved relation between dust, gas, and metal abundance with the TYPHOON survey}",
      journal = {\mnras},
         year = 2024,
        month = nov,
       volume = {535},
       number = {1},
        pages = {729-752},
          doi = {10.1093/mnras/stae2298},
archivePrefix = {arXiv},
       eprint = {2410.02222},
 primaryClass = {astro-ph.GA},
       adsurl = {https://ui.adsabs.harvard.edu/abs/2024MNRAS.535..729P}
}

@ARTICLE{Clark2021,
       author = {{Clark}, Christopher J.~R. and {Roman-Duval}, Julia C. and {Gordon}, Karl D. and {Bot}, Caroline and {Smith}, Matthew W.~L.},
        title = "{The Quest for the Missing Dust. I. Restoring Large-scale Emission in Herschel Maps of Local Group Galaxies}",
      journal = {\apj},
         year = 2021,
        month = nov,
       volume = {921},
       number = {1},
          eid = {35},
        pages = {35},
          doi = {10.3847/1538-4357/ac16d4},
archivePrefix = {arXiv},
       eprint = {2107.14302},
 primaryClass = {astro-ph.GA},
       adsurl = {https://ui.adsabs.harvard.edu/abs/2021ApJ...921...35C}
}

@ARTICLE{Pattle2023,
       author = {{Pattle}, Kate and {Gear}, Walter and {Wilson}, Christine D.},
        title = "{The JCMT nearby galaxies legacy survey: SCUBA-2 observations of nearby galaxies}",
      journal = {\mnras},
         year = 2023,
        month = jun,
       volume = {522},
       number = {2},
        pages = {2339-2368},
          doi = {10.1093/mnras/stad652},
archivePrefix = {arXiv},
       eprint = {2302.14800},
 primaryClass = {astro-ph.GA},
       adsurl = {https://ui.adsabs.harvard.edu/abs/2023MNRAS.522.2339P}
}

@ARTICLE{Davies2010,
       author = {{Davies}, J.~I. and {Wilson}, C.~D. and {Auld}, R. and {Baes}, M. and {Barlow}, M.~J. and {Bendo}, G.~J. and {Bock}, J.~J. and {Boselli}, A. and {Bradford}, M. and {Buat}, V. and {Castro-Rodriguez}, N. and {Chanial}, P. and {Charlot}, S. and {Ciesla}, L. and {Clements}, D.~L. and {Cooray}, A. and {Cormier}, D. and {Cortese}, L. and {Dwek}, E. and {Eales}, S.~A. and {Elbaz}, D. and {Galametz}, M. and {Galliano}, F. and {Gear}, W.~K. and {Glenn}, J. and {Gomez}, H.~L. and {Griffin}, M. and {Hony}, S. and {Isaak}, K.~G. and {Levenson}, L.~R. and {Lu}, N. and {Madden}, S. and {O'Halloran}, B. and {Okumura}, K. and {Oliver}, S. and {Page}, M.~J. and {Panuzzo}, P. and {Papageorgiou}, A. and {Parkin}, T.~J. and {Perez-Fournon}, I. and {Pohlen}, M. and {Rangwala}, N. and {Rigby}, E.~E. and {Roussel}, H. and {Rykala}, A. and {Sacchi}, N. and {Sauvage}, M. and {Schulz}, B. and {Schirm}, M.~R.~P. and {Smith}, M.~W.~L. and {Spinoglio}, L. and {Stevens}, J.~A. and {Srinivasan}, S. and {Symeonidis}, M. and {Trichas}, M. and {Vaccari}, M. and {Vigroux}, L. and {Wozniak}, H. and {Wright}, G.~S. and {Zeilinger}, W.~W.},
        title = "{On the origin of M81 group extended dust emission}",
      journal = {\mnras},
         year = 2010,
        month = nov,
       volume = {409},
       number = {1},
        pages = {102-108},
          doi = {10.1111/j.1365-2966.2010.17774.x},
archivePrefix = {arXiv},
       eprint = {1010.4770},
 primaryClass = {astro-ph.CO},
       adsurl = {https://ui.adsabs.harvard.edu/abs/2010MNRAS.409..102D}
}

@ARTICLE{Viaene2017,
       author = {{Viaene}, S. and {Baes}, M. and {Tamm}, A. and {Tempel}, E. and {Bendo}, G. and {Blommaert}, J.~A.~D.~L. and {Boquien}, M. and {Boselli}, A. and {Camps}, P. and {Cooray}, A. and {De Looze}, I. and {De Vis}, P. and {Fern{\'a}ndez-Ontiveros}, J.~A. and {Fritz}, J. and {Galametz}, M. and {Gentile}, G. and {Madden}, S. and {Smith}, M.~W.~L. and {Spinoglio}, L. and {Verstocken}, S.},
        title = "{The Herschel Exploitation of Local Galaxy Andromeda (HELGA). VII. A SKIRT radiative transfer model and insights on dust heating}",
      journal = {\aap},
         year = 2017,
        month = mar,
       volume = {599},
          eid = {A64},
        pages = {A64},
          doi = {10.1051/0004-6361/201629251},
archivePrefix = {arXiv},
       eprint = {1609.08643},
 primaryClass = {astro-ph.GA},
       adsurl = {https://ui.adsabs.harvard.edu/abs/2017A&A...599A..64V}
}

@ARTICLE{Ho1996,
       author = {{Ho}, Luis C. and {Filippenko}, Alexei V. and {Sargent}, Wallace L.~W.},
        title = "{New Insights into the Physical Nature of LINERs from a Multiwavelength Analysis of the Nucleus of M81}",
      journal = {\apj},
         year = 1996,
        month = may,
       volume = {462},
        pages = {183},
          doi = {10.1086/177140},
       adsurl = {https://ui.adsabs.harvard.edu/abs/1996ApJ...462..183H}
}

@ARTICLE{Walter2008,
       author = {{Walter}, Fabian and {Brinks}, Elias and {de Blok}, W.~J.~G. and {Bigiel}, Frank and {Kennicutt}, Jr., Robert C. and {Thornley}, Michele D. and {Leroy}, Adam},
        title = "{THINGS: The H I Nearby Galaxy Survey}",
      journal = {\aj},
         year = 2008,
        month = dec,
       volume = {136},
       number = {6},
        pages = {2563-2647},
          doi = {10.1088/0004-6256/136/6/2563},
archivePrefix = {arXiv},
       eprint = {0810.2125},
 primaryClass = {astro-ph},
       adsurl = {https://ui.adsabs.harvard.edu/abs/2008AJ....136.2563W}
}

@ARTICLE{Bolatto2013,
       author = {{Bolatto}, Alberto D. and {Wolfire}, Mark and {Leroy}, Adam K.},
        title = "{The CO-to-H$_{2}$ Conversion Factor}",
      journal = {\araa},
         year = 2013,
        month = aug,
       volume = {51},
       number = {1},
        pages = {207-268},
          doi = {10.1146/annurev-astro-082812-140944},
archivePrefix = {arXiv},
       eprint = {1301.3498},
 primaryClass = {astro-ph.GA},
       adsurl = {https://ui.adsabs.harvard.edu/abs/2013ARA&A..51..207B}
}

@ARTICLE{Patterson2012,
       author = {{Patterson}, Maria T. and {Walterbos}, Rene A.~M. and {Kennicutt}, Robert C. and {Chiappini}, Cristina and {Thilker}, David A.},
        title = "{An oxygen abundance gradient into the outer disc of M81}",
      journal = {\mnras},
         year = 2012,
        month = may,
       volume = {422},
       number = {1},
        pages = {401-419},
          doi = {10.1111/j.1365-2966.2012.20616.x},
archivePrefix = {arXiv},
       eprint = {1202.0308},
 primaryClass = {astro-ph.CO},
       adsurl = {https://ui.adsabs.harvard.edu/abs/2012MNRAS.422..401P}
}

@ARTICLE{Stanghellini2014,
       author = {{Stanghellini}, Letizia and {Magrini}, Laura and {Casasola}, Viviana and {Villaver}, Eva},
        title = "{The radial metallicity gradient and the history of elemental enrichment in M 81 through emission-line probes}",
      journal = {\aap},
         year = 2014,
        month = jul,
       volume = {567},
          eid = {A88},
        pages = {A88},
          doi = {10.1051/0004-6361/201423423},
archivePrefix = {arXiv},
       eprint = {1403.5547},
 primaryClass = {astro-ph.GA},
       adsurl = {https://ui.adsabs.harvard.edu/abs/2014A&A...567A..88S}
}

@ARTICLE{Belfiore2023,
       author = {{Belfiore}, Francesco and {Leroy}, Adam K. and {Sun}, Jiayi and {Barnes}, Ashley T. and {Boquien}, M{\'e}d{\'e}ric and {Cao}, Yixian and {Congiu}, Enrico and {Dale}, Daniel A. and {Egorov}, Oleg V. and {Eibensteiner}, Cosima and {Glover}, Simon C.~O. and {Grasha}, Kathryn and {Groves}, Brent and {Klessen}, Ralf S. and {Kreckel}, Kathryn and {Neumann}, Lukas and {Querejeta}, Miguel and {Sanchez-Blazquez}, Patricia and {Schinnerer}, Eva and {Williams}, Thomas G.},
        title = "{Calibration of hybrid resolved star formation rate recipes based on PHANGS-MUSE H{\ensuremath{\alpha}} and H{\ensuremath{\beta}} maps}",
      journal = {\aap},
         year = 2023,
        month = feb,
       volume = {670},
          eid = {A67},
        pages = {A67},
          doi = {10.1051/0004-6361/202244863},
archivePrefix = {arXiv},
       eprint = {2211.08487},
 primaryClass = {astro-ph.GA},
       adsurl = {https://ui.adsabs.harvard.edu/abs/2023A&A...670A..67B}
}

@ARTICLE{Wall2019,
       author = {{Wall}, Renae E. and {Kilic}, Mukremin and {Bergeron}, P. and {Rolland}, B. and {Genest-Beaulieu}, C. and {Gianninas}, A.},
        title = "{GALEX absolute calibration and extinction coefficients based on white dwarfs}",
      journal = {\mnras},
         year = 2019,
        month = nov,
       volume = {489},
       number = {4},
        pages = {5046-5052},
          doi = {10.1093/mnras/stz2506},
archivePrefix = {arXiv},
       eprint = {1909.02617},
 primaryClass = {astro-ph.SR},
       adsurl = {https://ui.adsabs.harvard.edu/abs/2019MNRAS.489.5046W}
}

@ARTICLE{Yuan2013,
       author = {{Yuan}, H.~B. and {Liu}, X.~W. and {Xiang}, M.~S.},
        title = "{Empirical extinction coefficients for the GALEX, SDSS, 2MASS and WISE passbands}",
      journal = {\mnras},
         year = 2013,
        month = apr,
       volume = {430},
       number = {3},
        pages = {2188-2199},
          doi = {10.1093/mnras/stt039},
archivePrefix = {arXiv},
       eprint = {1301.1427},
 primaryClass = {astro-ph.SR},
       adsurl = {https://ui.adsabs.harvard.edu/abs/2013MNRAS.430.2188Y}
}

@ARTICLE{Makarova2026,
       author = {{Makarova}, Lidia N. and {Purytin}, Denis G. and {Tully}, R. Brent and {Anand}, Gagandeep S.},
        title = "{The Metal Content of Resolved Galaxies}",
      journal = {arXiv e-prints},
         year = 2026,
        month = apr,
          eid = {arXiv:2604.18183},
        pages = {arXiv:2604.18183},
          doi = {10.48550/arXiv.2604.18183},
archivePrefix = {arXiv},
       eprint = {2604.18183},
 primaryClass = {astro-ph.GA},
       adsurl = {https://ui.adsabs.harvard.edu/abs/2026arXiv260418183M}
}

@ARTICLE{Sommovigo2025,
       author = {{Sommovigo}, Laura and {Algera}, Hiddo},
        title = "{Realistic multitemperature dust: how well can we constrain the dust properties of high-redshift galaxies?}",
      journal = {\mnras},
         year = 2025,
        month = jul,
       volume = {540},
       number = {4},
        pages = {3693-3708},
          doi = {10.1093/mnras/staf897},
archivePrefix = {arXiv},
       eprint = {2505.20105},
 primaryClass = {astro-ph.GA},
       adsurl = {https://ui.adsabs.harvard.edu/abs/2025MNRAS.540.3693S}
}

@ARTICLE{Nersesian2019,
       author = {{Nersesian}, A. and {Xilouris}, E.~M. and {Bianchi}, S. and {Galliano}, F. and {Jones}, A.~P. and {Baes}, M. and {Casasola}, V. and {Cassar{\`a}}, L.~P. and {Clark}, C.~J.~R. and {Davies}, J.~I. and {Decleir}, M. and {Dobbels}, W. and {De Looze}, I. and {De Vis}, P. and {Fritz}, J. and {Galametz}, M. and {Madden}, S.~C. and {Mosenkov}, A.~V. and {Tr{\v{c}}ka}, A. and {Verstocken}, S. and {Viaene}, S. and {Lianou}, S.},
        title = "{Old and young stellar populations in DustPedia galaxies and their role in dust heating}",
      journal = {\aap},
         year = 2019,
        month = apr,
       volume = {624},
          eid = {A80},
        pages = {A80},
          doi = {10.1051/0004-6361/201935118},
archivePrefix = {arXiv},
       eprint = {1903.05933},
 primaryClass = {astro-ph.GA},
       adsurl = {https://ui.adsabs.harvard.edu/abs/2019A&A...624A..80N}
}

@ARTICLE{Nersesian2026,
       author = {{Nersesian}, A. and {Xilouris}, E.~M. and {Bianchi}, S. and {Galliano}, F. and {Jones}, A.~P. and {Baes}, M. and {Casasola}, V. and {Cassar{\`a}}, L.~P. and {Clark}, C.~J.~R. and {Davies}, J.~I. and {Decleir}, M. and {Dobbels}, W. and {De Looze}, I. and {De Vis}, P. and {Fritz}, J. and {Galametz}, M. and {Madden}, S.~C. and {Mosenkov}, A.~V. and {Tr{\v{c}}ka}, A. and {Verstocken}, S. and {Viaene}, S. and {Lianou}, S.},
        title = "{Old and young stellar populations in DustPedia galaxies and their role in dust heating: (Corrigendum)}",
      journal = {\aap},
         year = 2026,
        month = feb,
       volume = {706},
          eid = {C2},
        pages = {C2},
          doi = {10.1051/0004-6361/202558721e},
       adsurl = {https://ui.adsabs.harvard.edu/abs/2026A&A...706C...2N}
}

@ARTICLE{Verstocken2020,
       author = {{Verstocken}, Sam and {Nersesian}, Angelos and {Baes}, Maarten and {Viaene}, S{\'e}bastien and {Bianchi}, Simone and {Casasola}, Viviana and {Clark}, Christopher J.~R. and {Davies}, Jonathan I. and {De Looze}, Ilse and {De Vis}, Pieter and {Dobbels}, Wouter and {Galliano}, Fr{\'e}d{\'e}ric and {Jones}, Anthony P. and {Madden}, Suzanne C. and {Mosenkov}, Aleksandr V. and {Tr{\v{c}}ka}, Ana and {Xilouris}, Emmanuel M.},
        title = "{High-resolution, 3D radiative transfer modelling. II. The early-type spiral galaxy M 81}",
      journal = {\aap},
         year = 2020,
        month = may,
       volume = {637},
          eid = {A24},
        pages = {A24},
          doi = {10.1051/0004-6361/201935770},
archivePrefix = {arXiv},
       eprint = {2004.03615},
 primaryClass = {astro-ph.GA},
       adsurl = {https://ui.adsabs.harvard.edu/abs/2020A&A...637A..24V}
}

@ARTICLE{Salpeter1955,
       author = {{Salpeter}, Edwin E.},
        title = "{The Luminosity Function and Stellar Evolution.}",
      journal = {\apj},
         year = 1955,
        month = jan,
       volume = {121},
        pages = {161},
          doi = {10.1086/145971},
       adsurl = {https://ui.adsabs.harvard.edu/abs/1955ApJ...121..161S}
}

@ARTICLE{Clark2026,
       author = {{Clark}, C.~J.~R. and {Verstocken}, S. and {Bianchi}, S. and {Fritz}, J. and {Viaene}, S. and {Smith}, M.~W.~L. and {Baes}, M. and {Casasola}, V. and {Cassara}, L.~P. and {Davies}, J.~I. and {De Looze}, I. and {De Vis}, P. and {Evans}, R. and {Galametz}, M. and {Jones}, A.~P. and {Lianou}, S. and {Madden}, S. and {Mosenkov}, A.~V. and {Xilouris}, M.},
        title = "{DustPedia: Multiwavelength photometry and imagery of 875 nearby galaxies in 42 ultraviolet-microwave bands (Corrigendum)}",
      journal = {\aap},
         year = 2026,
        month = aug,
       volume = {712},
          eid = {C4},
        pages = {C4},
          doi = {10.1051/0004-6361/202558613e},
       adsurl = {https://ui.adsabs.harvard.edu/abs/2026A&A...712C...4C}
}

@ARTICLE{Gordon2014,
       author = {{Gordon}, Karl D. and {Roman-Duval}, Julia and {Bot}, Caroline and {Meixner}, Margaret and {Babler}, Brian and {Bernard}, Jean-Philippe and {Bolatto}, Alberto and {Boyer}, Martha L. and {Clayton}, Geoffrey C. and {Engelbracht}, Charles and {Fukui}, Yasuo and {Galametz}, Maud and {Galliano}, Frederic and {Hony}, Sacha and {Hughes}, Annie and {Indebetouw}, Remy and {Israel}, Frank P. and {Jameson}, Katherine and {Kawamura}, Akiko and {Lebouteiller}, Vianney and {Li}, Aigen and {Madden}, Suzanne C. and {Matsuura}, Mikako and {Misselt}, Karl and {Montiel}, Edward and {Okumura}, K. and {Onishi}, Toshikazu and {Panuzzo}, Pasquale and {Paradis}, Deborah and {Rubio}, Monica and {Sandstrom}, Karin and {Sauvage}, Marc and {Seale}, Jonathan and {Sewi{\l}o}, Marta and {Tchernyshyov}, Kirill and {Skibba}, Ramin},
        title = "{Dust and Gas in the Magellanic Clouds from the HERITAGE Herschel Key Project. I. Dust Properties and Insights into the Origin of the Submillimeter Excess Emission}",
      journal = {\apj},
         year = 2014,
        month = dec,
       volume = {797},
       number = {2},
          eid = {85},
        pages = {85},
          doi = {10.1088/0004-637X/797/2/85},
archivePrefix = {arXiv},
       eprint = {1406.6066},
 primaryClass = {astro-ph.GA},
       adsurl = {https://ui.adsabs.harvard.edu/abs/2014ApJ...797...85G}
}

@ARTICLE{RemyRuyer2013,
       author = {{R{\'e}my-Ruyer}, A. and {Madden}, S.~C. and {Galliano}, F. and {Hony}, S. and {Sauvage}, M. and {Bendo}, G.~J. and {Roussel}, H. and {Pohlen}, M. and {Smith}, M.~W.~L. and {Galametz}, M. and {Cormier}, D. and {Lebouteiller}, V. and {Wu}, R. and {Baes}, M. and {Barlow}, M.~J. and {Boquien}, M. and {Boselli}, A. and {Ciesla}, L. and {De Looze}, I. and {Karczewski}, O. {\L}. and {Panuzzo}, P. and {Spinoglio}, L. and {Vaccari}, M. and {Wilson}, C.~D.},
        title = "{Revealing the cold dust in low-metallicity environments. I. Photometry analysis of the Dwarf Galaxy Survey with Herschel}",
      journal = {\aap},
         year = 2013,
        month = sep,
       volume = {557},
          eid = {A95},
        pages = {A95},
          doi = {10.1051/0004-6361/201321602},
archivePrefix = {arXiv},
       eprint = {1309.1371},
 primaryClass = {astro-ph.CO},
       adsurl = {https://ui.adsabs.harvard.edu/abs/2013A&A...557A..95R}
}

@ARTICLE{Chang2026,
       author = {{Chang}, Zhengxue and {Zhou}, Jianjun and {Lamperti}, Isabella and {Saintonge}, Am{\'e}lie and {Zeng}, Xiangyun and {Wang}, Shumin and {Esimbek}, Jarken and {Shi}, Ruimin and {Qiu}, Jianjie and {Zhao}, Na and {He}, Yuxin and {Li}, Qiang and {Li}, Linlin and {Huang}, Long and {Shen}, Hailiang and {Liu}, Xiaona},
        title = "{Cold Dust Properties and Submillimeter Excess in 33 Nearby Spiral Galaxies}",
      journal = {\apjs},
         year = 2026,
        month = mar,
       volume = {283},
       number = {1},
          eid = {15},
        pages = {15},
          doi = {10.3847/1538-4365/ae3750},
       adsurl = {https://ui.adsabs.harvard.edu/abs/2026ApJS..283...15C}
}

%%%%%%%%%%%%%%%%%%%%%%%%%%%%%%%%%%%%%%%%%%%%%%%%%%%%%%%%%%%%%%%%%%
%\appendix
%\section{Supplementary Tables}
%\onecolumn

%%%%%%%%%%%%%%%%%%%%%%%%%%%%%%%%%%%%%%%%%%%%%%%%%%%%%%%%%%%%%%%%%%
% Don't change these lines!
\bsp % typesetting comment
\label{lastpage}
\end{document}